\documentstyle[eqsecnum,prd,aps,epsf]{revtex}

\draft
\begin{document}

\newcommand{\beq}{\begin{equation}}
\newcommand{\eeq}{\end{equation}}
\newcommand{\beqn}{\begin{eqnarray}}
\newcommand{\eeqn}{\end{eqnarray}}
\newcommand{\pa}{\partial}
\newcommand{\vp}{\varphi}
\newcommand{\varep}{\varepsilon}
\newcommand{\ep}{\epsilon}

\twocolumn[\hsize\textwidth\columnwidth\hsize\csname
@twocolumnfalse\endcsname

\begin{center}
{\large\bf{Fully general relativistic simulation of 
coalescing binary neutron stars: Preparatory tests }}
~\\
~\\
Masaru Shibata \\
~\\
{\em Department of Physics, 
University of Illinois at Urbana-Champaign, Urbana, IL 61801, USA \\
{\rm and}\\
Department of Earth and Space Science,~Graduate School of
 Science,~Osaka University,\\
Toyonaka, Osaka 560-0043, Japan 
}\\
\end{center}

\begin{abstract}
~\\
We present our first successful 
numerical results of 3D general relativistic simulations 
in which the Einstein equation as well as the hydrodynamic 
equations are fully solved. 
This paper is especially devoted to simulations of 
test problems such as spherical dust collapse, 
stability test of perturbed spherical stars, and preservation of 
(approximate) equilibrium states of 
rapidly rotating neutron star 
and/or corotating binary neutron stars. 
These test simulations confirm that simulations of coalescing 
binary neutron stars are feasible in a numerical relativity code. 
It is illustrated that using our numerical code, 
simulations of these problems, in particular 
those of corotating binary neutron stars, can be performed 
stably and fairly accurately for a couple of dynamical 
timescales. 
These numerical results indicate that our formulation 
for solving the Einstein field equation and hydrodynamic 
equations are robust and 
make it possible to perform a realistic simulation of 
coalescing binary neutron stars for a long time 
from the innermost circular orbit up to formation of a black 
hole or neutron star. 
\end{abstract}
\pacs{04.25.Dm, 04.30.-w, 04.40.D}
\vskip2pc]

\baselineskip 5.8mm

\section{Introduction}

The coalescence of binary neutron stars is one of the 
most promising sources for planned kilometer size laser 
interferometers such as LIGO\cite{LIGO,KIP}, VIRGO\cite{VIRGO}, 
GEO\cite{GEO} and TAMA\cite{TAMA}, which will be in operation 
within the next five years. When a signal of 
gravitational waves is detected from 
binary neutron stars, it will provide not only 
the first chance to observe highly relativistic objects 
in a dynamical motion, but also a wide variety of physical 
information of binary neutron stars including 
their mass, spin, radius and 
innermost stable circular orbit (ISCO) \cite{KIP}. 
The signal of gravitational waves from such compact objects 
will be analyzed using matched filter techniques 
to extract the physical information. 
To apply this technique, 
theoretical templates of gravitational waveforms are needed \cite{KIP}.
This fact has motivated an intense theoretical effort for preparing 
such templates. 

The orbital evolution 
of binary neutron stars can be divided into three stages; 
inspiraling stage, intermediate stage, and dynamical stage. 
For the study of the inspiraling stage 
in which their orbital radius is 
much larger than the stellar radius ($R$) and neutron stars  
are in quasi-periodic states 
gradually decreasing the orbital radius as a result of 
emission of gravitational waves, 
a post Newtonian approximation is powerful. 
Much effort has been paid for obtaining a template 
of high-order post Newtonian corrections, 
providing many recent satisfactory results \cite{blanchet}. 

When the orbital radius of the binary 
neutron stars decreases to a few $R$ 
as a consequence of gravitational wave emission, 
the effect of the multipole moments of each neutron star induced 
by the tidal field from the companion star 
cannot be ignored. Even at this 
stage, the emission timescale of gravitational waves 
is still much longer than the orbital period. 
In this intermediate stage between the 
inspiraling and dynamical stages, the binary 
can be assumed to be in a quasi-hydrostatic equilibrium state. 
For the theoretical study, it is adequate to obtain the 
quasi-equilibrium configuration 
taking into account the effects of the deformation of the 
neutron stars but assuming that emission of gravitational waves 
is negligible. 
In this research field, an effort has been paid recently, 
yielding gradually successful results \cite{BGM,Uryu,MM}. 

With the further emission of gravitational waves, the binary 
neutron stars approach the ISCO. 
Upon reaching the ISCO, 
they behave in a dynamical manner. 
For the theoretical study of such a dynamical stage, 
no approximation for general relativity is applicable 
because the system is dominated by highly general 
relativistic gravity and hydrodynamic effect. 
In this respect, numerical relativistic simulation is the only 
promising method for the theoretical study. 

Numerical relativity also plays an important role for 
a theoretical study on 
the origin of $\gamma$-ray bursts (GRBs) because 
the short rise times of the bursts imply that their 
central sources have to be relativistic objects \cite{piran}. Recently, 
at least some of GRBs have been turned out to be of cosmological 
origin \cite{grb}. In cosmological GRBs, 
the central sources must provide 
a large amount of the energy $\agt 10^{51}$ ergs in a very short timescale 
of order msec--sec. 
It has been suggested that 
the merger of binary neutron stars could be a likely candidate 
for the powerful central source \cite{piran}. 
The typical scenario is 
based on the assumption that a system composed of a rotating 
black hole and a surrounding 
massive disk is formed after the merger. 
To clarify whether such scenario is correct, 
numerical simulations have been performed including effects of 
emission by gravitational radiation and/or neutrino and 
adopting a realistic equation of state for the neutron stars 
\cite{ruffert}. 
So far, the results in the numerical simulations 
have not supported such scenario \cite{ruffert}; i.e., 
the evidence that 
the massive disk is formed around the black hole has 
not been found. However, 
all the simulations have been performed in the Newtonian or 
post Newtonian approximations. Needless to say, 
general relativistic effects can play 
a very important role in the mergers between two neutron stars. 
To obtain the true answer, therefore, 
a fully general relativistic simulation is necessary. 

Much effort has been paid toward constructing 
a reliable numerical relativity code which makes it 
possible to clarify 
the evolution of merging binary neutron stars and 
the gravitational waveform emitted by them. 
Several projects in the world such as 
those by Nakamura and Oohara \cite{supple,ON} and 
by Washington University group \cite{waimo} are 
in progress, but no satisfactory results 
have been reported yet.

To perform numerical simulation of coalescing 
binary neutron stars for a long time from 
the ISCO to formation of a black hole or new neutron star, 
it is necessary to choose appropriate gauge conditions 
which make it possible to perform the long-timescale simulation 
stably and to extract gravitational waves accurately. 
In a previous paper \cite{gw3p2}, we performed fully 
general relativistic simulations of coalescing binary clusters 
using collisionless particles as the matter source of the Einstein  
equation. For the simulation, we used the 
approximate minimum distortion gauge and approximate maximal 
slicing conditions as the spatial and time 
coordinate conditions (see Sec II.B). 
We found that these gauge conditions are robust enough to allow 
for stable and long-timescale simulations of merging clusters 
as well as for the fairly accurate extraction of gravitational waves. 
In this paper, we perform simulations adopting 
the same gauge conditions and formulation for 
the Einstein equation and incorporating a solver of 
the relativistic hydrodynamic equations. 
We demonstrate the robustness of our formulation 
presenting successful numerical results 
of 3D hydrodynamic simulations. 

In particular, this paper is devoted to simulations of test 
problems. The purpose here is to ensure that numerical 
simulations of coalescing binary neutron stars for a long 
timescale from the ISCO to the formation of a black hole or neutron star 
are feasible in our formulation and numerical code. 
The test problems presented in this paper are a 
spherical dust collapse, stability test of 
spherical stars in equilibrium states, excitation of 
quadrupole oscillations of perturbed spherical stars, 
preservation of stable rapidly rotating stars in 
equilibrium states, and preservation of  
corotating binary neutron stars in an 
approximate quasi-equilibrium orbit. 
We show that stable and fairly accurate 
simulations for these test problems are feasible with our code, 
indicating the feasibility of forthcoming realistic simulations of 
coalescing binary neutron stars.

The paper is organized as follows. In Sec. II, 
our formulation for solving the Einstein field equations 
as well as relativistic hydrodynamic equations are described. 
We also describe the gauge conditions and  
the numerical method employed in this paper briefly. 
In Sec. III, we describe test problems which should 
be carried out to check the accuracy and performance of 
a numerical relativity code for solving the coupled equations 
composed of the Einstein and hydrodynamic equations.  
Then, we present the numerical results for the test simulations. 
In Sec. IV, we present numerical results of merger 
of corotating binary neutron stars as an example. 
Sec. V is devoted to a summary. 
Throughout this paper, we adopt the units $G=c=M_{\odot}=1$ 
where $G$, $c$ and $M_{\odot}$ denote the gravitational 
constant, speed of light and the solar mass, respectively. 
Hence, the units of length, time, mass, and density are 
$1.477$km, $4.927\times 10^{-6}$sec, $1.989 \times 10^{33}$g, and 
$6.173 \times 10^{17}{\rm g/cm}^3$. 
Latin and Greek indices denote spatial components ($1-3$) 
and spacetime components ($0-3$), respectively. 
As spatial coordinates, we use the Cartesian coordinates 
$x^k=(x, y, z)$ with $r=\sqrt{x^2+y^2+z^2}$. 
$\delta_{ij}(=\delta^{ij})$ denotes the Kronecker delta. 

\section{Formulation}
\subsection{Basic equations}

Our code solves the coupled equations of the Einstein equation and 
relativistic hydrodynamic equations. 
Our formulation for solving the Einstein equation has been 
described in detail in previous papers \cite{SN,gw3p,gw3p2}. 
Since we adopt the same formulation only changing the matter source, 
we here only review the basic equations.

We write the line element in the form 
\beqn
ds^2&=&g_{\mu\nu}dx^{\mu}dx^{\nu} \nonumber \\
&=&(-\alpha^2+\beta_k\beta^k)dt^2
+2\beta_i dx^i dt+\gamma_{ij}dx^i dx^j ,
\eeqn
where $g_{\mu\nu}$, $\alpha$, 
$\beta^i~(\beta_i=\gamma_{ij}\beta^j)$, 
and $\gamma_{ij}$ are the 4D metric, 
lapse function, shift vector, and 3D spatial metric, respectively. 
Following previous papers \cite{SN,gw3p,gw3p2}, 
we define the quantities to be solved in numerical computation as 
\beqn
&& \gamma={\rm det}(\gamma_{ij}) \equiv e^{12\phi},\\
&& \tilde \gamma_{ij} \equiv e^{-4\phi}\gamma_{ij} 
~({\rm i.e.},~{\rm det}(\tilde \gamma_{ij})=1), \\
&& \tilde A_{ij} \equiv 
e^{-4\phi} \Bigl(K_{ij}-{1 \over 3} \gamma_{ij} K_k^{~k} \Bigr),
\eeqn
where $K_{ij}$ is the extrinsic curvature, $K_k^{~k}$ its trace. 
We note that indices of $\tilde A_{ij}$ and/or $\tilde A^{ij}$ are 
raised and lowered in terms of $\tilde \gamma_{ij}$ and 
$\tilde \gamma^{ij}$. In the numerical computation, we solve 
for $\tilde \gamma_{ij}$, $\tilde A_{ij}$, $\phi$ and $K_k^{~k}$  
instead of $\gamma_{ij}$ and $K_{ij}$. 
Hereafter, we use $\nabla_{\mu}$, $D_i$ and $\tilde D_i$ 
as the covariant 
derivatives with respect to $g_{\mu\nu}$, $\gamma_{ij}$ and 
$\tilde \gamma_{ij}$, respectively. 

As the matter source of the Einstein equation, 
we adopt a perfect fluid. In this 
case, the energy momentum tensor is written as 
\beq
T_{\mu\nu}=(\rho+\rho\varep+P)u_{\mu}u_{\nu}+P g_{\mu\nu},
\eeq
where $\rho$, $\varep$, $P$, and $u_{\mu}$ are 
the rest mass density, specific internal energy density, 
pressure and four-velocity, respectively. 
Hereafter, we assume an equation of state in the form, 
$P=(\Gamma-1)\rho\varep$, where $\Gamma$ is a constant. 

The hydrodynamic equations are composed of the 
continuity, Euler and energy (or entropy) 
equations, which are derived from  
\beqn
&&\nabla_{\mu} (\rho u^{\mu})=0,\\
&&\gamma_i^{~\mu} \nabla_{\nu} T^{~\nu}_{\mu}=0,\\
&&u^{\mu} \nabla_{\nu} T^{~\nu}_{\mu}=0. 
\eeqn
We write their explicit forms as 
\beqn
&&\pa_t \rho_* + \pa_i (\rho_* v^i )=0,\label{eqrho}\\
&&\pa_t (\rho_* \hat u_k)+ \pa_i (\rho_* \hat u_k v^i ) 
\nonumber \\
&&~~=-\alpha e^{6\phi}\pa_k P 
-\rho_* \biggl[w h \pa_k \alpha - \hat u_j\pa_k \beta^j
\nonumber \\
&& \hskip 1cm 
+{\alpha e^{-4\phi} \hat u_i \hat u_j \over 2 w h} 
\pa_k \tilde \gamma^{ij}
-{2\alpha h (w^2-1) \over w} \pa_k \phi \biggr],\label{euler}\\
&&\pa_t e_* + \pa_i (e_* v^i )=0,\label{energy}
\eeqn
where $\pa_{\mu}=\pa/\pa x^{\mu}$, 
$\rho_*=\rho w e^{6\phi}$, $h=1+\varep+P/\rho$, $w=\alpha u^0$, 
$\hat u_k=h  u_k$, 
$e_*=(\rho\varep)^{1/\Gamma} w e^{6\phi}$, and 
$v^i (\equiv u^i/u^0)$ is written as 
\beq
v^i = -\beta^i + {\alpha \tilde \gamma^{ij} \hat u_j 
\over w h e^{4\phi}}. \label{eqvelo}
\eeq
In numerical simulation, we solve 
Eqs. (\ref{eqrho})--(\ref{energy}) 
to evolve $\rho_*$, $\hat u_k$ and $e_*$. 

The volume integral of $\rho_*$ in the three hypersurface, 
\beq
M_* \equiv \int d^3x \rho_*,
\eeq
denotes the total rest mass of the system, 
which should be conserved with time. 
Although the equation for $e_*$ has the same form as that for 
$\rho_*$, the volume integral of $e_*$ is not conserved 
in the presence of shocks. 
This implies that using Eq. (\ref{energy}) 
we cannot obtain the correct solution when shock is formed. 
Thus, the artificial viscosity terms are added in 
Eqs. (\ref{euler}) and (\ref{energy}) for some problems 
in which shock is formed and plays an important role 
during the evolution of the system (see Appendix A). 

Once $\hat u_i$ is obtained, $w (=\alpha u^0)$ 
is determined from the normalization relation of the four-velocity, 
which can be written as 
\beq
w^2=1+e^{-4\phi} \tilde \gamma^{ij} \hat u_i \hat u_j
\biggl[1+ {\Gamma e_*^{\Gamma} \over \rho_* (w e^{6\phi})^{\Gamma-1}}
\biggr]^{-2}.  \label{eqforw}
\eeq

The Einstein equation is split into the constraint 
and evolution equations. 
The Hamiltonian and momentum constraint equations are 
written as 
\beqn
&& R_k^{~k}- \tilde A_{ij} \tilde A^{ij}+
{2 \over 3} (K_k^{~k})^2=16\pi E,
\label{ham}\\
&& D_i \tilde A^i_{~j}-{2 \over 3}D_j K_k^{~k}=8\pi J_j, \label{mom}
\eeqn
where 
\beqn
&& E \equiv T^{\mu\nu}n_{\mu}n_{\nu} =\rho_* h w e^{-6\phi} - P,\\
&& J_i \equiv -T^{\mu\nu} n_{\mu}\gamma_{\nu i}
=\rho_* e^{-6\phi} \hat u_i, 
\eeqn
$n_{\mu}=(-\alpha, 0)$, and 
$R_{ij}$ is the Ricci tensor with respect to $\gamma_{ij}$. 

Following our previous works, 
we write the evolution equations for the geometric variables 
in the form \cite{gw3p,gw3p2} 
\beqn
&&(\pa_t - \beta^l \pa_l) \tilde \gamma_{ij} \nonumber \\
&&\hskip 5mm  =-2\alpha \tilde A_{ij} 
+\tilde \gamma_{ik} \beta^k_{~,j}+\tilde \gamma_{jk} \beta^k_{~,i}
-{2 \over 3}\tilde \gamma_{ij} \beta^k_{~,k}, \label{heq} \\
&&(\pa_t - \beta^l \pa_l) \tilde A_{ij} 
= e^{ -4\phi } \biggl[ \alpha \Bigl(R_{ij}
-{1 \over 3}e^{4\phi}\tilde \gamma_{ij} R_k^{~k} \Bigr) \nonumber \\
&& \hskip 2.5cm -\Bigl( D_i D_j \alpha - {1 \over 3}e^{4\phi}
\tilde \gamma_{ij} D_k D^k \alpha \Bigr)
\biggr] \nonumber \\
&& \hskip 2.5cm +\alpha (K_k^{~k} \tilde A_{ij} 
- 2 \tilde A_{ik} \tilde A_j^{~k}) \nonumber \\
&& \hskip 2.5cm +\beta^k_{~,i} \tilde A_{kj}+\beta^k_{~,j} 
\tilde A_{ki}
-{2 \over 3} \beta^k_{~,k} \tilde A_{ij} \nonumber \\
&& \hskip 2.5cm-8\pi\alpha \Bigl( 
e^{-4\phi} S_{ij}-{1 \over 3} \tilde \gamma_{ij} S_k^{~k}
\Bigr), \label{aijeq} \\
&&(\pa_t - \beta^l \pa_l) \phi = {1 \over 6}\Bigl( 
-\alpha K_k^{~k} + \beta^k_{~,k} \Bigr), \label{peq} \\
&&(\pa_t - \beta^l \pa_l) K_k^{~k} 
=\alpha \Bigl[ \tilde A_{ij} \tilde A^{ij}+{1 \over 3}(K_k^{~k})^2
\Bigr] \nonumber \\
&& \hskip 2.5cm -D_k D^k \alpha +4\pi \alpha (E+ S_k^{~k}), 
\label{keq}
\eeqn
where $Q_{,i}=\pa_i Q$ for an arbitrary variable $Q$, and 
\beq
S_{ij} \equiv T^{\mu\nu}\gamma_{\mu i}\gamma_{\nu j}
=\rho_* e^{-6\phi} 
(w h)^{-1} \hat u_i \hat u_j + e^{4\phi}\tilde \gamma_{ij} P. 
\eeq
In calculating $R_{ij}$ and $R_k^{~k}$ 
in Eq. (\ref{aijeq}), we have terms of the type as 
$\delta^{kl} \tilde \gamma_{ik,lj}$ and 
$\delta^{kl}\tilde \gamma_{jk,li}$. 
For evaluation of such terms, 
we define the auxiliary variable 
$F_i=\delta^{jl}\pa_l \tilde \gamma_{ij}$ \cite{Nakamura,gw3p,gw3p2} 
and solve the evolution equation 
\beqn
(\pa_t - \beta^l \pa_l)F_i&& = 2\alpha 
\Bigl\{ f^{kj} \tilde A_{ik,j}
+f^{kj}_{~~,j} \tilde A_{ik} \nonumber \\
&&~-{1 \over 2} \tilde A^{jl} h_{lj,i} 
+6\phi_{,k} \tilde A^k_{~i}-{2\over 3}(K_k^{~k})_{,i} \Bigr\} 
\nonumber \\
&&~-2\delta^{jk} \alpha_{,k} \tilde A_{ij} 
+ \delta^{jl} \beta^k_{~,l}h_{ij,k} \nonumber \\
&&~+(\tilde \gamma_{il}\beta^l_{~,j}+\tilde \gamma_{jl}\beta^l_{~,i}
-{2\over 3}\tilde \gamma_{ij} \beta^l_{~,l})_{,k}\delta^{jk} \nonumber \\
&&~-16\pi \alpha J_i,
\eeqn
where $h_{ij}=\tilde \gamma_{ij}-\delta_{ij}$, and 
$f^{ij}=\tilde \gamma^{ij}-\delta^{ij}$. 
Then, $\delta^{kl}\tilde \gamma_{ik,lj}$ is evaluated as $F_{i,j}$. 

We define the total angular momentum of 
the system as
\beqn
J \equiv && \lim_{r \rightarrow \infty} 
{1 \over 8\pi} \oint (x\tilde A_{yj} -y \tilde A_{xj})e^{6\phi} dS^j
\nonumber \\
=&& \int d^3x e^{6\phi} \Bigl[  x J_y-y J_x
+{1 \over 8\pi}(\tilde A^x_{~y}-\tilde A^y_{~x}) \nonumber \\
&&\hskip 2cm 
-{1 \over 16\pi}\tilde A_{ij} (x\pa_y-y\pa_x)\tilde \gamma^{ij}
\nonumber \\
&&\hskip 2cm 
+{1 \over 12\pi} (x\pa_y-y\pa_x)K_k^{~k} \Bigr], \label{eqj}
\eeqn
where we use the Gauss's law and 
Eq. (\ref{mom}) to derive the final expression. 
We also use the following quantity to roughly estimate the 
angular momentum inside a coordinate radius $r$ as 
\beq
J(r)=\int_{|x^i| < r}  d^3x e^{6\phi} (x J_y - y J_x). 
\eeq

\subsection{Gauge conditions}

As in a previous paper \cite{gw3p2}, we 
adopt an approximate maximal slice (AMS) condition and 
an approximate minimum distortion (AMD) gauge condition as 
the time and spatial gauge conditions, respectively, 
for most of the simulations in this paper. 
In some test simulations carried out 
in Sec. III, we also use the zero shift 
vector gauge condition $\beta^k=0$ for comparison between 
two results obtained in different gauge conditions.

To impose the AMS condition, we solve the following 
parabolic type equation for $\ln \alpha$ 
at each time step until an approximate convergence is achieved:  
\beqn
\pa_{\lambda} \ln \alpha && = 
D_k D^k \ln \alpha + (D_k \ln \alpha) (D^k \ln \alpha) 
-4\pi (E+S_k^{~k}) \nonumber \\
&&~- \tilde A_{ij} \tilde A^{ij} -
{1 \over 3}(K_k^{~k})^2
+f_{\alpha} K_k^{~k} \rho_*^{1/2}.\label{eqalp2}
\eeqn
Here $\lambda$ denotes a control parameter and 
$f_{\alpha}$ is a constant for which we assign a constant of $O(1)$. 
Assuming that the convergence is achieved and that 
the right-hand side of Eq. (\ref{eqalp2}) becomes zero, 
the evolution equation for $K_k^{~k}$ can be written as 
\beq
(\pa_t - \beta^l \pa_l) K_k^{~k} =
-f_{\alpha} \alpha K_k^{~k} \rho_*^{1/2}. \label{Keq}
\eeq
Thus, if $K_k^{~k}$ is zero initially and the 
convergence is completely achieved, the 
the maximal slice condition $K_k^{~k}=0$ is preserved. 
Even when the convergence is incomplete 
and $K_k^{~k}$ deviates from zero, 
the right-hand side of Eq. (\ref{Keq}) 
enforces $|K_k^{~k}|$ to approach to zero in the local dynamical 
timescale $\sim \rho_*^{-1/2}$. 
Hence, the condition $K_k^{~k}=0$ is expected to 
be satisfied approximately. 

To impose the AMD gauge condition, we solve the following 
simple elliptic type equations 
\beqn
&&\Delta P_i = S_i, \\
&&\Delta \eta= -S_i x^i, 
\eeqn
where $\Delta$ denotes the Laplacian in the flat 3D space, and 
\beq
S_i\equiv 16\pi\alpha J_i 
+2\tilde A_{ ij} (\tilde D^j \alpha - 6\alpha \tilde D^j \phi)
+{4 \over 3}\alpha \tilde D_i K_k^{~k}.
\eeq
{}From $P_i$ and $\eta$, we determine $\beta^i$ as 
\beq
\beta^j=\delta^{ji}\biggl[
{7 \over 8}P_i - {1 \over 8}(\eta_{,i}+P_{k,i} x^k) \biggr]. 
\eeq
Namely, $\beta^i$ satisfies an elliptic type equation
in the form 
\beq
\delta_{ij} \Delta \beta^i + {1 \over 3} \beta^k_{~,kj}=S_j. 
\eeq

As we described in a previous paper \cite{gw3p2}, 
if an action 
\beq
I=\int d^3x (\pa_t {\tilde \gamma_{ij}}) 
(\pa_t {\tilde \gamma_{kl}})
\tilde \gamma^{ik} \tilde \gamma^{jl} . 
\eeq
is minimized with respect to $\beta^i$, 
we obtain the equation of a minimum distortion (MD) gauge 
condition \cite{SY} for $\beta^i$ as
\beq
\tilde \gamma_{jk}  \tilde D^i \tilde D_i \beta^k
+{1 \over 3} \tilde D_j \tilde D_i \beta^i + \tilde R_{jk} \beta^k
=S_j, \label{eqMD}
\eeq
where $\tilde R_{jk}$ is the Ricci tensor with respect to 
$\tilde \gamma_{ij}$. 
Thus, the equation for $\beta^i$ in 
the AMD gauge condition is obtained by neglecting coupling terms 
between $\beta^i$ and $h_{ij}$ in Eq. (\ref{eqMD}). 
Since the neglected terms are expected to be small 
\cite{gw3p2}, we can expect that $I$ 
is approximately minimized in the AMD gauge condition. 

The other benefit in the AMD gauge condition is that 
$F_i$ is guaranteed to be small everywhere except in 
the strong field region just around a highly 
relativistic object \cite{gw3p2}. This implies that 
a transverse condition, 
$\delta^{ij} \pa_i \tilde \gamma_{jk} = 0$, approximately holds 
for $\tilde \gamma_{ij}$  in the wave zone, helping 
the accurate extraction of gravitational waves near the 
outer boundaries of the computational domain. 

\subsection{Initial value formalism}

Initial conditions are obtained by solving the constraint 
equations (\ref{ham}) and (\ref{mom}). 
In this paper, we restrict our attention to
initial conditions in which 
$h_{ij}(=\tilde \gamma_{ij}-\delta_{ij})=0$ and $K_k^{~k}=0$. 
Then, the basic equations for obtaining the initial 
data are the same as those in \cite{gw3p,gw3p2} 
as described below. 

Using the conformal factor $\psi \equiv e^{\phi}$, 
$\hat A_{ij}=\psi^6 \tilde A_{ij}$ and 
$\hat A^{ij}=\psi^6 \tilde A^{ij}$, the 
Hamiltonian and momentum constraint equations are rewritten 
in the form
\beqn
&& \Delta \psi = -2\pi E \psi^5 -{1 \over 8}
\hat A_{ij} \hat A^{ij}\psi^{-7}, \label{ham2} \\
&& \hat A^{~j}_{i~,j} = 8\pi J_i \psi^6.\label{mom2}
\eeqn
After we decompose $\hat A_{ij}$ in the standard manner as 
\beq
\hat A_{ij}=W_{i,j}+W_{j,i}-{2 \over 3}\delta_{ij} \delta^{kl}
W_{k,l},
\eeq
we set $W_i$ as \cite{gw3p,gw3p2}
\beq
W_i={7 \over 8}B_i - {1 \over 8}(\chi_{,i}+B_{k,i} x^k), 
\eeq
where $\chi$ and $B_i$ denote scalar and vector functions.  
Then, Eq. (\ref{mom2}) can be decomposed into 
two simple elliptic type equations
\beqn
&&\Delta B_i = 8\pi J_i \psi^6, \nonumber \\
&&\Delta \chi= -8\pi J_i x^i \psi^6.
\eeqn
Since $J_i \psi^6 (=\rho_* \hat u_i)$ is non-zero only in 
the strong field region,  
the solution of the momentum constraint equation is accurately 
obtained. 

In addition to the constraint equations, we solve an elliptic 
type equation for $\alpha$ 
to impose the maximal slice condition initially. 
In the conformally flat 3D space, the equation is written in the form
\beq
\Delta (\alpha\psi) = 2\pi \alpha \psi^5 (E+2 S_k^{~k})
+{7 \over 8}\alpha \psi^{-7}\hat A_{ij} \hat A^{ij} .
\eeq

{}From the initial condition, the total 
gravitational mass and angular momentum 
of the system at $t=0$ are calculated from 
\beqn
&&(M_g)_0=\int d^3x \biggl( E \psi^5 +{1 \over 16\pi\psi^7}
\hat A_{ij} \hat A^{ij}\biggr), \\
&&J_0=\int d^3x (x J_y - y J_x)\psi^6. \label{eqj0}
\eeqn
Note that Eq. (\ref{eqj}) reduces to Eq. (\ref{eqj0}) 
in the 3D space in which 
$\tilde \gamma_{ij}=\delta_{ij}$ and $K_k^{~k}=0$. 

\subsection{Boundary conditions}

In this paper, we assume $\pi$-rotation symmetry around 
the $z$-axis as well as 
a plane symmetry with respect to the $z=0$ plane. 
Hence, we solve equations in 
a quadrant region $L \geq x \geq -L$ and $L\geq y,~z \geq 0$ 
where $L$ denotes the location of the outer boundaries. 
We impose the boundary condition in the $y=0$ plane such as 
\beqn
Q(x,0,z)&&=Q(-x,0,z),\\
Q^A(x,0,z)&&=-Q^A(-x,0,z), \nonumber \\
Q_A(x,0,z)&&=-Q_A(-x,0,z),\\
Q^z(x,0,z)&&=Q^z(-x,0,z),\nonumber \\
Q_z(x,0,z)&&=Q_z(-x,0,z),\\
Q_{AB}(x,0,z)&&=Q_{AB}(-x,0,z),\\
Q_{Az}(x,0,z)&&=-Q_{Az}(-x,0,z),\\
Q_{zz}(x,0,z)&&=Q_{zz}(-x,0,z),
\eeqn
where $A$ and $B=x$ or $y$, and $Q$, $Q^i(Q_i)$, and $Q_{ij}$ 
denote arbitrary 
scalar, vector and tensor quantities, respectively. 

At the outer boundaries, we impose an approximate outgoing 
boundary condition for $h_{ij}$ and $\tilde A_{ij}$ \cite{gw3p} 
as 
\beq
r Q_{ij}(u)={\rm const},\label{eqboud}
\eeq
where we set $u=\alpha t - e^{2\phi} r$. 
(Even if we simply set $u = t - r$, 
the results do not change significantly.) 
More explicitly, Eq. (\ref{eqboud}) is rewritten in the form 
\beq
Q_{ij}(t,r)=\biggl(1-{\delta r \over r}\biggr) 
Q_{ij}(t-\delta t,r-\delta r), \label{outer}
\eeq
where $\delta t$ is a time step, and 
$\delta r = \alpha e^{-2\phi} \delta t$. 
$Q_{ij}(t-\delta t,r-\delta r)$ is linearly 
interpolated from the nearby 8 grid points. 
We also note that the numerical results are not sensitive to the 
boundary condition of $\tilde A_{ij}$ and 
we have not found significant change in the numerical 
results even when we impose the boundary condition 
with the radial fall-off as $O(r^{-3})$. 
A possible explanation for this result is that the spatial derivative of 
$\tilde A_{ij}$, which appears in the evolution equations 
for $\tilde A_{ij}$ and $F_i$, 
does not play an important role for evolution of the system. 
On the other hand, numerical results and stability of 
the numerical system are significantly dependent of 
the boundary condition for $\tilde \gamma_{ij}$. 
The condition defined by Eq. (\ref{outer}) is 
one of the best conditions among those we have tried so far.

For $\phi$ and $K_k^{~k}$, 
we impose the following boundary conditions 
at the outer boundaries 
\beqn
&&(r\phi )_{,r}=0, \\
&&K_k^{~k}=0.
\eeqn
For $F_i$, we have tried a number of boundary conditions such as 
$F_i=O(r^{-3})$, $F_i=0$, and $\pa_j F_i=$const at $x_j=L$, 
and have found that the results are weakly dependent of the 
boundary condition. 
We have found that the condition $F_i=0$ 
is preferable for a long-timescale numerical evolution, 
but for extracting gravitational waves near the outer 
boundaries, the condition $F_i=O(r^{-3})$ or $\pa_j F_i=$const 
is preferable. In the spherical cases, we use 
the condition $F_i=0$, but we change 
the condition case by case for other cases. 

It should be noted that all the outer boundary conditions 
described above are only approximate. This implies that 
numerical errors such as spurious back reflection and 
incoming of gravitational waves 
will be generated near the outer boundaries. 
For precise numerical simulation, apparently, 
we have to adopt more sophisticated outer boundary conditions 
as have been proposed and 
investigated by a couple of groups \cite{bishop}. 
Imposing precise outer boundary conditions is 
one of important future issues.

\subsection{Grid and time step}

Throughout this paper, we use a uniform grid, i.e., 
$\delta x=\delta y =\delta z=$const.
We take $(2N+1, N+1, N+1)$ grid points in $(x,y,z)$ direction, 
respectively (i.e., $N=L/\delta x$). 

The time step $\delta t$ must satisfy the stability condition 
restricted by the Courant criterion for geometric variables. 
If we neglect the other two directions, 
the geometric Courant condition in the $x^i$ direction 
is written as 
\beq
\delta t < [\alpha \gamma_{ii}^{-1/2}+\beta^i]^{-1} \delta x. 
\eeq
Since $[\alpha \gamma_{ii}^{-1/2}+\beta^i]^{-1}$ is expected to 
be greater than unity for most cases, 
we simply set a geometric time step as 
\beq
\delta t_g =C_g \delta x,
\eeq
where $C_g$ is a constant which we choose typically as $0.3$. 
Also, $\delta t$ must be sufficiently small so that 
the matter distribution cannot change by a large fraction amount 
in one time step. The timescale is shortest when 
the matter distribution changes in a dynamical timescale as in 
the case when the matter collapses to a singularity. 
Thus, we simply set
\beq
\delta t= {\rm min}( C_m \sqrt{{3\pi \over 32 \rho_*}},\delta t_g), 
\label{eqdt}
\eeq
where $C_m$ is a constant for which we choose $0.02-0.04$. 
Note that the first term in the right-hand side of Eq. (\ref{eqdt}) 
denotes the time for the collapse to a singularity of a spherical, 
homogeneous dust in the Newtonian limit. 

In the case when the density is so high that a black hole seems 
to be formed, the first term of Eq. (\ref{eqdt}) is smaller than 
the second term, but besides such a highly relativistic case, 
the second term is smaller and determines the time 
step. We note that the hydrodynamic Courant condition is 
less severe than the geometric one \cite{SP}, 
so that we do not consider it.

\subsection{Brief summary of numerical methods and 
evolution scheme}

For solving the equations for geometric variables as well as 
for determining the apparent horizon, 
the same methods as employed 
in previous papers \cite{gw3p,gw3p2} are used here. 
The numerical method for solving the hydrodynamic equations is 
briefly discussed in Appendix A. 

The evolution scheme for the geometric and fluid variables 
from a time slice at $t$ to the next time 
slice at $t+\delta t$ is as follows 
(see the schematic figure 1): 
We put $\tilde \gamma_{ij}$, $F_i$, $\phi$, 
$\alpha$, $\beta^k$, $\rho_*$, $\hat u_i$, $v^i$, and $e_*$ 
on $t^{(0)}$, $t^{(1)},\cdots, t^{(n)}$ and 
$\tilde A_{ij}$ and $K_k^{~k}$ 
on $t^{(-1/2)}$, $t^{(1/2)}$, $t^{(3/2)},\cdots$, 
and $t^{(n-1/2)}$, where $t^{(n)}$ denotes the coordinate 
time at the $n$-th time step, 
$t^{(n+1/2)}\equiv (t^{(n)}+t^{(n+1)})/2$, 
and $n$ is a positive integer. Namely, we use the leapfrog 
method \cite{recip} for evolution of the geometric variables. 

For a given set of the geometric variables 
$\tilde \gamma_{ij}$, $F_i$, $\phi$, 
$\alpha$, $\beta^k$ and fluid variables 
$\rho_*$, $\hat u_i$, $v^i$, and $e_*$ at $t^{(n)}$, and 
other geometric variables $\tilde A_{ij}$ and $K_k^{~k}$ 
at $t^{(n-1/2)}$ (the stage (1) of Fig. 1), first, 
$\tilde A_{ij}$ and $K_k^{~k}$ are 
evolved from $t^{(n-1/2)}$ to 
$t^{(n+1/2)}$ using the evolution equations (\ref{aijeq}) and 
(\ref{keq}) (the stage (2) of Fig. 1). 
Since the right-hand side of these evolution equations 
includes $\tilde A_{ij}$ and $K_k^{~k}$ which are 
defined only at $t^{(n-1/2)}$, we use the 
linear extrapolation method at each spatial grid point as 
\beqn
(K_k^{~k})^{(n)}&&={3 \over 2}(K_k^{~k})^{(n-1/2)}
-{1 \over 2}(K_k^{~k})^{(n-3/2)},
\nonumber \\
(\tilde  A_{ij})^{(n)}&&={3 \over 2}(\tilde A_{ij})^{(n-1/2)}
-{1 \over 2}(\tilde A_{ij})^{(n-3/2)}, \nonumber 
\eeqn
to preserve the second order accuracy in time. 
We note that in adopting the linear 
extrapolation, we implicitly make use of the fact 
that the time steps at $t^{(n-1)}$ 
and $t^{(n)}$ are approximately equal. 

Once we obtain $(K_k^{~k})^{(n+1/2)}$ and 
$(\tilde  A_{ij})^{(n+1/2)}$, 
the geometric variables $\tilde \gamma_{ij}$, 
$\phi$ and $F_i$ are 
evolved to $t^{(n+1)}$ (the stage (3) of Fig. 1). 
We also use the extrapolation method such as 
\beqn
\alpha^{(n+1/2)}&&={3 \over 2} \alpha^{(n)}
-{1 \over 2} \alpha^{(n-1)},\nonumber \\
(\beta^{i})^{(n+1/2)}&&={3 \over 2}(\beta^i)^{(n)}-{1 \over 2}
(\beta^i)^{(n-1)},\nonumber 
\eeqn
because they are necessary to preserve the second 
order accuracy in time. 

Next, the hydrodynamic equations for 
$\rho_*$, $\hat u_i$, and $e_*$ are evolved to 
$t^{(n+1)}$. For solving the evolution equations, 
we use a Runge-Kutta method of second order accuracy \cite{recip}; 
i.e., the hydrodynamic equations are solved from $t^{(n)}$ to 
$t^{(n+1/2)}$ in the first step (the stage (4) of Fig. 1), 
and using the fluid 
variables defined both at $t^{(n)}$ and at $t^{(n+1/2)}$, 
those at $t^{(n+1)}$ are subsequently 
obtained in the second step (the stage (5) of Fig. 1). 
Since there appear 
$\tilde \gamma_{ij}$, $\phi$, $\alpha$, $\beta^i$, $w$, $h$, 
and $P$ in the right-hand side of the relativistic 
Euler equation (\ref{euler}) and we need those at $t^{(n+1/2)}$ in 
the second step for solving the equation, we 
use the extrapolation and interpolation, 
\beqn
\alpha^{(n+1/2)}&&={3 \over 2} \alpha^{(n)}
-{1 \over 2} \alpha^{(n-1)},\nonumber \\
(\beta^{i})^{(n+1/2)}&&={3 \over 2}(\beta^i)^{(n)}-{1 \over 2}
(\beta^i)^{(n-1)},\nonumber \\
\phi^{(n+1/2)}&&={1 \over 2}[\phi^{(n+1)}+
\phi^{(n)}],\nonumber \\
(\tilde \gamma_{ij})^{(n+1/2)}&&={1 \over 2}
[(\tilde \gamma_{ij})^{(n+1)}
+(\tilde \gamma_{ij})^{(n)}], \nonumber \\
w^{(n+1/2)}&&={3 \over 2} w^{(n)}-{1 \over 2}w^{(n-1)}.\nonumber
\eeqn
$P$ and $h$ at $t^{(n+1/2)}$ are calculated from 
$\rho_*$, $e_*$, $w$ and $\phi$ at $t^{(n+1/2)}$ obtained 
in the first step. 
(Note that $w^{(n+1)}$, 
$\alpha^{(n+1)}$ and $(\beta^{i})^{(n+1)}$ have 
not yet been obtained by this stage.) 
Once the fluid variables are evolved to 
$t^{(n+1)}$, Eq. (\ref{eqforw}) is 
solved at each spatial grid point for obtaining $w^{(n+1)}$ 
using the Newton-Raphson method \cite{recip}. Subsequently, we 
can obtain $P$ and $h$ at $t^{(n+1)}$. 

In the final step, 
we determine $\alpha^{(n+1)}$ and $(\beta^i)^{(n+1)}$ 
by imposing the gauge conditions (the stage (6) of Fig. 1). 
In solving their equations, we again use the extrapolation as 
\beqn
(K_k^{~k})^{(n+1)}&&={3 \over 2}(K_k^{~k})^{(n+1/2)}
-{1 \over 2}(K_k^{~k})^{(n-1/2)},
\nonumber \\
(\tilde  A_{ij})^{(n+1)}&&={3 \over 2}(\tilde A_{ij})^{(n+1/2)}-
{1 \over 2}(\tilde A_{ij})^{(n-1/2)}. \nonumber 
\eeqn
Since all the quantities are evolved by this stage, we 
can derive $v^i$ at $t^{(n+1)}$ using Eq. (\ref{eqvelo}) 
without any extrapolation.

\section{Tests and results}

Our current priority in numerical relativity is to perform 
simulations of the merger of binary neutron stars. 
Before carrying out such simulations successfully, 
it is necessary to confirm the accuracy and performance 
of our numerical code for many different problems. 
In particular, the following issues have to 
be addressed: 
{({\it i})} the merger will take place 
for a couple of orbital periods 
from the time when the binary just enters inside the ISCO 
to formation of a black hole or neutron star. 
Can we carry out the simulation 
stably for a couple of orbital periods ? 
{({\it ii})} the final product of the 
merger will either be a black hole or a neutron star. If the merged object 
is unstable against gravitational collapse, 
a black hole is formed. Can we 
judge the stability of the merged object against 
the gravitational collapse ? 
{({\it iii})} the formation of a black hole will be signaled by 
the appearance of an apparent horizon. Can we 
determine the apparent horizon during the simulations ? 
{({\it iv})} can we extract waveforms of gravitational 
waves ? 

To answer these questions, 
we have performed simulations for a wide variety of test problems: 
\begin{enumerate}
\item Spherical collapse of dust ($P=0$) to a 
black hole: We compare 
the results with those obtained in a 1D (spherical symmetric) 
simulation. We also check 
whether the apparent horizon can be found at a correct time 
and location. This test confirms that 
our code can simulate the formation of a black hole accurately. 
\item Stability of spherical stars: 
We check that the stability of spherical 
stars can be judged in our code 
preparing both stable and unstable stars as initial conditions.  
We also check whether our code can provide a correct output 
on the fundamental radial oscillation frequency of 
perturbed spherical stars. 
This test is useful to confirm that 
we can determine the stability of the merged object 
against gravitational collapse. 
\item Quadrupole oscillations of 
perturbed spherical stars and emission of gravitational waves: 
We give a quadrupole perturbation to a spherical 
stable star, and check whether we can obtain the frequency  
of the fundamental mode ($f$-mode) oscillation and extract 
the waveform of gravitational waves near the outer boundaries. 
This test is useful to confirm that the 
extraction of gravitational waves near the outer boundaries 
is feasible. 
\item Preservation of rapidly rotating stars in equilibrium states: 
We check that 
(approximate) equilibrium states of rapidly rotating, stable 
stars can be preserved for a couple of the rotation periods. 
The simulations are carried out choosing rigidly and 
rapidly rotating stars at mass-shedding limits. 
This test is useful to confirm that the coordinate twisting 
due to the rotation of the stars 
is sufficiently suppressed to a level 
adequate to carry out stable and accurate simulations 
in our AMD gauge condition. 
\item Preservation of a 
corotating binary neutron star in a quasi-equilibrium state: 
We prepare a mildly relativistic corotating binary neutron star 
in an approximate quasi-equilibrium state 
obtained assuming a conformally flat 3D geometry \cite{bcsst}. 
Although we ignore $h_{ij}$ which 
is necessary to obtain a true quasi-equilibrium configuration, 
it is at most a second post Newtonian quantity from the 
post Newtonian point of view \cite{AS} 
and for mildly relativistic 
binaries, the error is expected to be small. This test 
confirms that an approximate 
quasi-equilibrium state of a binary neutron star can be preserved 
for more than one orbital period. 
\end{enumerate}
In the following subsections, 
we present the numerical results of the tests 1--5 separately.

In the tests 2--5, we prepare the equilibrium and/or  
quasi-equilibrium states as the initial conditions 
adopting a polytropic equation of state $P=K\rho^{\Gamma}$ with 
$\Gamma=5/3$ and/or 2. 
For $\Gamma=5/3$ and 2, we fix $K$ as $10$ and $200/\pi$, 
respectively, to mimic 
a relation between the rest mass $M_*$ and 
the central density $\rho_c$ for neutron stars. 

In Fig. 2, we show $M_*$ and the circumferential radius $R$ 
of the spherical equilibrium stars as a function of $\rho_c$ 
for $\Gamma=5/3$ and 2, respectively. For $\Gamma=5/3$, 
$M_*$ ($M_g$) reaches a maximum value $\simeq 1.552$ ($1.487$) 
at $\rho_c \simeq 1.85 \times 10^{-3}$, and 
for $\Gamma=2$, $M_*$ ($M_g$) $\simeq 1.435$ 
($1.306$) at $\rho_c \simeq 5.0 \times 10^{-3}$. 
Beyond the critical densities, the star is unstable.

Although we adopt particular units fixing $K$, 
the mass, length, and density may be rescaled 
using the following rule 
\beqn
&& M_*(M_g) \rightarrow M_* C^{\bar n/2} (M_g C^{\bar n/2}), 
~R \rightarrow R C^{\bar n/2}, \nonumber \\
&&\rho_c \rightarrow \rho_c C^{-\bar n}~{\rm and}~
J \rightarrow J C^{\bar n}
~{\rm for}~K \rightarrow KC,
\eeqn
where $\bar n=1/(\Gamma-1)$ and $C$ is an arbitrary constant. 
Namely, the invariant quantities are only the dimensionless 
quantities such as $M_*K^{-\bar n/2}$ ($M_gK^{-\bar n/2}$), 
$R K^{-\bar n/2}$, $\rho_c K^{\bar n}$, 
$M_*/R$($M_g/R$), and $J/M_*^2$ ($J/M_g^2$). 


\subsection{Test 1: Spherical collapse of dust}
 
We consider a time symmetric, conformally flat initial condition 
for the dust sphere, 
and adopt the following density profile  
\beq
\rho_*=A\biggl[1+\exp\Bigl({r^2-r_0^2 \over \delta r^2}\Bigr)
\biggr]^{-1},
\eeq
where we choose 
$r_0=4M_g$ and $\delta r^2=0.18M_g^2$. $A$ is adjusted so that 
the gravitational mass of the system is $1$ 
($A \simeq 4.287 \times 10^{-3}$). In this test, 
we assign the negligible specific internal energy and 
pressure, i.e., $\varep \ll 1 $ and $P \ll \rho$. 
Throughout this subsection, every 
quantity is shown in the unit $M_g=1$ (and $G=1=c$). 

We perform the 3D numerical simulation as well as 
1D (spherical symmetric) simulation and compare two results. 
The simulations are carried out in two 
different spatial gauge conditions, 
the MD and $\beta^i=0$ gauge conditions. 
We note that in a spherical symmetric case with a 
conformally flat initial condition, the AMD gauge condition 
is identical with the MD gauge condition because the condition 
$\pa_t \tilde \gamma_{ij}=0$ holds in both 
cases throughout the whole evolution. 
Hence, the comparison can be done without 
any coordinate transformation. We use the 
maximal slice condition in the 1D case and AMS condition 
in the 3D case. Since $K_k^{~k}$ can be kept nearly equal to 
zero in the AMS condition, we can consider them 
as the same slice conditions. 
We have used a grid with $N=60$ and $\delta x=0.15$ 
in the 3D simulation. 

First, we show numerical results for the $\beta^k=0$ case. 
In Fig. 3, we show $\alpha$ and $\rho_*$ 
at $r=0$ as a function of time. The dotted and solid 
lines are the 1D and 3D results, respectively. 
In Fig. 4, we also plot $\alpha$ along the $x$-axis 
at selected times ($t=12.0$, 17.2 and 20.4) for 
the 1D (the solid lines) and 3D (the filled circles) results, 
respectively. We note that 
in the 1D simulation, the apparent horizon is found at 
$t \sim 18.6$. In the 3D simulation, it is also found nearly at 
the equal time and the same location. These results show that 
the 1D and 3D results agree well at least 
up to the formation of a black hole and confirm 
that it is possible to investigate a 
black hole formation in a spherical symmetric spacetime 
accurately with our formulation and numerical code. 

In the late phase ($t \agt 23$), 
the accuracy of the 3D results deteriorates. 
This is because the error increases in 
$\tilde \gamma_{ij}$ rapidly as a consequence of 
the so-called horizon stretching around the location of the 
apparent horizon.

Next, we present numerical results in the MD (and AMD) gauge conditions. 
In Fig. 5, we show $\alpha$ and $\rho_*$ 
at $r=0$ as a function of time. 
In this case, the 1D (the dotted lines) and 
3D (the solid lines) results agree well 
by $t \sim 20$. Fortunately, we could determine 
the apparent horizon because it is formed $t \sim 18.6$, 
but the accuracy deteriorates soon after the formation. 
This illustrates that 
the MD gauge condition is less appropriate than $\beta^k=0$ gauge 
condition for evolution of the late phase of 
the gravitational collapse in the simulations performed 
under identical grid number and spacing. 
The reason is apparently related to the drawback of 
the MD gauge conditions 
pointed out in \cite{gw3p2}: In the 
MD or AMD gauge conditions with $K_k^{~k} \simeq 0$, 
the coordinates spread outward and the proper distance between 
two neighboring grids increases, i.e., 
$\pa_i \beta^i>0$ ($\beta^r > 0$) and $\pa_t \phi > 0$ 
(cf. Eq. (\ref{peq})), around $r=0$ 
during gravitational collapses. As a result, 
the black hole forming region cannot be well resolved. 
In Fig. 6, we show the time evolution of $\phi$ at $r=0$. 
It changes from $0.17$ to 0.9 by $t=20$. Thus, 
the proper distance between two neighboring grids increases 
by a factor $e^{2(\phi(t=20)-\phi(t=0))}\sim 4$. 
In the $\beta^i=0$ gauge condition with $K_k^{~k}=0$, 
on the other hand, 
$\phi(x^i)$ is constant in time and $h_{ij}$ for $r \sim 0$ is 
nearly equal to $0$, 
so that the physical grid spacing in the MD gauge conditions is 
$4$ times as large as that in the $\beta^i=0$ gauge condition 
around $r=0$ at $t\sim 20$. 

To overcome the coordinate spreading effect without 
changing the gauge condition, we have to 
take a large number of grid points around the black hole 
forming region. 
The solution to this problem may be to adopt an adaptive 
mesh refinement technique \cite{AMR} in which we can improve 
the resolution around the black hole forming region effectively. 
However, since the spreading factor increases rapidly as 
shown in Fig. 6, we have to improve the resolution 
also quickly when using such a technique. 
We do not think it a good idea to simply rely 
on such a technique. 
We consider it necessary to modify the gauge condition. 
A strategy of the modification and the numerical experiment will be 
presented in one of forthcoming papers \cite{rotstar}. 


\subsection{Test 2: Stability of spherical stars}

In this second test, we prepare spherical 
equilibrium stars of a polytropic equation of state 
of $\Gamma=5/3$. We use two stars for the test. 
One is a stable star of $\rho_c=10^{-3}$, $M_*=1.499$, 
and $M_g=1.440$, and the other 
is an unstable star of $\rho_c=2.4\times 10^{-3}$, 
$M_*=1.542$, and $M_g=1.478$. In both cases, 
the total rest mass is slightly less than the maximum value. 
In Fig. 2, we show with the filled circles and 
the open circle the locations of the two stars and 
the star at the critical density of stability, respectively. 

Here, we again use two spatial gauge conditions:
AMD gauge and $\beta^i=0$ gauge conditions. 
In Fig. 7, we show the 
time evolution of the density $\rho(=\rho_* e^{-6\phi}/w)$ and 
$\alpha$ at $r=0$ as a function of time for a 
stable configuration of $\rho_c=10^{-3}$. 
These simulations were performed with $N=50$ and $\delta x=0.4$. 
The solid and dotted lines are the results in the 
$\beta^i=0$ gauge condition, and the dashed line is in the 
AMD gauge condition. The dotted line also denotes the result 
for the case in which we 
give a perturbation to the equilibrium configuration 
by reducing $K$ of $0.5\%$ initially. In other cases, we 
give the equilibrium state without any changes. 
These figures clearly illustrate the feasibility of our code 
to judge the 
stability of the spherical stable star and to preserve 
it stably at least in a few oscillation periods irrespective of 
the gauge conditions. 

As shown with 
the approximate perturbation analysis in appendix B \cite{chandra}, 
the period of the fundamental radial oscillation is 
$\sim 10.5 \rho_c^{-1/2}$. For the pressure depleted case 
(the dotted line), the oscillation period is clearly recognized 
in Fig. 7. 
This shows that the frequency of the fundamental radial 
oscillation can be computed accurately in our code.

In Fig. 8, we show the time evolution of 
$\rho$ and $\alpha$ at $r=0$ for an unstable configuration of 
$\rho_c = 2.4 \times 10^{-3}$. 
The simulations are  performed with $N=50$ and $\delta x=0.3$. 
We prepare pressure depleted initial conditions 
in which we reduce $K$ by $0.5\%$ and $0.1\%$. 
As in the stable case, we use both the $\beta^i=0$ and AMD gauge 
conditions, and the numerical results are 
denoted by the solid and dashed lines, respectively. 
As shown in Fig. 8, the unstable star 
collapses into a black hole. 
As one would expect, the star collapses 
more quickly in the case when the depletion factor of the 
pressure is larger. These results confirm that our code 
provides correct results.

The time evolution of $\rho$ and $\alpha$ at $r=0$ in 
the two spatial gauge conditions approximately agrees with 
each other except for the late phase of the 
gravitational collapse during which the coordinate spreading 
effect is severe in the AMD (and MD) gauge conditions \cite{foot1}. 
The overall agreement (except for the late phase) 
is an important test of consistency because 
$\rho$ and $\alpha$ at $r=0$ 
are both gauge independent quantities.


\subsection{Test 3: Quadrupole oscillations of a 
perturbed spherical star}

For the third test, we prepare stable 
spherical stars in equilibrium states as in test 2. 
We adopt $\Gamma=5/3$ and $2$ in this test, and 
choose the stars in which  $\rho_c=10^{-3}$ 
for $\Gamma=5/3$ and $\rho_c=3\times 10^{-3}$ for 
$\Gamma=2$, respectively. 
For $\Gamma=2$ and $\rho_c=3\times 10^{-3}$, 
the mass and circumferential radius are $1.25$ and $6.99$, 
respectively (see filled circles in Fig. 2). 
We perform the simulations with $N=50$ and $\delta x =0.5$ for 
$\Gamma=5/3$ and with $N=50$ and $\delta x = 0.25$ for 
$\Gamma=2$, respectively. 
The test is performed in the AMD gauge condition and 
at the outer boundaries, we set $F_i=O(r^{-3})$. 

As the source of the initial quadrupole oscillation, we give a velocity 
perturbation of ``$+$" type  
\beq
u_i(t=0)=A\sqrt{{M_g \over R^3}}(-x,y,0 ),\label{plus}
\eeq
or of ``$\times$" type 
\beq
u_i(t=0)=A\sqrt{{M_g \over R^3}}(-y,-x,0 ),\label{cross}
\eeq
where $A$ is a constant which is set to be $0.082$. 

As a result of a time-varying mass 
quadrupole moment of the star, 
gravitational waves are emitted. This allows to 
check the feasibility of the gravitational wave extraction 
near the outer boundaries. 
To extract gravitational waveforms, we define \cite{gw3p2}
\beqn
&&h_+ \equiv r(\tilde \gamma_{xx} - \tilde \gamma_{yy})/2,\\
&&h_{\times} \equiv  r \tilde \gamma_{xy},
\eeqn
along the $z$-axis. 
Since we adopt the AMD gauge condition and 
prepare initial conditions in which $F_i=0$, $h_{ij}$ 
is approximately transverse and traceless 
in the wave zone \cite{gw3p2}. 
As a result, $h_+$ and $h_{\times}$ are expected to be  
appropriate measures of gravitational waves. 
They are also useful to 
find the maximum amplitude of gravitational waves because 
we treat the problems in which the amplitude is 
maximum along the $z$-axis.

In Figs. 9--11, we show a root mean square radius 
\beq
x^i_{\rm rms} \equiv \biggl[
{1 \over M_*} \int d^3x \rho_* (x^i)^2 \biggr]^{1/2}, 
\eeq
as a function of time for the $+$ and $\times$ 
mode perturbations for $\Gamma=5/3$ (Figs. 9 and 10), 
and for $+$ mode perturbation for $\Gamma=2$ (Fig. 11), 
respectively. 
In the case of the $+$ mode oscillation, 
$x_{\rm rms}$ (the solid line) 
and $y_{\rm rms}$ (the dotted line) oscillate with a 
characteristic period. 
For $\Gamma=5/3$ and 2, the oscillation period 
is $\sim 5.5 \rho_c^{-1/2}$ and 
$\sim 4.7\rho_c^{-1/2}$, respectively. As shown by 
perturbative studies on stellar pulsations \cite{kojima}, 
the angular frequencies of the $f$-mode oscillation for $\Gamma=5/3$ 
and 2, and $M_g/R \sim 0.1$ are approximately written as 
\beq
\omega\simeq \left\{
\begin{array}{ll}
\displaystyle 
1.44 \sqrt{{M_g  \over R^3}} & ~{\rm for}~~\Gamma=5 / 3,\\
~~ & ~~\\
\displaystyle 
1.22 \sqrt{{M_g  \over R^3}} & ~{\rm for}~~\Gamma=2. \\
\end{array}
\right.\label{koji}
\eeq
Or, according to an empirical formula \cite{Kok}, 
the angular frequency for the stars of $M_g / R \sim 0.1$ is also written, 
irrespective of $\Gamma$, as 
\beq
\omega \simeq 0.012+0.93 \sqrt{{M_g  \over R^3}}. \label{kokk}
\eeq
For $\Gamma=5/3$ and $\rho_c=10^{-3}$ 
(i.e., $M_g \simeq 1.44$ and $R \simeq 13.1$), the oscillation period 
is $\simeq 5.5 \rho_c^{-1/2}$ in both formulas. 
For $\Gamma=2$ and $\rho_c= 3 \times 10^{-3}$ (i.e., 
$M_g \simeq 1.25$ and $R=6.99$), 
the period is $\simeq 4.7 \rho_c^{-1/2}$ from Eq. (\ref{koji}) 
and $\simeq 5.0 \rho_c^{-1/2}$ from Eq. (\ref{kokk}). 
Thus, our numerical results agree with these perturbative results 
fairly accurately. 

On the other hand, the oscillation periods of $z_{\rm rms}$ (the 
dashed line) for the $+$ mode perturbation  
and of all the components of $x^i_{\rm rms}$ 
for the $\times$ mode perturbation are roughly 
$\sim 10\rho_c^{-1/2}$ for 
$\Gamma=5/3$ and $\sim 7\rho_c^{-1/2}$ for 
$\Gamma=2$.  
These are in agreement with the periods of the 
fundamental radial oscillation and different from 
those of the $f$-mode oscillation. 
This reflects the fact that they are not relevant for the 
$f$-mode oscillation at linear order. 

In Figs. 12 and 13, we show $h_+$ and $h_{\times}$ at 
$z_{\rm obs}=24.5$ (the solid lines) and $19.5$ (the dashed lines) 
for $\Gamma=5/3$ as a function of the retarded time 
for the $+$ and $\times$ mode perturbations. In Fig. 14, we 
also show  $h_+$ and $h_{\times}$ 
at $z_{\rm obs}=12.25$ (the solid lines) and $9.75$ 
(the dashed lines) for $\Gamma=2$ 
for the $+$ mode perturbation. 
As the oscillation frequency of the $f$-mode shows, the wavelength of 
gravitational waves is several times larger than $z_{\rm obs}$, 
so that we expect the extracted waveforms to be different from 
the asymptotic waveforms. 
However, they appear to constitute a fair 
description of the asymptotic waveforms, at worst qualitatively,
because 
({\it a}) the solid and dashed lines approximately agree,  i.e., 
$\tilde \gamma_{xx}-\tilde \gamma_{yy}$ and $\tilde \gamma_{xy}$ 
behave approximately as $f(t-z)/z$ along the $z$-axis where 
$f$ denotes a generic function, 
({\it b}) the oscillation period agrees well 
with that of the $f$-mode, 
({\it c}) the order of magnitude of $h_+$ and $h_{\times}$ 
roughly agrees with that derived by 
the quadrupole formula, i.e., $M_gv^2
\sim M_g^2 A^2/R \sim 10^{-3}$, and ({\it d}) 
for the perturbation given by Eq. (\ref{plus}), 
$h_{\times}$ remains nearly zero, and for Eq. (\ref{cross}), 
$h_+$ remains nearly zero, as expected by the 
quadrupole formula. 
These facts suggest that $h_+$ and $h_{\times}$ represent 
(at least approximately) gravitational waves 
emitted by the stellar oscillation 
and indicate the ability of our code to extract 
gravitational waves near the outer boundary. 
We expect that a simulation of higher resolution would produce 
even more precise gravitational waveforms. 


\subsection{Test 4: Stability of rapidly rotating stars} 

To carry out this test, we prepare 
rapidly and rigidly rotating neutron stars 
in (approximate) equilibrium states. 
We adopt $\Gamma=5/3$ and $2$, and select stable compact 
neutron stars. 

To prepare the rotating stars in 
approximate equilibrium states, we use a conformal flatness 
approximation. 
Then, the geometric and hydrostatic equations 
for solutions of equilibrium states are described as
\cite{bcsst}
\beqn
&& \Delta \psi = -2\pi (\rho h w^2-P) \psi^5 -{\psi^5 \over 8}
\delta^{ik}\delta^{jl}L_{ij} L_{kl}, \label{eq310} \\
&& \Delta (\alpha\psi) = 2\pi \alpha \psi^5 
[\rho h (3w^2-2)+5P] \nonumber \\
&& \hskip 1.5cm +{7\alpha\psi^5 \over 8}
\delta^{ik}\delta^{jl} L_{ij} L_{kl} ,\\
&& \delta_{ij} \Delta \beta^i + {1 \over 3} \beta^k_{~,kj}
-2L_{jk}\delta^{ki} \Bigl(
\pa_i \alpha - {6\alpha \over \psi} \pa_i \psi \Bigr) 
\label{quasibet}\nonumber \\
&& \hskip 3cm =16\pi\alpha \rho h w u_j,\\
&& {\alpha h \over w}={\rm const.},\label{eq313}
\eeqn
where 
\beqn
&&u_i=w\psi^4(\epsilon_{izk}\Omega x^k+\delta_{ij}\beta^j)/\alpha,\\ 
&&w^2=1+\psi^{-4}\delta^{ij}u_iu_j, \\
&&h=1+K\Gamma\rho^{\Gamma-1}/(\Gamma-1), \\
&&L_{ij}={1 \over 2\alpha}
\biggl(\delta_{jk} \pa_i \beta^k + \delta_{ik} \pa_j \beta^k 
- {2 \over 3}\delta_{ij}
\pa_k \beta^k\biggr), 
\eeqn
and $\Omega$ denotes the angular velocity of the rotation. 

Although the solutions obtained from 
Eqs. (\ref{eq310})--(\ref{eq313}) are not exact, 
we can still expect that they are excellent approximate solutions 
as illustrated in \cite{cook}. 
As shown in Fig. 15, this is the case: 
We show the gravitational mass $M_g$ as a 
function of the central density $\rho_c$ for 
rotating stars at mass-shedding limits 
obtained from exact equations (the solid lines) \cite{SS} and 
obtained by 
the conformal flatness approximation (the circles). 
We have found that the sequences of the circles almost 
coincide with the solid lines for mildly relativistic stars of 
$\rho_c \alt \rho_t$ where $\rho_t\sim 0.0015$ for $\Gamma=5/3$ and 
$\rho_t \sim 0.003$ for $\Gamma=2$. 
As the density of the rotating stars increases 
(i.e., $\rho_c \agt \rho_t$), the coincidence becomes worse 
gradually, but even for $\rho_c \sim 2\rho_t $ at which 
the star seems to be unstable, the difference between 
the sequences of circles and solid lines is small.

We perform the simulations for the rotating 
stars marked with the filled circles in Fig. 15. 
The relevant quantities for the rotating stars are shown in 
Table I. We adopt 
$\delta x=0.434$ for $\Gamma=5/3$ and 
$\delta x=0.202$ for $\Gamma=2$ 
as the grid spacing. With these grid spacings, the 
major and minor axes of the stars are initially 
covered by 40 and $23-24$ grid points, respectively. 

To suppress the coordinate twisting, 
the AMD gauge condition is adopted. 
The shift vector $\beta^k$ determined 
in this gauge condition for 
the conformal 3D metric ($\tilde \gamma_{ij}=\delta_{ij}$) 
agrees with that obtained from Eq. (\ref{quasibet}). This 
implies that  
when a simulation is started, the gauge condition is identical 
with that used for obtaining the approximate equilibrium states. 
Therefore, we can use the approximate equilibrium states 
as the initial conditions without any coordinate transformation. 

In this test, $F_i$ is set to be zero at the outer boundaries. 
The grid number $N$ is set to be $54$ typically. 
We also adopted $N=76$ without changing the grid spacing in 
some of the following simulations, but we did not 
find significant difference in the results. This indicates that 
the outer boundary condition adopted here is adequate. 

We have considered two kinds of initial conditions: In one case, 
we use the approximate equilibrium configurations without any change. 
In the other case, we initially decrease the pressure by 
reducing $K$ of $1\%$ (i.e., $\Delta K/K=1\%$). 
Even in the pressure depleted case, 
the stars should be stable because the gravitational mass 
is $\sim 3-5\%$ smaller than the maximum mass of 
the rotating stars (see Fig. 15). 
In Fig. 16, we show snapshots of the 
density contours lines for $\rho_*$ 
and the velocity field for $(v^x, v^y)$ 
in the equatorial plane (left) and in the $y=0$ plane (right) 
at selected times for $\Delta K=0$ and 
for $\Gamma=2$ as an example (for $\Gamma=5/3$, we have found that 
similar figures can be drawn). 
We also show $\alpha$ and $\rho$ at $r=0$ as 
a function of $t/{\rm P}$ in Fig. 17, $x_{\rm rms}$ and 
$z_{\rm rms}$ as a function of $t/{\rm P}$ in Fig. 18, 
and $J/J_0$ as a function of $t/{\rm P}$ in Fig. 19 
for $\Delta K=0$ (the solid lines) and $\Delta K/K=1\%$
(the dotted lines) and for $\Gamma=5/3$ and 2. 
Because of numerical dissipation at the stellar surface, 
the total angular momentum of the system 
($J$ in Eq. (\ref{eqj})) 
decreases by $\sim 2\%$ by $t \sim 2{\rm P}$ in all the 
simulations (see Fig. 19). As a result, $\alpha$ at $r=0$ and 
$x^i_{\rm rms}$ decrease and $\rho$ at $r=0$ 
increases with the time evolution. 
Also, the stars suffer slight non-axisymmetric (quadrangular shape) 
deformation near the surface (see Fig. 16) 
because we use the Cartesian coordinates for rotating stars 
of a spheroidal shape. 
However, besides these slight changes, 
the stars remain almost in the stationary states 
for more than two rotational periods 
(i.e., by the time when we terminated these simulations).

Since the initial conditions adopted are only approximate 
equilibrium states, 
some oscillations are induced with time evolution. However, 
the amplitude is very small and cannot be distinguished from the 
numerical errors. 
Thus, the states of the rotating stars are close to the 
true equilibrium ones. Actually, 
the absolute value of each component of $h_{ij}$ remains small 
and of order $\sim 0.05$. 
These results re-confirm that the conformal flatness approximation 
is really a good approximation for obtaining axisymmetric 
rotating stars in equilibrium states. 

For $\Delta K/K=1\%$, 
$\alpha$ and $\rho$ at $r=0$ and $x^i_{\rm rms}$ 
oscillate with the time evolution. 
The oscillation period is roughly $0.9{\rm P}$ 
for both $\Gamma=5/3$ and $2$. We deduce that the period 
denotes a fundamental quasi-radial oscillation period of the 
rotating stars. 

We emphasize that the simulations can be stably carried out 
for more than two orbital periods 
without any instabilities and with $h_{ij}$ being kept small. 
These results clearly demonstrate that our formulation for solving 
the Einstein equation is robust even for systems of 
non-zero angular momentum and that 
the coordinate twisting is sufficiently suppressed 
to a level adequate for long-timescale simulations 
in our AMD gauge condition. 

In this paper, we only perform test simulations using 
stable rotating stars. It is very interesting to perform 
simulations adopting rapidly rotating supramassive neutron stars 
to investigate the stability and 
the fate of unstable stars \cite{CST}. 
Such rapidly rotating supramassive neutron stars may be 
frequently formed as a result of accretion onto a neutron star 
from a companion star in normal binary systems \cite{CST5}. 
However, such simulations are 
beyond the scope of this paper. 
More detailed analysis of the stability of rapidly rotating 
stars and of the final fate of unstable stars 
will be presented in \cite{rotstar}. 


\subsection{Test 5: Corotating binary in an approximate 
quasi-equilibrium state} 

In the final test, we adopt a mildly relativistic corotating binary 
neutron star in an approximate quasi-equilibrium 
state as initial condition. As mentioned in Sec. I, 
our purpose in future is to carry out 
simulations of coalescing binary neutron stars from the 
ISCO to formation of a black hole or new neutron star. 
In the early stage in which the binary is 
near the ISCO, the radial velocity 
of each star is expected to be small, and 
we can consider the binaries as on approximate 
quasi-equilibrium orbits rather than on plunging orbits. 
As the radial velocity gradually increases, the 
orbit changes from the near inspiraling one to the plunging one. 
For a successful simulation of 
a coalescing binary neutron star, therefore, 
the nearly quasi-equilibrium orbit has to be maintained 
at least for $\sim 1$ orbital period stably. 
In this test, we show this feasible with our code. 

The configuration of a binary neutron star is again 
obtained in the assumption of the conformally flat 3D metric and 
the maximal slice condition $K_k^{~k}=0$ \cite{bcsst}. 
We prepare a binary in which the surfaces of two stars come into 
contact. We adopt $\Gamma=5/3$ in this test.  
The equation of state for such small $\Gamma \alt 2$ 
is not so stiff that 
binary neutron stars in contact orbits are 
stable against a hydrodynamic instability \cite{bcsst}. 
Since the binary neutron star is not very compact 
(see Table II) and is also far from the ISCO, 
it can remain on the stable orbit for a time comparable 
with the emission timescale of gravitational waves. 

Approximate quasi-equilibrium states of binary 
neutron stars are obtained solving Eqs. (\ref{eq310}--\ref{eq313}). 
The equations are solved using a numerical method 
similar to that employed in \cite{shibaPN}. 

In Fig. 2, we denote, by the cross (of higher density, 
$\rho_{\rm max}=10^{-3}$), the 
relation between the maximum density $\rho_{\rm max}$ and 
half of the total rest mass $M_*/2$ for the corotating binary 
neutron star in an approximate quasi-equilibrium state 
which is used as an initial condition of the simulation. 
We compared the numerical results 
on the relation between the maximum density and rest 
mass with those by Baumgarte et al.\cite{bcsst}, 
and found that they agree within $2\%$ error. 
In Table II, we also list the relevant quantities of the 
binary neutron star. We define an 
orbital radius $a\equiv M_g^{1/3} \Omega^{-2/3}$, 
and we define an 
approximate ratio of the emission timescale of 
gravitational waves to the orbital period as 
\beq
R_{\tau}  = {5 \over 128\pi}\biggl({a \over M_g }\biggr)^{5/2}
={5 \over 128\pi}(M_g \Omega)^{-5/3},
\eeq
where we have used the Newtonian expression of the energy and orbital 
period, and the quadrupole formula for the energy luminosity of 
gravitational waves \cite{ST}. 
Since our purpose is to check the feasibility 
of our code to preserve the approximate 
quasi-equilibrium state stably, 
this test should be performed for a binary in which 
$R_{\tau} >1$. 

In the numerical simulation, we adopt the AMD gauge condition 
to sufficiently suppress the coordinate twisting. 
As we discussed in Sec.III.D, the gauge condition at $t=0$ is 
identical with that used for obtaining the approximate 
quasi-equilibrium state, so that we do not have to carry out  
coordinate transformation at $t=0$.

In this subsection, we adopt the grid spacing as $\delta x =0.927$, 
varying the grid number $N$ as 76 and 116. 
With this setup, the major diameter of each star 
is covered by 30 grid points initially. 
For $N=76$ and 116, $L (=N \delta x)$ 
is only $\sim 0.2$ and $0.3$ times as large as the 
wavelength of gravitational waves $\lambda_{\rm gw} \equiv  
\pi/\Omega$. For a precise simulation, 
the outer boundary should be located in the wave zone so that  
$L \gg \lambda_{\rm gw}$. This is 
because we should impose an outgoing boundary condition for 
$\tilde \gamma_{ij}$ and $\tilde A_{ij}$, and 
in the system of binary neutron stars, 
the existence of gravitational waves 
plays an important role for evolution of the system. 
We have found that the present incomplete treatment for 
the outgoing boundary conditions seems to produce numerical errors 
(see below further discussion). 
However, for imposing the boundary condition in the wave zone 
with the uniform grid, 
it is necessary to take a very large grid number 
$N \geq 400$. That is not feasible in the present computational 
resources on supercomputers. 
In this paper, we perform simulations overlooking the deficits. 
Thus, the radiation reaction of gravitational waves is not 
precisely taken into account in the following simulation. 
A large simulation of $L > \lambda_{\rm gw}$ 
is one of future issues. 

In Fig. 20, we show snapshots of the density contour lines for 
$\rho_*$ and velocity field for $(v^x, v^y)$ 
in the equatorial plane at selected times for $N=116$ while 
in Fig. 21, we show $x^i_{\rm rms}$ 
as a function of $t/{\rm P}$. 
We note that if the binary remains in a quasi-equilibrium state, 
the orbital period should be kept to $\sim$P, 
the curves for $x_{\rm rms}$ and $y_{\rm rms}$ should be 
complete sine curves, and $z_{\rm rms}$ should be a constant.
From Figs. 20 and 21, it is evident 
that the state of binary neutron star fluctuates from 
the initial state with time. 
The fluctuations seem to be mainly due to the numerical error 
discussed below. Besides the spurious numerical effect, 
the binary neutron star is kept in an approximate 
quasi-equilibrium circular orbit for $\agt 1$ orbital periods stably. 

There appear to be two candidates for the numerical error. 
One is the numerical dissipation of the 
angular momentum at the stellar surface \cite{SON} which 
was already pointed out in test 4. 
This is a simple consequence of insufficient resolution. 
As a result, the orbital radius decreases. 
The other candidate is the incomplete treatment of the 
outgoing boundary conditions. 
As mentioned before, we impose an 
approximate outgoing boundary 
condition for $\tilde \gamma_{ij}$ and $\tilde A_{ij}$ 
not in the wave zone. From this incomplete 
treatment, the angular momentum seems to go out and come in 
inaccurately from the outer boundaries. 
As a result, the orbital radius 
increases and decreases. To illustrate the fact, 
$J/J_0$ as a function of 
$t/{\rm P}$ is depicted in Fig. 22. From this figure, 
we can recognize 
that the angular momentum increases and decreases with time 
by $\sim \pm 3\%$ of the total angular momentum. 
We note that the angular momentum should be monotonically 
dissipated by gravitational radiation by about $2\%$ of 
the total angular momentum in one 
orbital period according to the estimation by 
the quadrupole formula. However, the numerical results 
do not reflect this effect accurately. 

Despite these errors, however, the binary neutron star 
is preserved in the approximate quasi-equilibrium orbit 
for more than one orbital periods {\it stably}. This 
seems to imply that we are choosing an adequate gauge condition 
and formulation of the Einstein equation. 
We therefore expect that 
if we could perform a simulation taking a sufficient number of 
grid points to impose the outer boundary conditions 
in the wave zone and to improve the resolution, 
it would be possible to perform the simulation not only stably, 
but also accurately. 

During the evolution, $h_{ij}$ deviates from zero and 
gradually reaches a finite amplitude. The maximum 
absolute value of each component is of order 
$\sim 0.05$ by $\sim $P. (This is roughly 
consistent with what is inferred from a post Newtonian study 
\cite{AS} in which $h_{ij}$ has magnitude of 
order $\sim (M_g/a)^2$.) 
Since $h_{ij}$ becomes non-zero, it could slightly affect 
the quasi-equilibrium configuration of binary neutron stars. 
However, this does not seem to be a serious perturbation 
for the present mildly relativistic case.

In Fig. 23, we show $h_+$ and $h_{\times}$ as a function 
of $t/{\rm P}$. 
The solid and dotted lines are extracted at $z_{\rm obs}=106.6$ 
and 85.3 for $N=116$, and the dashed line at $z_{\rm obs}=69.5$ 
for $N=76$. As mentioned before, we 
extract them at $\sim 0.2-0.3 \lambda_{\rm gw}$, 
so that they only indicate approximate 
asymptotic waveforms. Nevertheless, 
we find that they behave in the periodic manner and that 
the period approximately agrees with the orbital period. 
Furthermore, the solid and dotted lines agree very well, 
implying that the waves propagate at the speed of light. 
These facts indicate that the approximate waveforms 
constitute a fair description of the asymptotic ones. 

There are, however, differences from what is expected in 
the asymptotic waveforms. 
The first one is found in the 
long wavelength modulation (apparent especially in $h_+$), 
which should not appear in the correct waveforms. 
The modulation is larger for the simulation with smaller $N$, 
indicating that it is caused in the outer boundaries. 
The second difference is found in the irregular, non-periodic waveforms 
in the early stage for $t \alt z_{\rm obs}$. 
These waves are emitted because we set an approximate 
quasi-equilibrium state as the initial condition neglecting 
$h_{ij}$. The first radiation will be the 
relaxation of the system to the true quasi-equilibrium state. 
To avoid this shortcoming, we have to prepare more realistic 
quasi-equilibrium states adopting a formalism in which $h_{ij}$ 
is appropriately taken into account. 
The third difference is found in the amplitude: 
For the case where two point 
particles of equal mass are in a circular orbit, 
the maximum amplitudes of $h_+$ and $h_{\times}$ in the 
post Newtonian approximation can be written as \cite{BIWW}
\beq
M_g \tilde x\biggl[1-{17 \over 8}\tilde x+2\pi \tilde x^{3/2} - 
{15917 \over 2880}\tilde x^2\biggr] \label{eqpnw}
\eeq
where $\tilde x=(M_g\Omega)^{2/3}$. Using this formula, 
the maximum value of $h_+$ and $h_{\times}$ in the binary 
of $M_g=2.92$ and $\tilde x=0.0887$ is $0.24$. (Note that 
in the quadrupole formula, it is $0.26$.) 
Since the convergence of the post Newtonian formula for 
$\tilde x \sim 0.1$ is not very good, we should take into account 
an error of $\sim 10\%$. Even if we consider such error, 
the amplitude obtained in the numerical simulation 
is found to be somewhat smaller than that derived 
from Eq. (\ref{eqpnw}). 
The reason seems again due to the incomplete treatment of the 
outgoing boundary condition at $L \sim 0.2-0.3\lambda_{\rm gw}$. 

In summary, we can perform simulations of corotating binary 
neutron stars in an approximate quasi-equilibrium state 
stably and fairly accurately, and extract 
gravitational waves with $\sim 10\%$ error even in the 
present restricted grid numbers. 
However, to improve the accuracy for gravitational waveforms, 
it is necessary 
to adopt more sophisticated boundary conditions \cite{bishop}, to take 
a larger number of grid points and to prepare a more realistic 
quasi-equilibrium state as the initial condition. 
These issues will be addressed in future simulations. 


\section{Merger of corotating binary neutron stars: 
A demonstration}

In this section, we show numerical results of merger 
between two neutron stars as demonstration that such 
simulations are feasible. 
We again adopt corotating binary 
neutron stars of $\Gamma=5/3$ in contact and in 
approximate quasi-equilibrium orbits. 
Two binaries shown in Table II (or in Fig. 2 with 
crosses) are prepared as the initial conditions: One is the same as 
that adopted in Sec. III.E, and the other is 
less relativistic one in which the initial maximum density of each star 
is $6\times 10^{-4}$. To accelerate the merger, 
we artificially reduce 
the angular momentum by $5\%$ in the initial stage; 
i.e.,  we reduce the initial values of 
$u_x$ and $u_y$ by $5\%$ uniformly. 
In this section, we adopt $N=116$. 

In Figs. 24 and 25, we show snapshots of the 
density contours lines for $\rho_*$ 
and the velocity field for $(v^x, v^y)$ 
in the equatorial plane 
for $\rho_{\rm max}(t=0)=10^{-3}$ and $6\times 10^{-4}$, respectively. 
Since we artificially reduce the angular momentum of the 
system at $t=0$, the neutron stars approach each other 
to merge soon after the simulations start. 
For both types of initial data at 
$t \agt 0.5 {\rm P}$ (${\rm P}$ denotes the orbital period of the 
quasi-equilibrium states without the angular momentum depletion), 
the neutron stars begin to merge forming spiral arms, and 
by $t \sim 1.5 {\rm P}$, new neutron stars which 
are nearly axisymmetric are formed. 
In Figs. 26 and 29, we show snapshots of the 
density contour lines for $\rho_*$ 
in the $y=0$ plane at $t=1.62{\rm P}$ for 
$\rho_{\rm max}(t=0)=10^{-3}$, and at $t=1.59{\rm P}$ for 
$\rho_{\rm max}(t=0)=6\times 10^{-4}$. 
In Fig. 28, we also 
show the angular velocity $\Omega\equiv (x v^y - y v^x)/(x^2+y^2)$ 
along the $x$ and $y$-axes in the equatorial plane at $t=1.62{\rm P}$ 
for $\rho_{\rm max}(t=0)=10^{-3}$ and at $t=1.59{\rm P}$ for 
$\rho_{\rm max}(t=0)=6\times 10^{-4}$, respectively. 
These results show that the final products are rapidly and 
{\it differentially} rotating, highly flattened neutron stars. 
For the case $\rho_{\rm max}(t=0)=10^{-3}$, the central density of 
the merged object for $t \simeq 1.6$P is $\sim 1.4 \times 10^{-3}$, 
which is nearly the maximum allowed density along the sequence of 
stable neutron stars of $K=10$ and $\Gamma=5/3$ (cf. Figs. 2 and 15). 
This indicates that the new neutron star is expected to be 
located near the critical point of stability against 
gravitational collapse and that 
a black hole could be formed in the merger of 
more massive neutron stars. 

In Fig. 29, we show fraction of the rest mass inside a coordinate 
radius defined as
\beq
{M_*(r) \over M_*}\equiv {1 \over M_*} \int_{|x^i| < r} 
\rho_* d^3x 
\eeq
as a function of time for 
$\rho_{\rm max}(t=0)=10^{-3}$ (the solid lines) and 
$\rho_{\rm max}(t=0)=6\times 10^{-4}$ (the dashed lines). 
The figures show that $\simeq 96\%$ and $\simeq 97\%$ of 
the matter are inside $r =18 (\simeq 6M_g)$ for 
$\rho_{\rm max}(t=0)=10^{-3}$ and $r = 20.4 (\simeq 7.5 M_g)$ 
for $\rho_{\rm max}(t=0)=6\times 10^{-4}$ 
forming the new neutron stars. Thus, only a small fraction of 
the matter can spread outward 
to form a halo around the central objects. 
We also find that $J(r=18) \sim 0.85J_0$ 
for $\rho_{\rm max}(t=0)=10^{-3}$ and 
$J(r=24.4=9M_g) \sim 0.85 J_0$ 
for $\rho_{\rm max}(t=0)=6\times 10^{-4}$. 
In both cases, $J/M_g^2$ of the newly formed neutron 
stars seems roughly $\sim 1$, and that appears to be one of 
the reasons for which 
the neutron stars do not collapse to be a black hole.

In Fig. 30, we show $h_+$ and $h_{\times}$ as a function of 
$t/{\rm P}$ for the two cases. 
For $\rho_{\rm max}(t=0)=10^{-3}$ and $6\times 10^{-4}$, 
$z_{\rm obs} \simeq 106.6$ and 126.2, respectively. 
In both cases, $z_{\rm obs} \simeq 0.3 \lambda_{\rm gw}$. 
In the early phase, their amplitudes gradually increase 
as their orbital radii decrease. The amplitudes reach their 
maxima when the neutron stars merge. Since 
the binaries change their configuration to nearly 
axisymmetric ones soon after the mergers, 
the amplitudes drop quickly, 
leaving quasi-periodic waves of small amplitude which 
result from small non-axisymmetric perturbations. 
Since the equation of state is not stiff enough to 
allow for the bar-mode instability to set in \cite{StF}, 
the amplitudes will monotonically fall 
in the emission timescale of gravitational waves. 
For $\rho_{\rm max}(t=0)=10^{-3}$, 
the amplitude rises and falls more sharply 
than for $\rho_{\rm max}(t=0)=6\times 10^{-4}$, 
because a higher density merged object is formed 
more rapidly due to the 
stronger relativistic gravity. However, 
the overall waveforms of the two cases closely resemble each other 
and no significant difference is found besides the 
difference of their amplitude and frequency. 
This indicates that 
the waveforms only weakly depends on the compaction 
of original neutron stars if black holes are not formed 
after the merger.

It is interesting to consider the evolution after the 
formation of a newly formed, differentially 
rotating massive neutron star of $J/M_g^2 \sim 1$ 
such as those obtained in the above simulations. 
Since the mass of such neutron stars ($M_g \sim 2.5-3M_{\odot}$) 
is much larger than the maximum allowed mass of the original 
neutron stars of zero angular momentum ($M_g \sim 1.5M_{\odot}$), 
the new star is strongly supported by the 
rapid rotation. Hence, if the angular momentum is 
slightly dissipated or transported outward, 
they could collapse to be black holes eventually. 
As discussed in \cite{BS}, there are many 
processes which contribute to the angular momentum dissipation 
and the angular momentum redistribution; e.g., 
neutrino emission, magnetic radiation, viscous 
dissipation, and gravitational radiation. Since 
each process can affect others in complicated manner, 
it is difficult to give a precise analysis. Here, 
we present a rough and conservative upper limit of the 
timescale for collapse to be a black hole. 

Since it will be cooled mainly by the neutrino emission \cite{BS}, 
the new neutron star will first 
contract on a neutrino emission timescale 
$\tau_{\nu}\sim 10-100$ sec after the formation. 
According to \cite{BS}, however, 
the contraction will not lead to a black hole 
because $J/M_g^2$ will remain $\sim 1$ even after a large 
amount of neutrino emission \cite{BS}. 
Moreover, even in the case $J/M_g^2 <1$, 
the maximum allowed mass of neutron stars after the cooling 
may not change significantly 
if a realistic equation of state is taken into 
account \cite{Goussard}. 
The bulk viscosity could play a dominant role when 
the merged object has high temperature \cite{FI} 
and high non-axisymmetry. 
Although the temperature could be sufficiently high just after 
its formation, the merged object seems to remain 
nearly axisymmetric during the early stage as suggested by the 
simulations in this paper. 
Therefore, the neutron star could not collapse in $\tau_{\nu}$. 

Nevertheless, the newly formed neutron star
is only in a quasi-equilibrium state and is continuously driven away 
from the equilibrium state. As a result, the neutron star will 
eventually collapse to a black hole either because of 
magnetic fields or because of viscosity. 
An important role in fact could be played by magnetic fields 
if the neutron stars before the 
merger had magnetic fields of $\sim 10^{12}$ Gauss. 
It seems likely that the strength of 
magnetic fields of the newly formed neutron star could become 
larger than their initial strength $\agt 10^{13}$ Gauss 
due to the compression of matter during the merger 
and/or its differential rotation \cite{piran}. 
The angular momentum of the merged object 
could then be dissipated by the magnetic dipole 
radiation rather efficiently. 
The dissipation timescale 
of the angular momentum is estimated as \cite{ST2}  
\beqn
\tau_{B} \sim {M_g c^3 \over B^2 R^4 \Omega^2} 
&& \sim 2 \times 10^{7} {\rm sec}
\biggl({M_g \over 3M_{\odot}}\biggr) 
\biggl({10^{13} {\rm Gauss} \over B}\biggr)^2 \nonumber \\
&& \hskip 2cm \times \biggl( {10^6 {\rm cm} \over R} \biggr)^2
\biggl( {0.3c \over R \Omega} \biggr)^2,
\eeqn
where $B$ denotes the typical strength of the magnetic field. 
Thus, the newly formed neutron star could collapse to a black hole 
within a year if $B$ is larger than $10^{13}$ gauss. 

If the magnetic field is not very large, 
on the other hand, 
the dissipation by the shear viscosity may play a dominant role 
for evolution of the newly formed neutron star. 
If the shear viscosity is present, 
the differential rotation is forced 
into rigid rotation in viscous timescale. 
In this case, the angular momentum around the central region is 
transported to the outer parts and 
the centrifugal force around the 
central region is weakened. Because it 
has a very large mass which is probably the nearly maximum allowed 
mass for the differential rotation law, 
the neutron star might collapse as a result of 
the outward transport of the angular momentum. 
If the microscopic viscosity is dominant, 
the dissipation timescale can be estimated as 
\beqn
\tau_{v} \sim {M_g \over \eta_s R} 
&&\sim 3 \times 10^{8} {\rm sec}
\biggl({M_g \over 3M_{\odot}}\biggr) 
\biggl({10^{15}{\rm g/cm^3} \over \rho_c}\biggr)^{9/4} \nonumber \\
&&\hskip 2cm \times \biggl({T \over 10^8 {\rm K} }\biggr)^2
\biggl({10^6 {\rm cm} \over R}\biggr).\label{vistime}
\eeqn
where we use $\eta_s=347\rho^{9/4}T^{-2}$ \cite{FI} as 
the shear viscosity of neutron star matter, and $T$ denotes the 
temperature which could be dropped by neutrino emission 
from $10^{11}$K to $10^{8-9}$K in $\sim 1$ year \cite{ST3}. 
It should be noted that even if the magnitude of 
magnetic fields is not very large, they could 
be at the origin of an effective viscosity. This is because the 
differential rotation of the star could cause a shearing 
instability resulting in the change of 
the magnetic field configuration and outward 
transport of the angular momentum \cite{balbus}. 
The effective viscous timescale, thus, might be much shorter 
than that of Eq. (\ref{vistime}). 

We also have to consider the 
effect of gravitational wave emission 
due to the bar-mode \cite{StF,BS} or r-mode \cite{rmode}, 
by which the non-axisymmetric perturbation may 
grow contributing the dissipation of 
the angular momentum or the redistribution of the differential rotation. 
Such effects are likely to be important, in particular 
after the newly formed neutron star contracts due to the neutrino 
emission, because in such stage the ratio of the 
rotational kinetic energy to 
the binding energy could become large ($\agt 0.14$) 
due to the contraction \cite{BS}
and/or the equation of state could be stiff 
($\Gamma \agt 1.8$ \cite{StF}) 
enough to allow for the bar-mode deformations. 
Since there are too many uncertain aspects such as 
the initial amplitude of the non-axisymmetric perturbation 
and interaction with viscosity \cite{det}, 
the dissipation timescales cannot be estimated in a simple manner. 
However, it is likely that such effects 
contribute to the dissipation of the angular momentum and 
make the timescale shorter. We can then conclude that the 
newly formed supramassive neutron star (if it could be formed) 
will eventually collapse to a black hole 
in the timescale $\sim 10^{\pm 1}$ years as a result of one of the 
above dissipation processes. 

The argument presented here 
suggests that the strength of the magnetic field 
of the newly formed neutron star is one of key parameters for 
the evolution. It may well be important to 
investigate the evolution of the magnetic field 
during the merger 
incorporating a solver of the magnetic field equation.

\section{summary}

In this paper, we present our first successful 
results of numerical 
simulations carried out using fully general relativistic 
3D hydrodynamic code. 
We have performed a wide variety of simulations for 
test problems toward more realistic simulations of 
coalescing binary neutron stars and have 
confirmed it possible to obtain 
the solutions for the test problems fairly accurately. 
In particular, we have illustrated that it is possible to 
preserve an approximate quasi-equilibrium state of 
a mildly relativistic binary neutron star 
as well as to perform simulations of merger between 
two corotating neutron stars to be new neutron stars 
stably for a couple of orbital periods in our numerical code. 
These results indicate 
that numerical simulations of binary neutron stars 
for a long time from their ISCO to mergers are feasible. 

Since we could not take a sufficiently large number of grid 
points to impose the outer boundaries in the wave zone as well as 
to resolve each neutron stars precisely, 
numerical errors are unavoidably included in the results. 
In particular, we feel that 
radiation reaction effect due to gravitational wave 
emission could not be taken 
into account precisely. For a more accurate computation of 
the radiation reaction effect, we will have to adopt 
a numerical technique such as a nested grid technique 
which effectively makes the computational region bigger 
or to impose sophisticated boundary conditions at 
the outer boundaries \cite{bishop}, unless computational power 
is improved soon. 
These issues should be pursued in future works. 
We emphasize, however, that besides the error induced by 
the incomplete treatment of the outgoing 
boundary conditions as well as 
the numerical dissipation due to the restricted resolution, 
the simulation can be performed stably and fairly accurately. 
If we can accept such small error 
(for example, say $\alt 5\%$ error in the angular momentum), 
the code can be used for investigation of 
many interesting problems even at present. 

In this paper, we have paid attention only to 
test problems. Since we consider binary neutron stars of 
mildly large compaction parameter and small $\Gamma$,
the final products of the merger are not black holes. 
A black hole may be more easily formed for larger $\Gamma$ and 
preliminary simulations indicate that this is the case. 
One of the most interesting and important issues in 
numerical relativity is to clarify the criterion for 
formation of black holes. In a forthcoming paper \cite{SU}, 
we will perform simulations 
of corotating binary neutron stars of a large $\Gamma \sim 2$, 
in which a black hole could be formed more easily. 

It is well known, however, that the corotating velocity field is not 
realistic for binary neutron stars 
because the viscosity of the neutron stars is not large enough 
to achieve the corotation \cite{KBC}. Instead, the 
irrotational velocity field is considered to be 
more realistic one \cite{KBC}. 
For making a realistic model of the final phase of binary 
neutron stars, it is necessary to perform simulations for 
irrotational binary neutron stars. 

Fortunately, several groups have just recently developed 
numerical methods for obtaining an approximate 
quasi-equilibrium state of irrotational binary neutron stars 
in a conformal flatness approximation \cite{BGM,Uryu,MM}, 
providing more realistic models of binary neutron stars. 
Namely, an initial condition has already been prepared. 
In the next step \cite{SU}, 
we will also carry out simulations 
adopting the approximate quasi-equilibrium states 
of irrotational binary neutron stars as the initial condition.

\acknowledgments

The author thanks T. Baumgarte, Y. Kojima, T. Nakamura, K. Oohara, 
L. Rezzolla, M. Sasaki, and S. Shapiro 
for helpful conversations and discussions. 
He also thanks T. Baumgarte and L. Rezzolla for reading this 
manuscript and providing valuable comments. 
Numerical computations were performed on the 
FACOM VPP 300R and VX/4R machines 
in the data processing center of the 
National Astronomical Observatory of Japan. 
This work was supported by a Grant-in-Aid (Nos. 08NP0801 and 
09740336) of 
the Japanese Ministry of Education, Science, Sports and Culture, 
and JSPS Fellowships for Research Abroad. 

\appendix
\section{Numerical method for solving hydrodynamic equations}

To compute the advection in the hydrodynamic equations, 
in this paper, 
we use the second order scheme by van Leer \cite{van} as 
Oohara and Nakamura used in their works \cite{ON,ON1} or 
as we used in a semi relativistic simulation \cite{shibaPNS}. 
As we have shown in Sec. III and IV, it is possible 
to perform simulations of many interesting problems 
stably and fairly accurately for several dynamical timescales 
using this scheme. 
If we are interested in problems in which 
shocks play a very important role, we should use 
one of the modern shock-capturing schemes as adopted in 
\cite{waimo,spain}. However, 
in coalescences of binary neutron stars, it seems 
that shocks do not play a crucially important 
role and that they do not change the results drastically 
\cite{ruffert}. We expect that 
the method employed here is suitable for our purpose, 
although there is still room for further improvement. 

Since we solve the entropy equation instead of the energy equation, 
it is impossible to capture 
shocks without adding artificial viscosity. Hence, 
we add an artificial viscosity using a method similar to that 
adopted by Hawley, Smarr and Wilson \cite{HSW}. 
The following artificial viscosity was included 
only in the case when we studied 
merger of binary neutron stars. 

For simplicity, 
we include the artificial viscosity like a bulk viscosity and 
it consequently appears in the evolution equation where 
the pressure does. As the form of the viscous 
pressure, we choose 
\beq
P_{\rm vis}=\left\{
\begin{array}{ll}
\displaystyle 
C_{\rm vis} 
{e_*^{\Gamma} \over (w e^{6\phi})^{\Gamma-1}} (\delta v)^2 
& {\rm for}~\delta v  < 0,\\
0 & {\rm for}~\delta v \geq 0 , \\
\end{array}
\right.
\eeq
where $\delta v =2\delta x \pa_k v^k$ and 
$C_{\rm vis}$ is a constant which we phenomenologically 
set below. 

When including the artificial viscosity, 
we change the pressure gradient term $\pa_k P$ in the right-hand 
side of Eq. (\ref{euler}) into $\pa_k (P+P_{\rm vis})$. 
If we simply regard $P_{\rm vis}$ as an additional pressure, 
we should add the following term is the right-hand side of 
Eq. (\ref{energy}):
\beq
-{1 \over \Gamma}(\rho \epsilon)^{-1+1/\Gamma} P_{\rm vis}
[\pa_t(we^{6\phi})+\pa_k(we^{6\phi}v^k)]. 
\eeq
However, as pointed out by Hawley et al.\cite{HSW}, 
the first term, 
which includes a time derivative, could cause a numerical instability. 
Hence, we neglect the first term, and only include 
the second one. 

$C_{\rm vis}$ is determined in a phenomenological manner. 
In merger of binary neutron stars, two neutron stars of 
nearly equal mass finally collide and a shock will be produced. 
The situation may be crudely 
compared with the 1D wall shock problem 
in which we consider the evolution of a fluid 
that initially has an uniform density and pressure, 
but a velocity field as 
\beq
v^x=\left\{
\begin{array}{ll}
\displaystyle  
V_0 & {\rm for}~x  < 0,\\
-V_0 & {\rm for}~x > 0 , \\
\end{array}\right.
\eeq
where $V_0$ is a positive constant, and $v^y=v^z=0$. 
The solution of the 1D wall shock problem in special 
relativity can be analytically calculated. We have performed 
test simulations of this problem, and the value of $C_{\rm vis}$ which 
well reproduces the analytical solution was used in 
numerical computations.  

To mimic the collision of two neutron stars, we 
set initial conditions as $V_0=0.1-0.4$ and $P/\rho =0.1$ 
for the test. 
In Fig. 31, we show the results for $V_0=0.4$ and 
$C_{\rm vis}=10$ at 
$t=0.4$. In this test, we set $\rho=1$ and $\delta x =0.005$. 
The solid lines denote the analytical solution, and the 
open circles are the numerical results. Although well known 
spurious oscillation of $\rho$, $P$, and $v^x$ is 
generated \cite{HSW}, the numerical results agree with 
the analytical solution fairly well. 

\section{Radial oscillation period of spherical stars}

We estimate an approximate radial oscillation period 
of spherical stars using 
the method presented by Chandrasekhar \cite{chandra}. 

If we write the line element as 
\beq
ds^2=-e^{\nu}dt^2+e^{\Phi}d \bar r^2+\bar r^2d\Omega,
\eeq
then the angular frequency $\sigma$ of the radial oscillation 
for a polytropic star of radius $R$ is written as \cite{chandra}
\beqn
\sigma^2=\biggl[&&
4 \int^R_0e^{(\Phi+\nu)/2} \bar r {dP \over d\bar r}\xi^2 d\bar r 
\nonumber \\
&&~+\int^R_0 e^{(\Phi+3\nu)/2} {\Gamma P \over \bar r^2}
\biggl\{ {d \over d\bar r}(\bar r^2 e^{-\nu/2} \xi)\biggr\}^2 d\bar r 
\nonumber \\
&&~-\int^R_0 e^{(\Phi+\nu)/2} \biggl( {dP \over d\bar r}\biggr)^2
{\bar r^2\xi^2 \over \rho h} d\bar r \nonumber \\
&&~+8\pi\int^R_0 e^{(3\Phi+\nu)/2} 
\rho P h \bar r^2 \xi^2 d\bar r\biggr] 
\nonumber \\
&& \times \biggl[ \int^R_0 e^{(3\Phi-\nu)/2} \rho h \bar r^2 \xi^2 d\bar r
\biggr]^{-1}, 
\eeqn
where $\xi$ denotes the $\bar r$-component of the 
Lagrangian displacement. 
If we substitute an eigenfunction $\xi$ for the eigenvalue 
equation, we can obtain a correct $\sigma$. Here, we evaluate 
$\sigma$ approximately by substituting a trial 
eigenfunction for $\xi$ 
without solving an eigenvalue equation for $\xi$. 
Following Chandrasekhar \cite{chandra}, 
we substitute the approximate eigenfunctions as 
\beq
\xi=\bar r e^{b \nu},
\eeq
where $b$ is a constant. 
In Fig. 32, we show $\sigma \rho_c^{-1/2}$ as a function of $\rho_c$ for 
polytropic star of $(\Gamma, K)=(5/3,10)$ and $(2, 200/\pi)$ 
for $b=0$ (the solid line), 1/2 (the dashed line) and 
$-1/2$ (the long dashed line). We note that beyond a 
critical density, $\sigma^2$ becomes negative in each model, 
implying that the star becomes unstable for some 
oscillation modes. Even though we use the trial function for 
$\xi$, we can approximately find the real critical points 
of stability ($\rho_c \sim 1.85 \times 10^{-3}$ for $\Gamma=5/3$ 
and $\sim 5.0 \times 10^{-3}$ for $\Gamma=2$). 
We also find that the oscillation frequency in the 
Newtonian limit ($\rho_c \rightarrow 0$) agrees well with 
that obtained in the Newtonian theory 
($\sigma \rho_c^{-1/2}=1.377$ for $\Gamma=5/3$, 
and $2.187$ for $\Gamma=2$ \cite{cox}). 
Thus, we can recognize that this approximate method is fairly reliable 
for estimating $\sigma$ approximately. 

From the figure, we can judge $\sigma \sim 0.60 \rho_c^{1/2}$ for 
$(\Gamma,\rho_c)=(5/3,10^{-3})$ and 
$\sigma \sim 0.91 \rho_c^{1/2}$ for 
$(\Gamma,\rho_c)=(2,3\times 10^{-3})$, and the resulting 
oscillation periods are $\sim 10.5 \rho_c^{-1/2}$ 
and $\sim 6.9 \rho_c^{-1/2}$, respectively. 
These are in good agreement with the numerical results in 
Sec. III.B and III.C.

\vskip 5mm
\noindent 
{\bf Table I.~} The list of the central density $\rho_c$, 
total rest mass $M_*$, gravitational mass $M_g$, 
$J/M_g^2$, and rotation period ${\rm P}$ for 
rotating neutron stars of $\Gamma=5/3$ and $2$ 
at mass-shedding limits. 

\vskip 5mm
\noindent
\begin{center}
\begin{tabular}{|c|c|c|c|c|c|c|c|} \hline
\hspace{1mm} $\rho_{c} $ \hspace{1mm} &
\hspace{1mm} $M_*$ \hspace{1mm} & \hspace{1mm} $M_g$\hspace{1mm} 
& \hspace{1mm} $J/M_g^2$ \hspace{1mm} 
& \hspace{1mm} ${\rm P} $ \hspace{1mm} & ~~$\Gamma$~~ \\ \hline
$ 9.94 \times 10^{-4}$ & 1.67 &  1.60 & 0.427 & 397 & 5/3 \\ 
\hline 
$ 2.77 \times 10^{-3}$ & 1.58 &  1.45 & 0.598 & 163 & 2 
\\ \hline
\end{tabular}
\end{center}

\vskip 5mm
\noindent 
{\bf Table II.~} The list of the maximum density, 
total rest mass $M_*$, gravitational mass $M_g$, 
$J/M_g^2$, coordinate separation $a$, orbital period ${\rm P}$, and 
ratio of the emission timescale of gravitational waves to 
the orbital period $R_{\tau}$ for corotating 
binary neutron stars in approximate quasi-equilibrium states.

\vskip 5mm
\noindent
\begin{center}
\begin{tabular}{|c|c|c|c|c|c|c|c|} \hline
\hspace{1mm} $\rho_{\rm max} $ \hspace{1mm} &
\hspace{1mm} $M_*$ \hspace{1mm} & \hspace{1mm} $M_g$\hspace{1mm} 
& \hspace{1mm} $J/M_g^2$ \hspace{1mm} & \hspace{1mm} $a$ \hspace{1mm}
& \hspace{1mm} ${\rm P} $ \hspace{1mm} 
& \hspace{2mm} $R_{\tau}$ \hspace{2mm}  \\ \hline
$ 6\times 10^{-4}$ & 2.84 & $2.72$ & 1.20 & 38.3 & 902 & 9.2 
\\ \hline
$ 10^{-3}$ & 3.07 & $2.92$ & 1.11 & 32.9 & 694 & 5.3
\\ \hline
\end{tabular}
\end{center}


\clearpage
\begin{figure}[t]
\epsfxsize=5.5in
\leavevmode
\epsffile{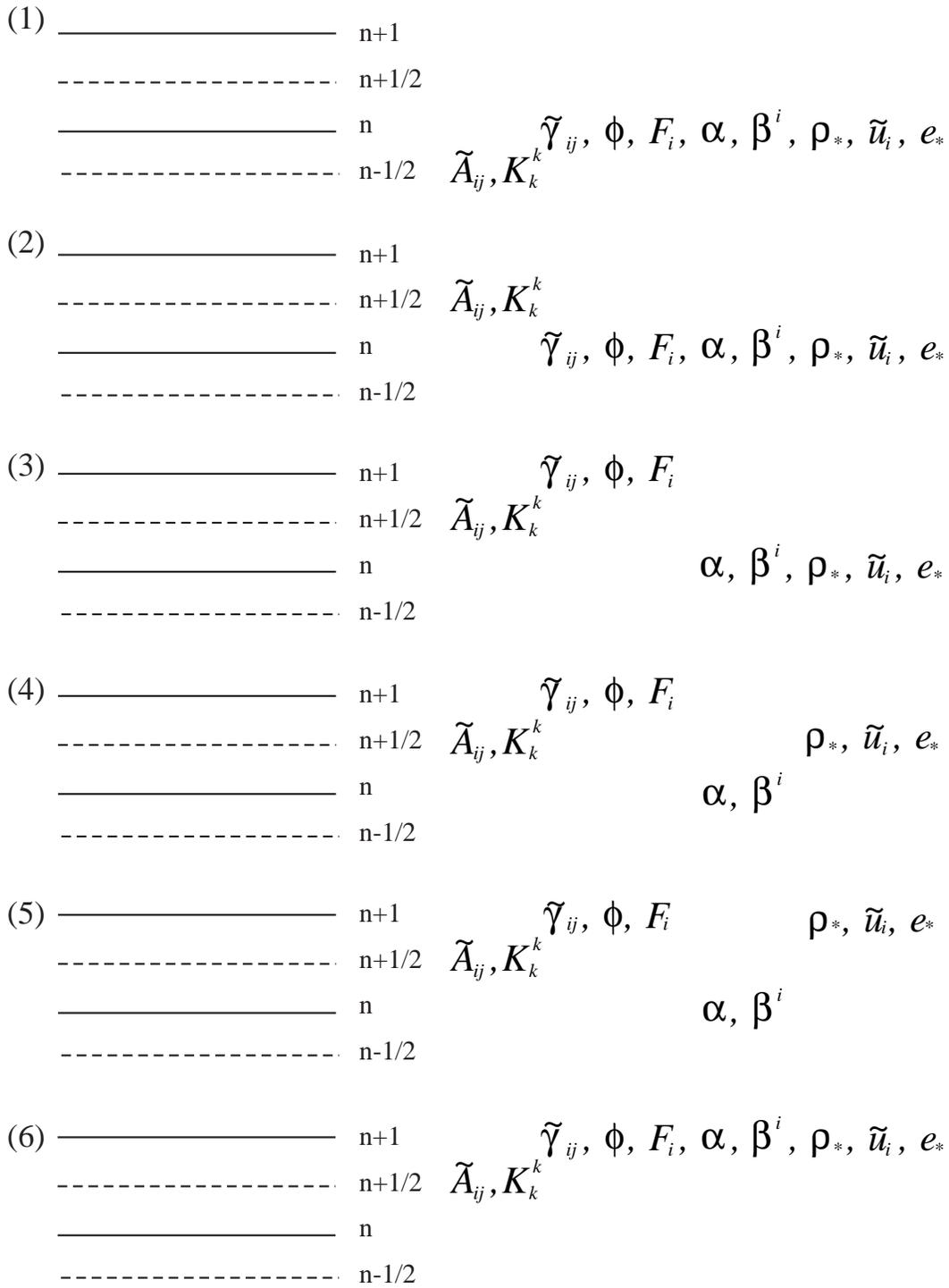}
~~\\
\caption{Schematic picture for a numerical scheme of 
time evolution of variables from $n$-th to 
($n+1$)-th time step. 
}
\end{figure}

\clearpage

\begin{figure}[t]
\epsfxsize=3.in
\leavevmode
\epsffile{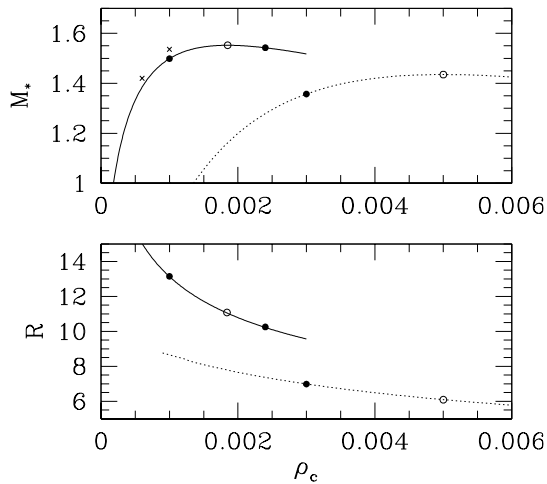}
\caption{Rest mass $M_*$ and circumferential radius $R$ 
as a function of central density $\rho_c$ for spherical 
polytropic stars of $K=10$ and $\Gamma=5/3$ (solid lines) and 
of $K=200/\pi$ and $\Gamma=2$ (dotted lines). 
The filled circles denote the equilibrium stars 
which are used in test simulations (the tests 2 and/or 3),  
and the open circles 
denote the critical configuration of stability against 
gravitational collapse. The cross in the figure for 
$M_*-\rho_c$ denotes 
the relation between $M_*/2$ and the maximum density for 
corotating binary 
neutron stars in approximate quasi-equilibrium states which are 
adopted as initial conditions in Sec. III.E and IV. 
}
\end{figure}

\begin{figure}[t]
\begin{center}
\epsfxsize=3.in
\leavevmode
\epsffile{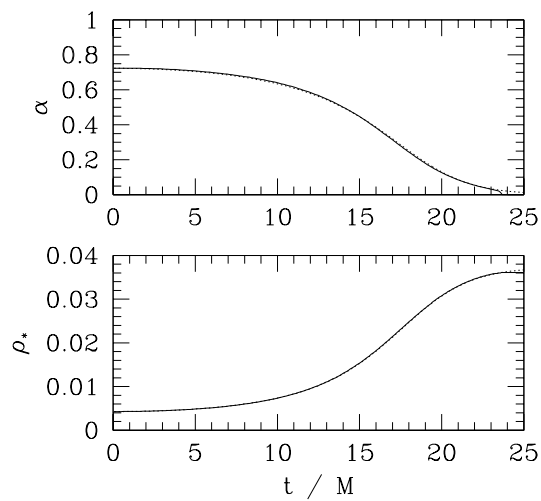}
\end{center}
\caption{$\alpha$ and $\rho_*$ at $r=0$ as a function of time 
for a spherical dust collapse with zero shift gauge condition. 
The dotted lines are the 1D results, and 
the solid lines are the 3D results. 
}
\end{figure}

\begin{figure}[t]
\epsfxsize=3.in
\leavevmode
\epsffile{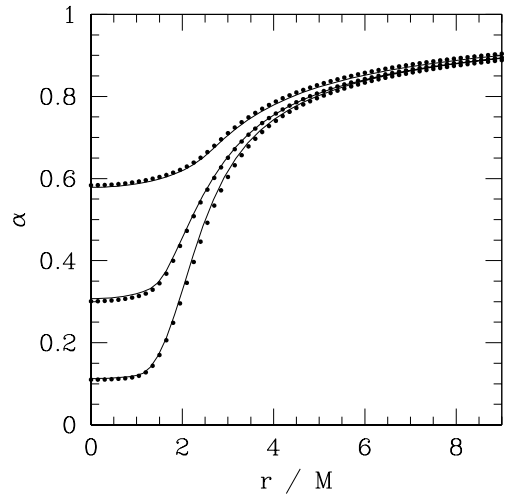}
\caption{$\alpha$ along the $x$-axis at selected times 
($t=12.0, 17.2$, and 20.4)
for a spherical dust collapse with zero shift gauge condition 
for the 1D results (the solid lines) and 3D results (the filled 
circles).
}
\end{figure}

\begin{figure}[t]
\epsfxsize=3.in
\leavevmode
\epsffile{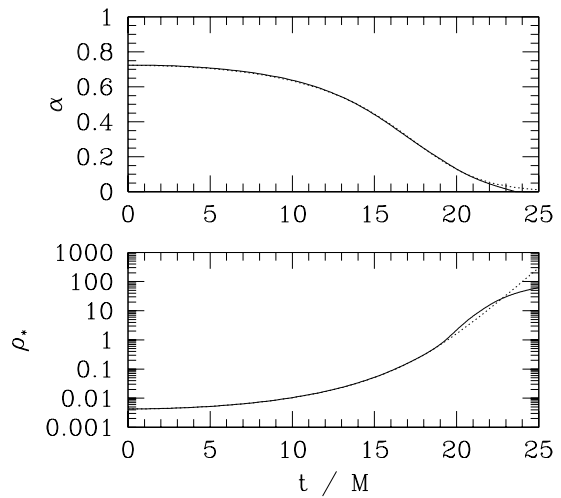}
\caption{The same as Fig. 3, but with the 
MD (or AMD) gauge condition. 
The dotted and solid lines denote the 1D and 3D results, respectively. 
}
\end{figure}

\begin{figure}[t]
\epsfxsize=3.in
\leavevmode
\epsffile{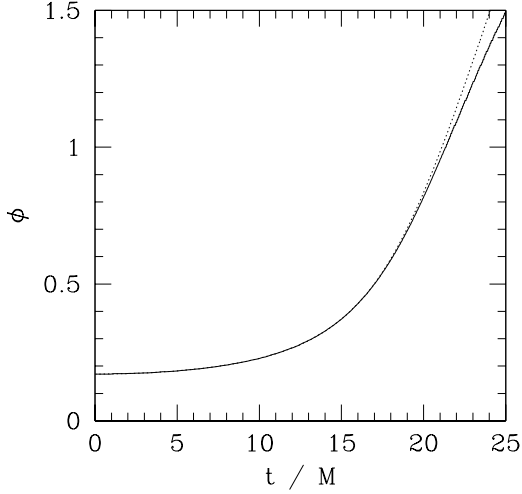}
\caption{$\phi$ as a function of time 
for a spherical dust collapse in the MD (or AMD) gauge condition. 
The dotted and solid lines denote the 1D and 3D results, respectively. 
}
\end{figure}

\begin{figure}[t]
\epsfxsize=3.in
\leavevmode
\epsffile{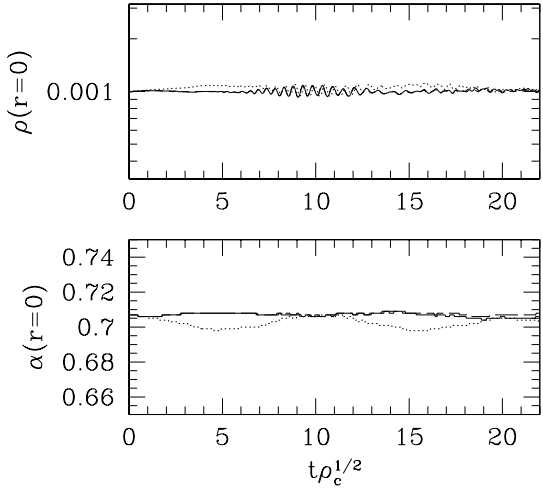}
\caption{$\rho$ and $\alpha$ at $r=0$ as a function of time 
($t \rho_c^{1/2}$) in 3D numerical evolution of a stable spherical 
polytropic star of $\rho_c=10^{-3}$, $K=10$ and $\Gamma=5/3$. 
The solid and dotted lines denote the results in zero shift gauge 
condition, and the dashed line in the AMD gauge condition. 
In the simulation shown with the dotted line, we reduce $K$ by a 
factor $0.5\%$ initially, and in other cases, we give equilibrium 
states without any perturbation. 
}
\end{figure}

\begin{figure}[t]
\epsfxsize=3.in
\leavevmode
\epsffile{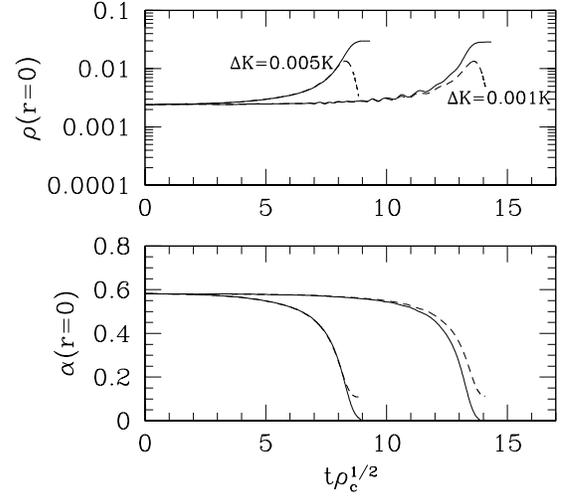}
\caption{$\rho$ and $\alpha$ at $r=0$ as a function of time 
($t \rho_c^{1/2}$) in 3D numerical evolution of an unstable spherical 
polytropic star of $\rho_c=2.4\times 10^{-3}$, $K=10$ and 
$\Gamma=5/3$. 
The solid and dashed lines denote the results in the zero shift gauge 
and AMD gauge conditions, respectively. 
In these simulations, we reduce $K$ by a 
factor $0.1\%$ or $0.5\%$ initially. 
}
\end{figure}

\begin{figure}[t]
\epsfxsize=3.in
\leavevmode
\epsffile{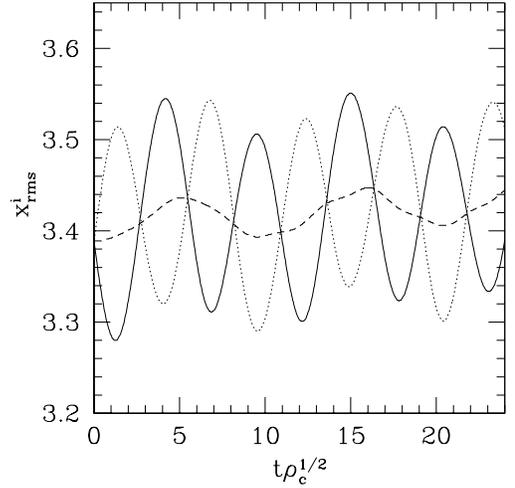}
\caption{$x^i_{\rm rms}$ and $\alpha$ at $r=0$ as a function of time 
($t \rho_c^{1/2}$) for a perturbed spherical 
star of $\rho_c=10^{-3}$, $K=10$ and $\Gamma=5/3$ 
with a quadrupole perturbation of $+$ mode. 
The solid, dotted and dashed lines denote $x_{\rm rms}$, 
$y_{\rm rms}$, and $z_{\rm rms}$, respectively. 
}
\end{figure}

\begin{figure}[t]
\epsfxsize=3.in
\leavevmode
\epsffile{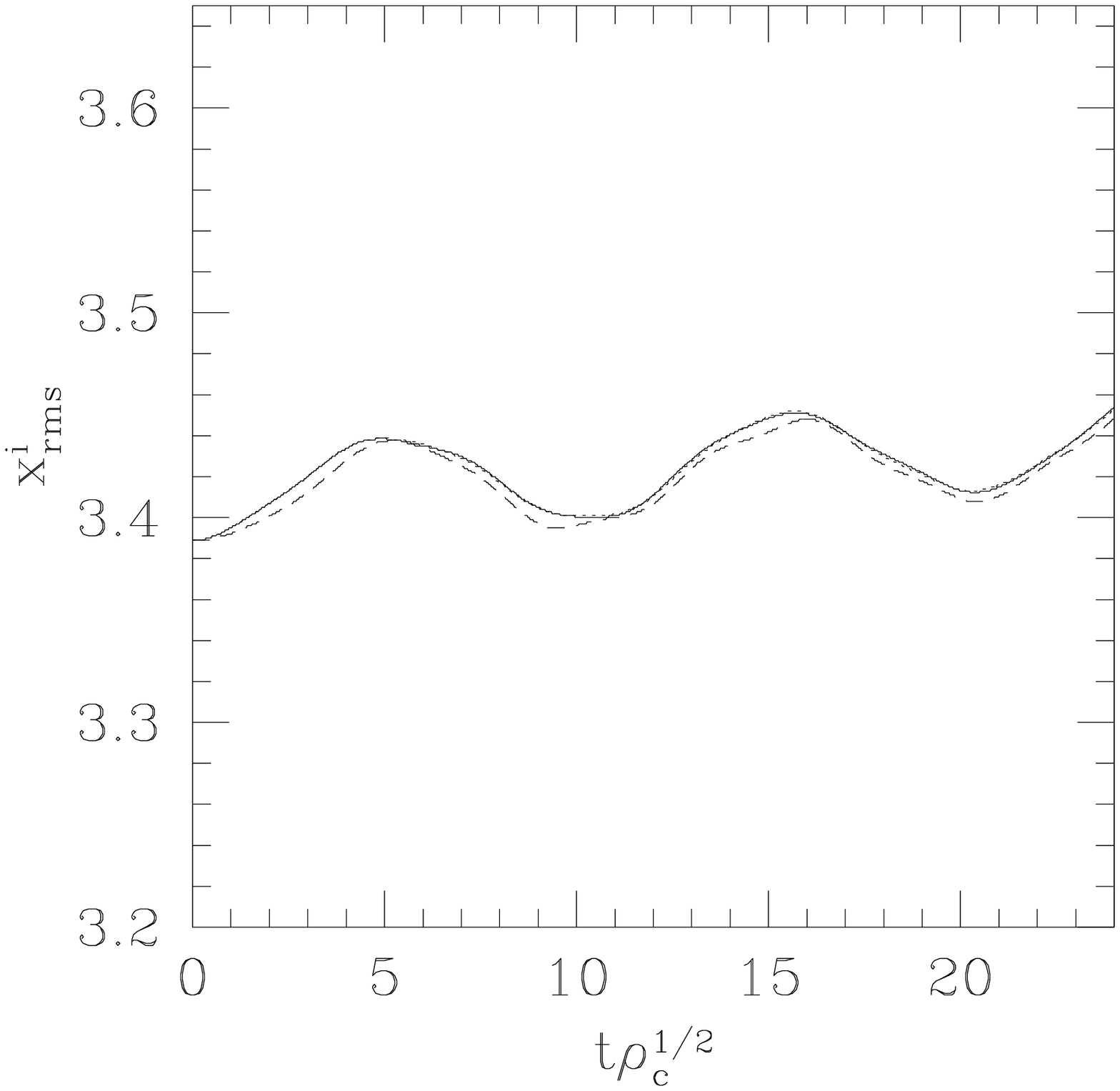}
\caption{The same as Fig. 9, but for $\times$ mode 
perturbation}
\end{figure}

\begin{figure}[t]
\epsfxsize=3.in
\leavevmode
\epsffile{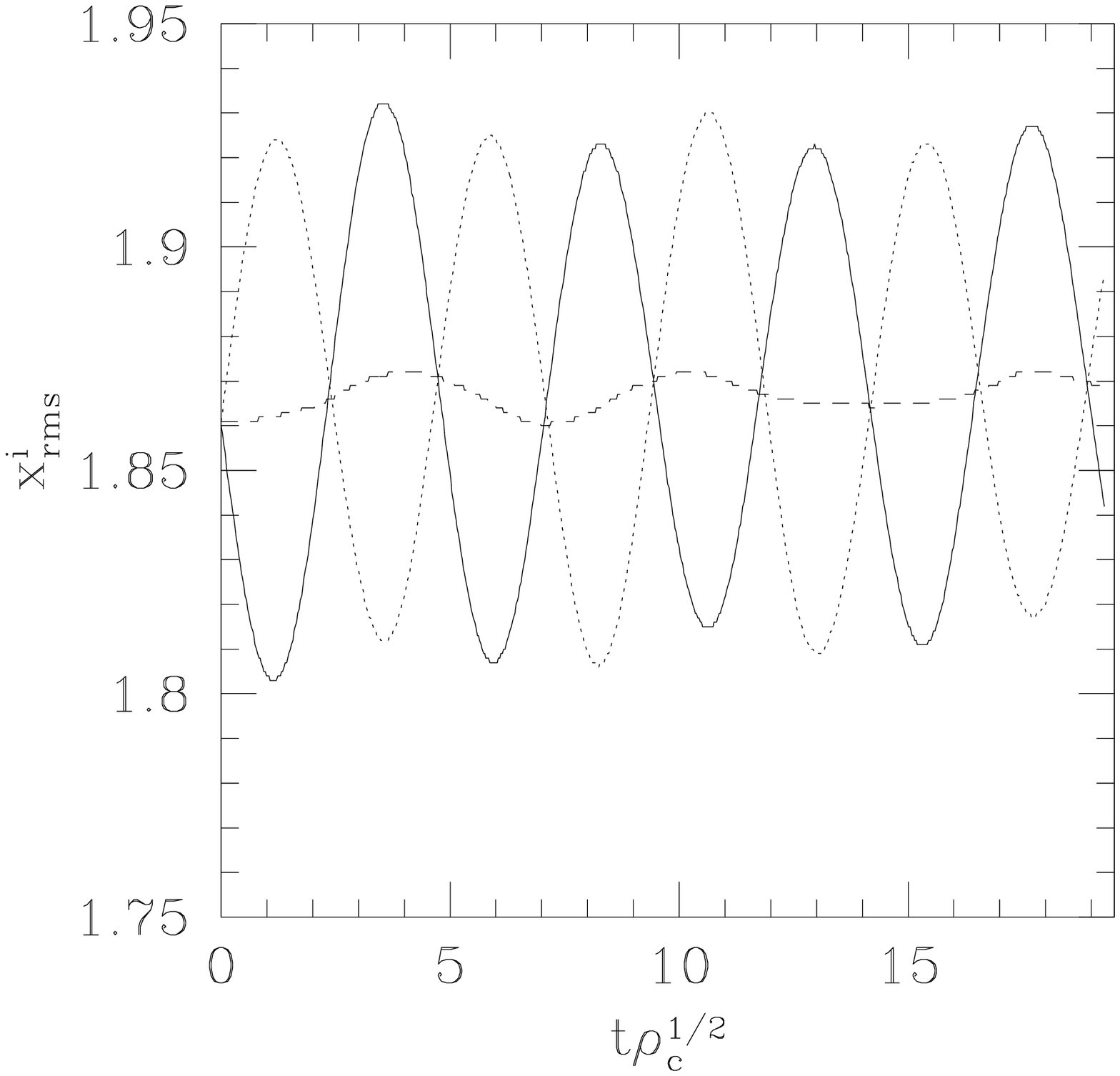}
\caption{The same as Fig. 9, but for 
$\rho_c=3\times 10^{-3}$, $\Gamma=2$, and $K=200/\pi$. }
\end{figure}

\begin{figure}[t]
\epsfxsize=3.in
\leavevmode
\epsffile{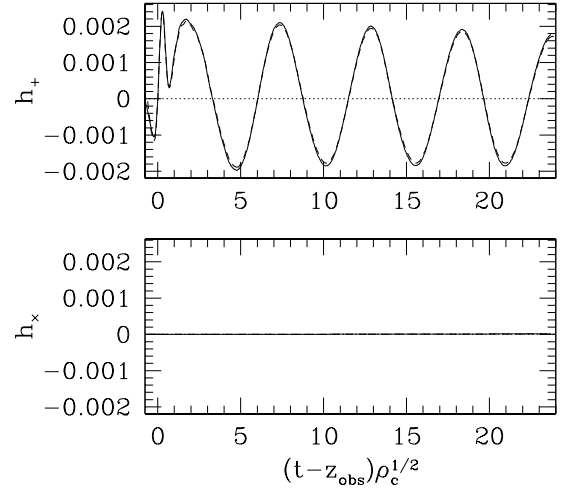}
\caption{$h_+$ and $h_{\times}$ as a function of a retarded 
time for a perturbed spherical 
star of $\rho_c=10^{-3}$, $K=10$ and $\Gamma=5/3$ 
with a quadrupole perturbation of $+$ mode.
The solid and dashed lines denote those extracted 
at $z_{\rm obs}=24.5$ and 19.5, respectively. 
}
\end{figure}

\begin{figure}[t]
\epsfxsize=3.in
\leavevmode
\epsffile{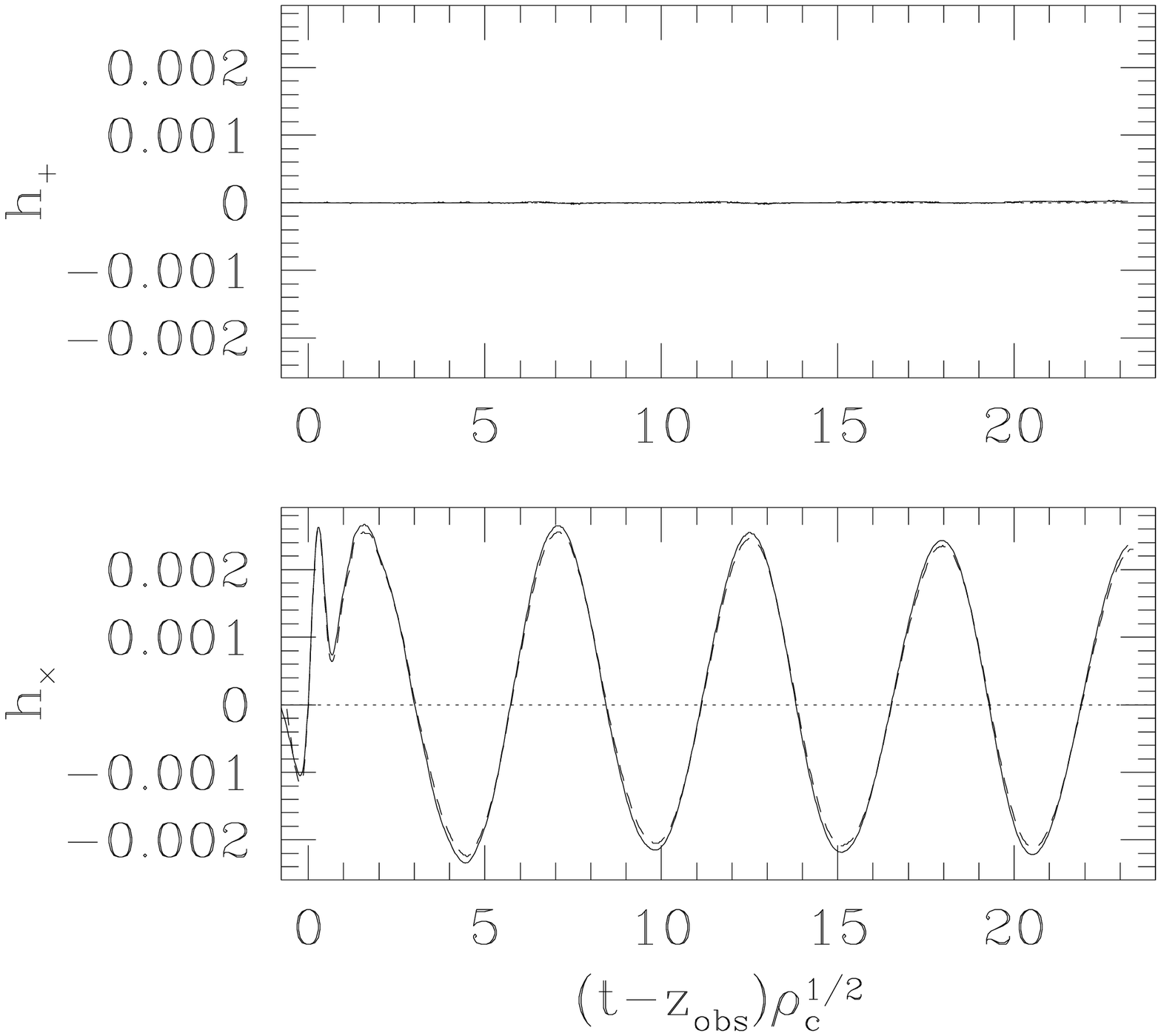}
\caption{The same as Fig. 12, but for $\times$ mode 
perturbation.}
\end{figure}

\begin{figure}[t]
\epsfxsize=3.in
\leavevmode
\epsffile{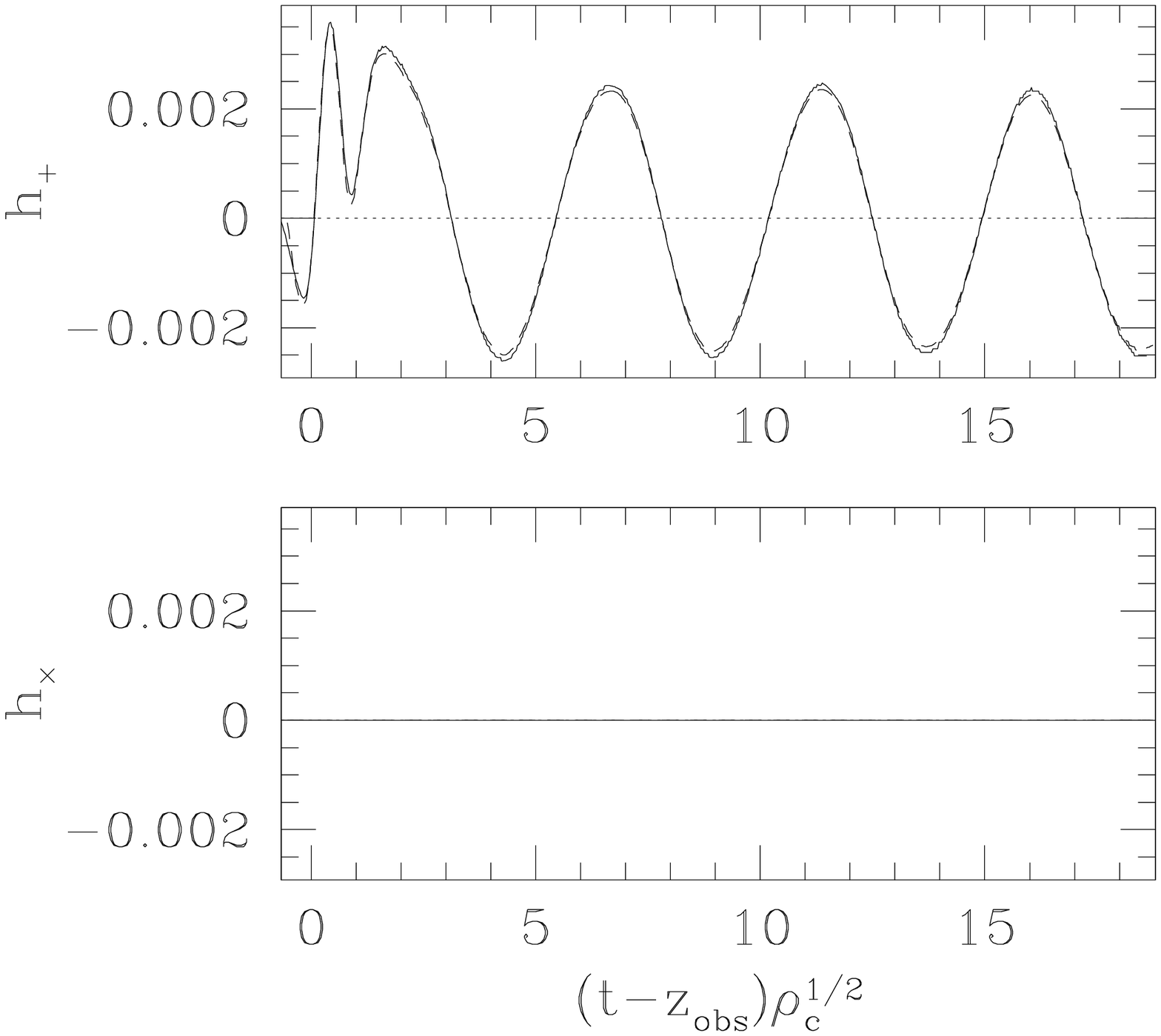}
\caption{The same as Fig. 12, but for 
$\rho_c=3\times 10^{-3}$, $\Gamma=2$, and $K=200/\pi$.}
\end{figure}

\begin{figure}[t]
\epsfxsize=3.in
\leavevmode
\epsffile{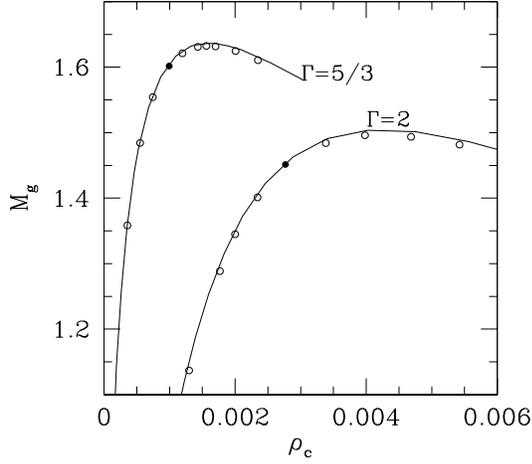}
\caption{The gravitational mass as a function of $\rho$ at $r=0$ 
($\rho_c$) for rotating stars of $\Gamma=5/3$ and $K=10$, and 
$\Gamma=2$ and $K=200/\pi$. 
The solid lines denote rotating stars at mass-shedding limits 
obtained from exact equations. 
The open and filled circles denote rotating stars at mass-shedding 
limits obtained using a conformal flatness approximation. 
The rotating stars we use in this paper are marked with 
the filled circles. }
\end{figure}


\clearpage

\begin{figure}[t]
\begin{center}
\epsfxsize=2.5in
\leavevmode
\epsffile{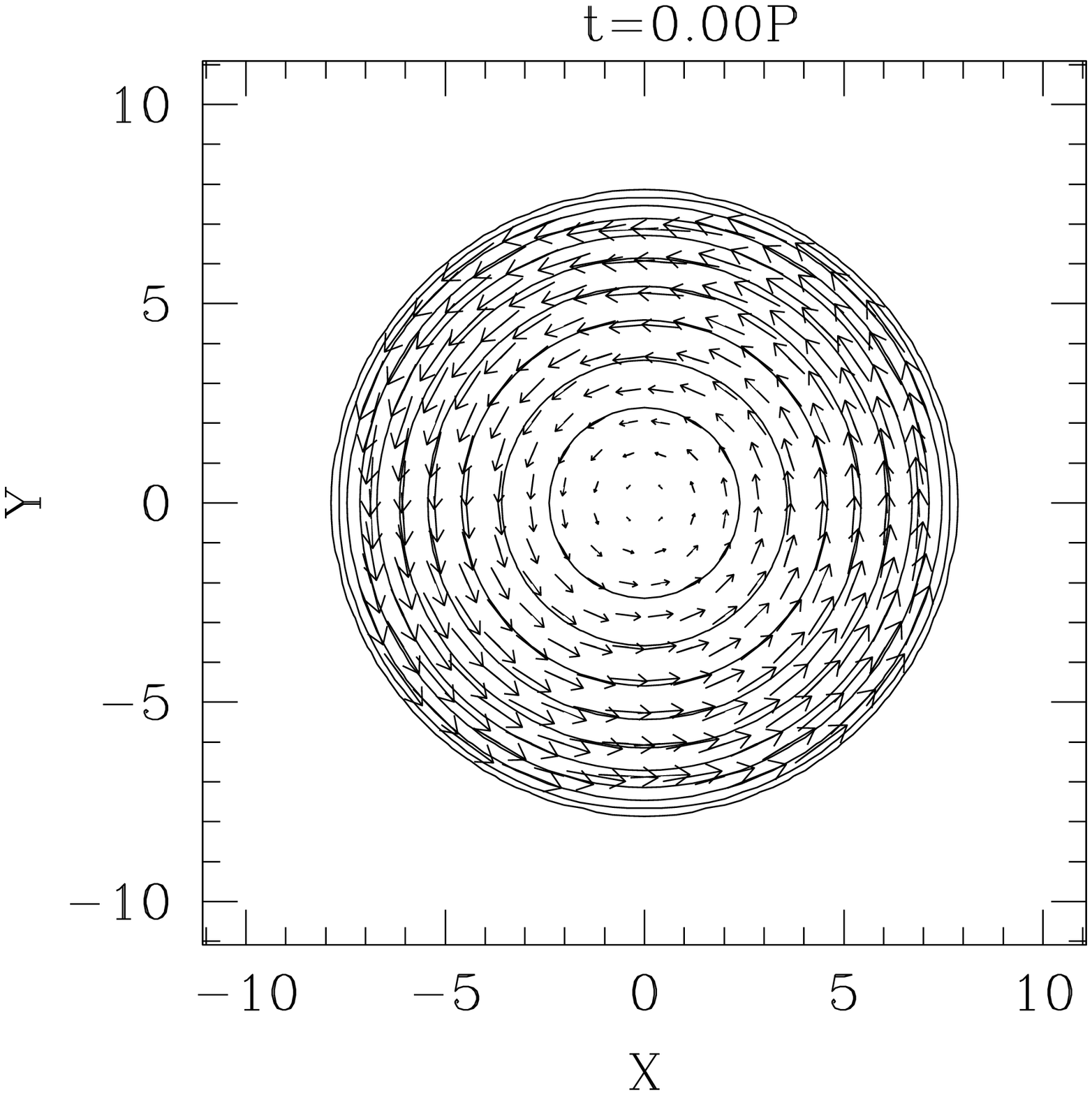}
\epsfxsize=2.5in
\leavevmode
\epsffile{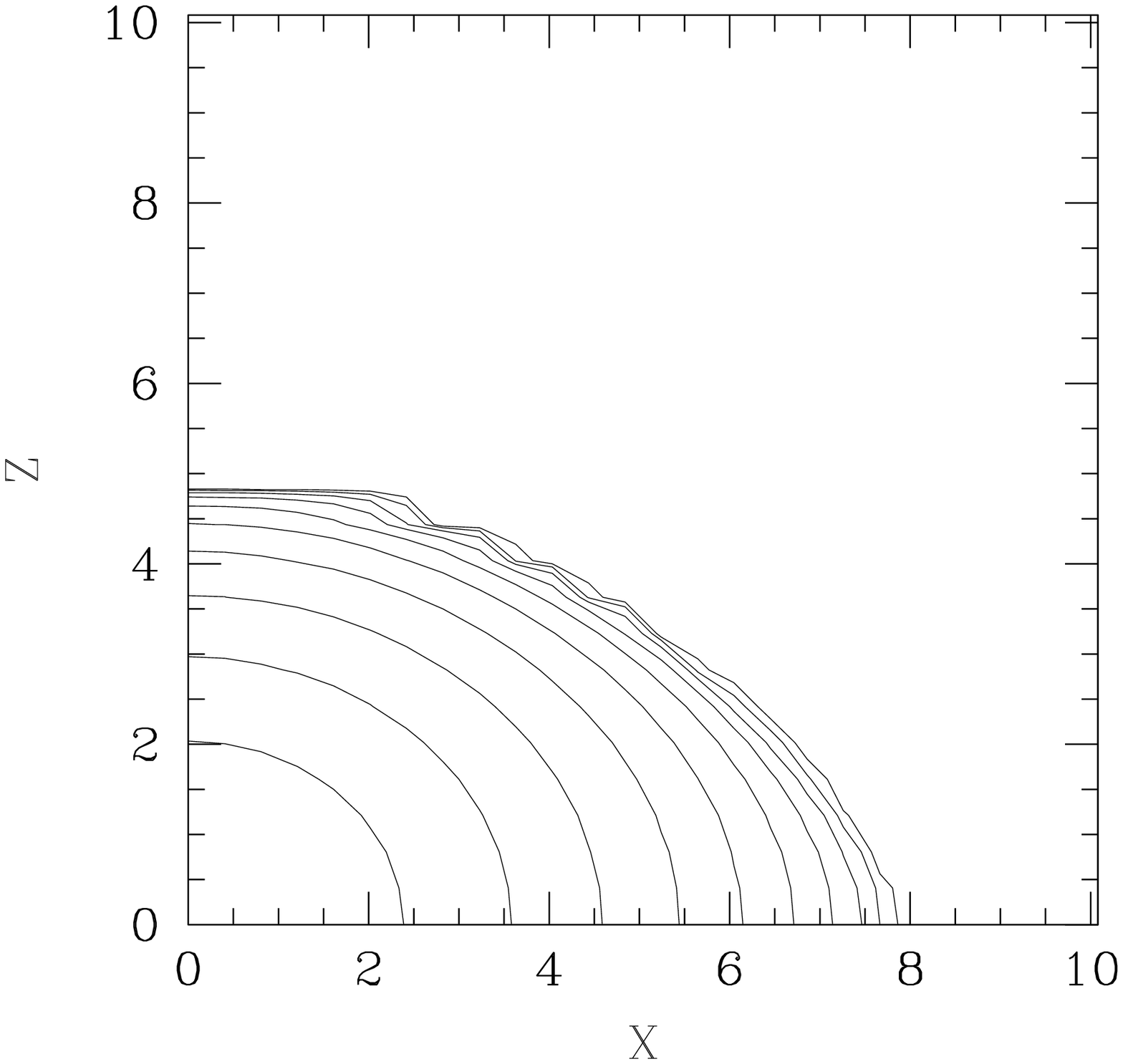}\\
\epsfxsize=2.5in
\leavevmode
\epsffile{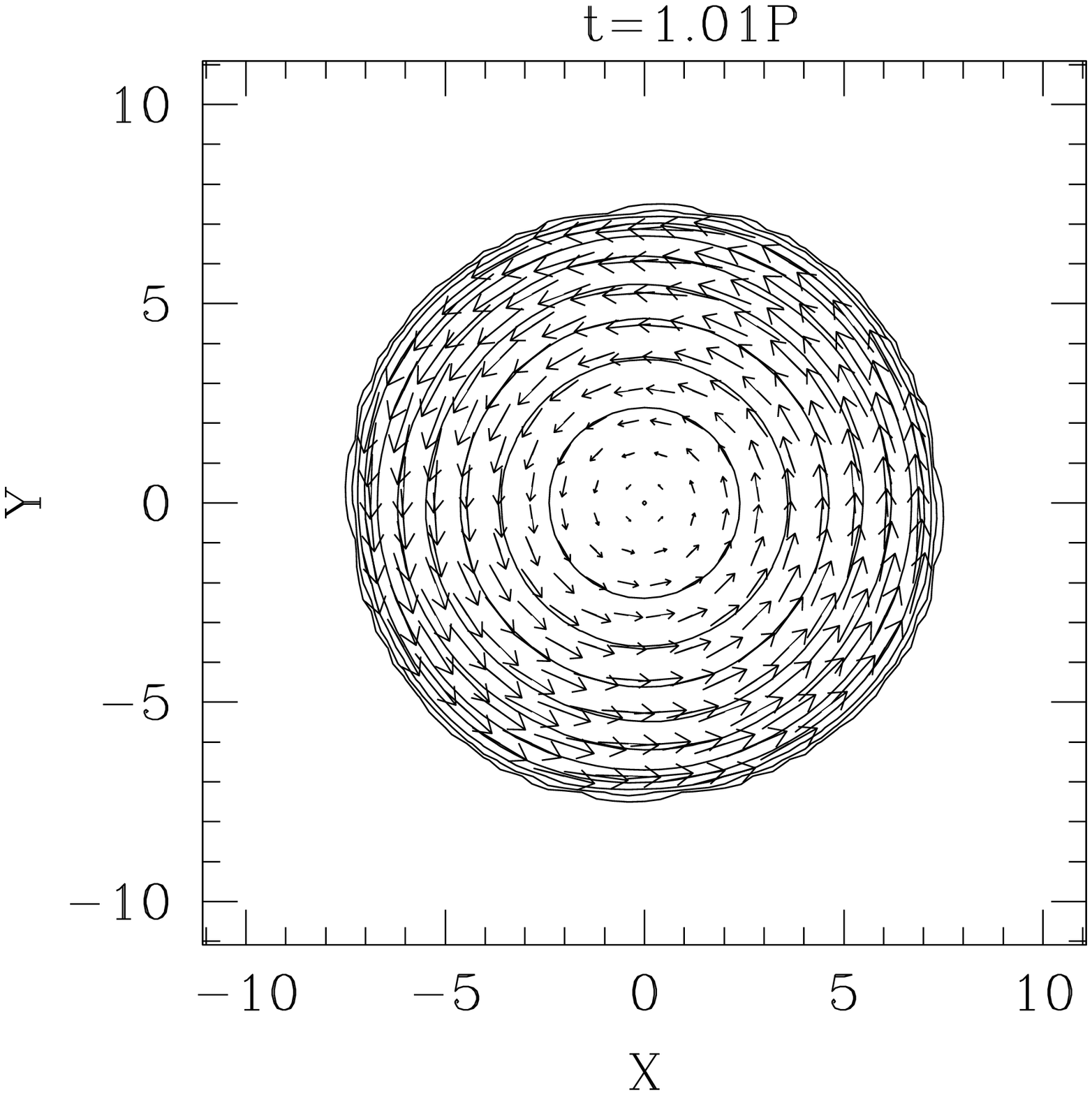}
\epsfxsize=2.5in
\leavevmode
\epsffile{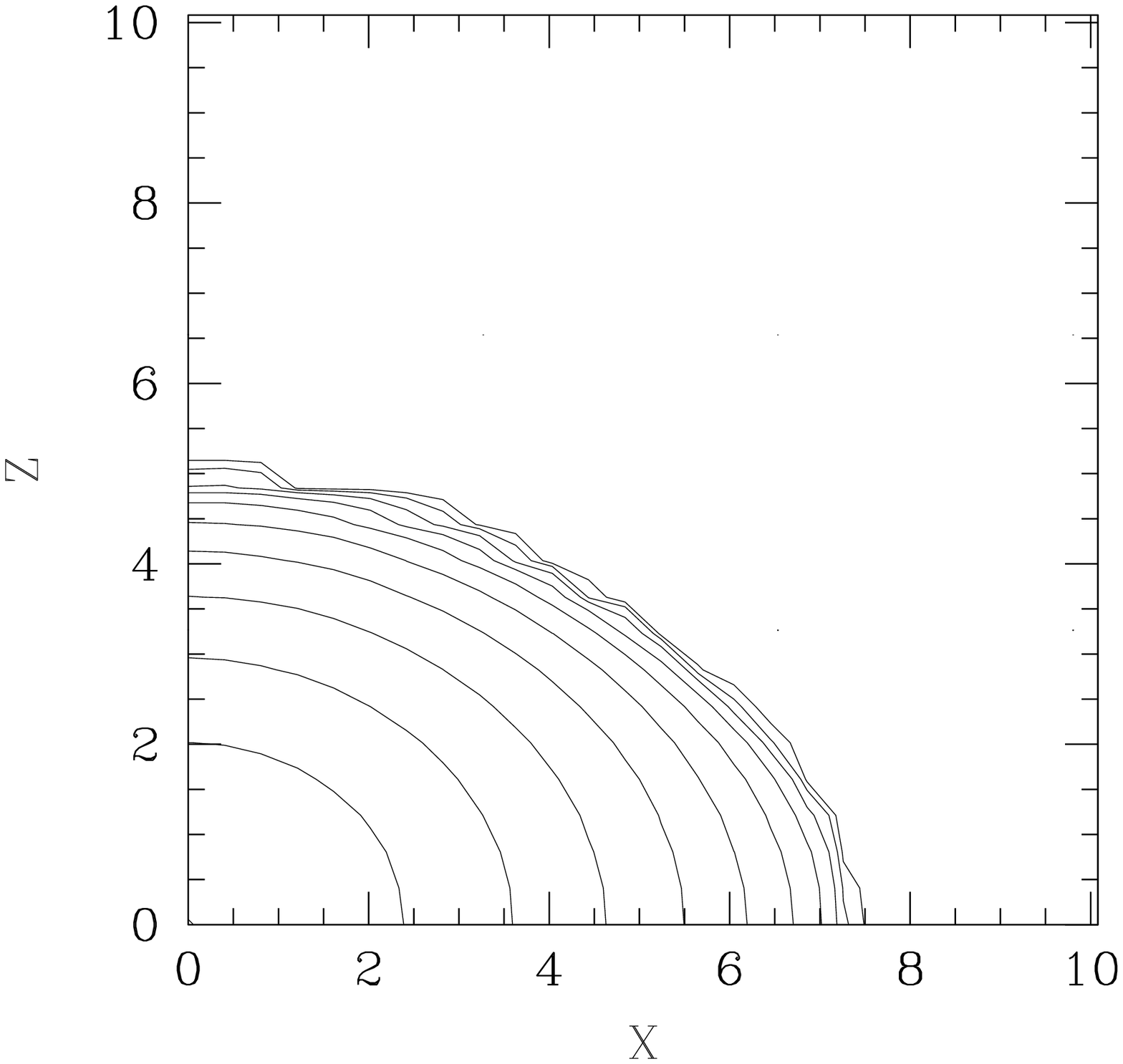}\\
\epsfxsize=2.5in
\leavevmode
\epsffile{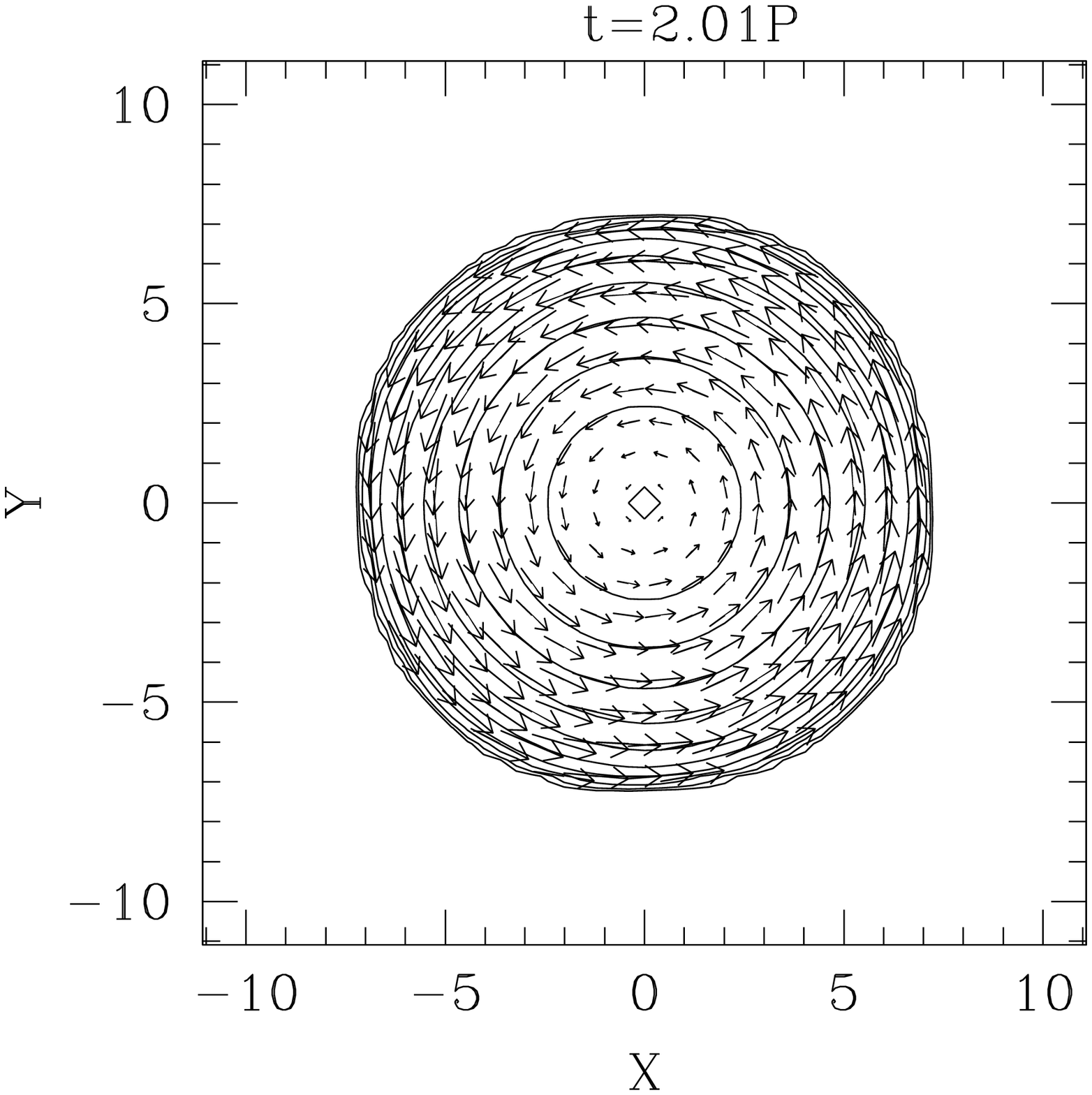}
\epsfxsize=2.5in
\leavevmode
\epsffile{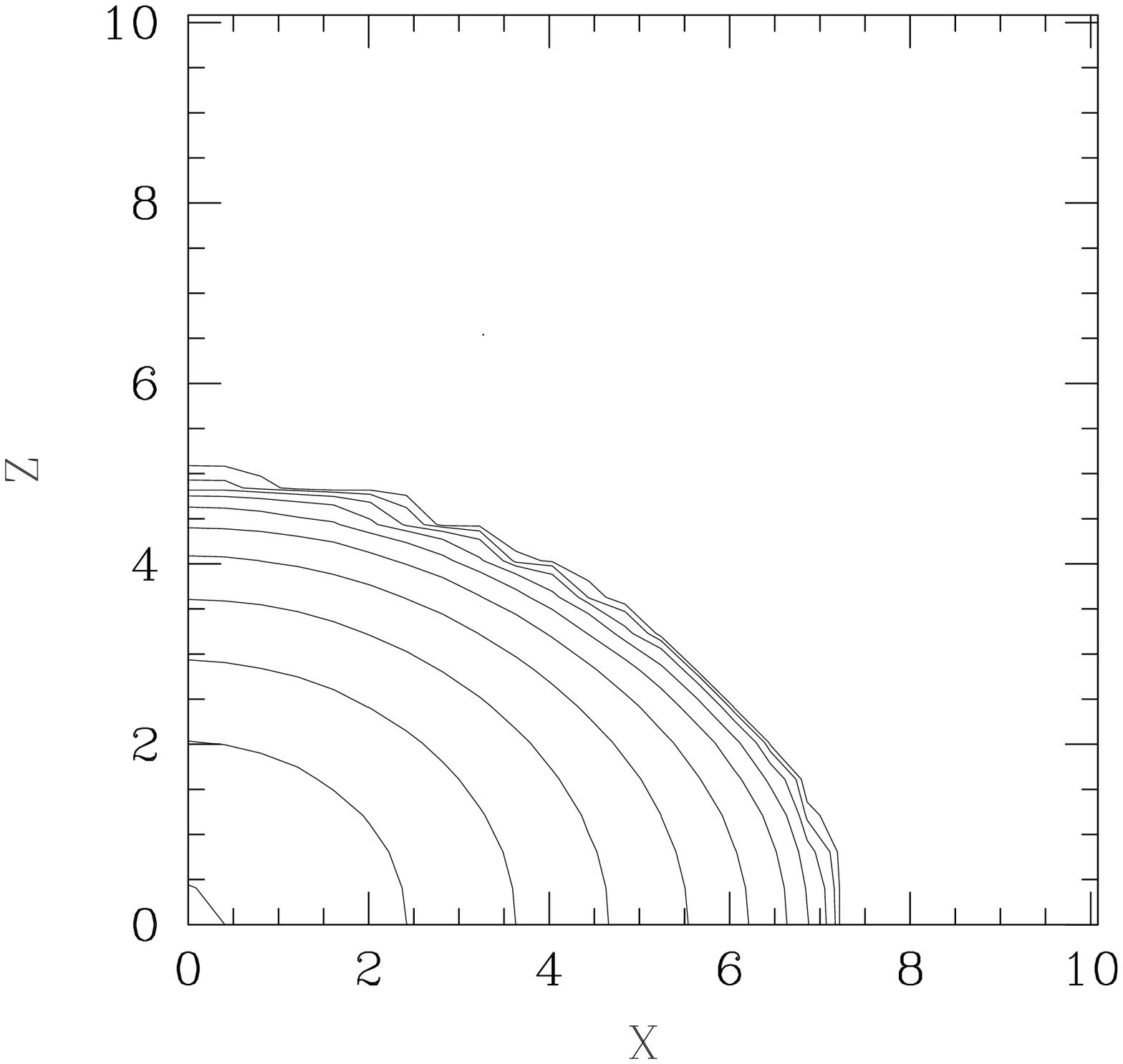}\\
\end{center}
\caption{Snapshots of the density contour lines for $\rho_*$ and 
the velocity flow for $(v^x,v^y)$ in the equatorial plane (left) 
and in the $y=0$ plane (right) for 
a rotating star at the mass-shedding limit and $\Gamma=2$ 
at selected times. 
The contour lines are drawn for 
$\rho_*/\rho_{*~c}=10^{-0.3j}$, where $\rho_{*~c}=0.0122$ denotes 
$\rho_*$ at $r=0$ and $t=0$, for $j=0,1,2,\cdots,10$. Vectors 
indicate the local velocity field and the maximum length denotes 
$\sim 0.26c$. 
}
\end{figure}

\clearpage

\begin{figure}[t]
\epsfxsize=3.in
\leavevmode
\epsffile{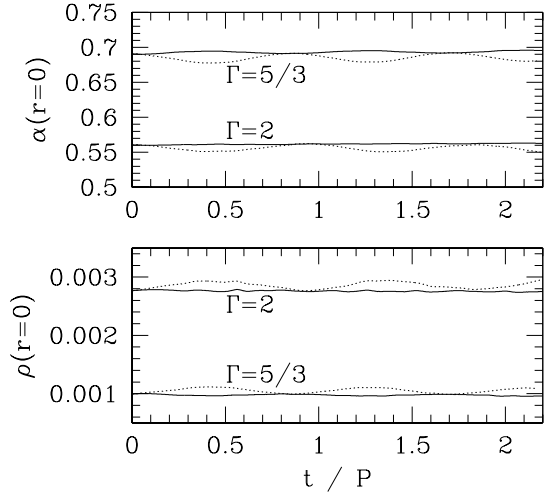}
\caption{$\alpha$ and $\rho(=\rho_* e^{-6\phi}/w)$ 
at $r=0$ as a function of $t/{\rm P}$ 
for $\Gamma=5/3$ and 2. 
The solid and dotted lines denote the results for 
initial conditions in which 
no perturbation is added and the pressure is depleted, 
respectively.  
}
\end{figure}

\begin{figure}[t]
\epsfxsize=3in
\leavevmode
\epsffile{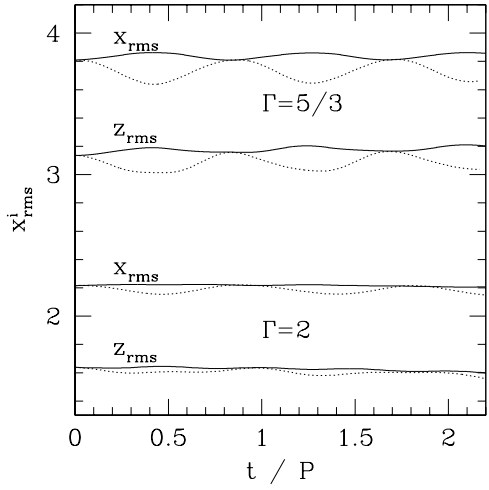}
\caption{The same as Fig. 17, but for 
$x_{\rm rms}$ and $z_{\rm rms}$ as a function of $t/{\rm P}$.  
}
\end{figure}

\begin{figure}[t]
\epsfxsize=3in
\leavevmode
\epsffile{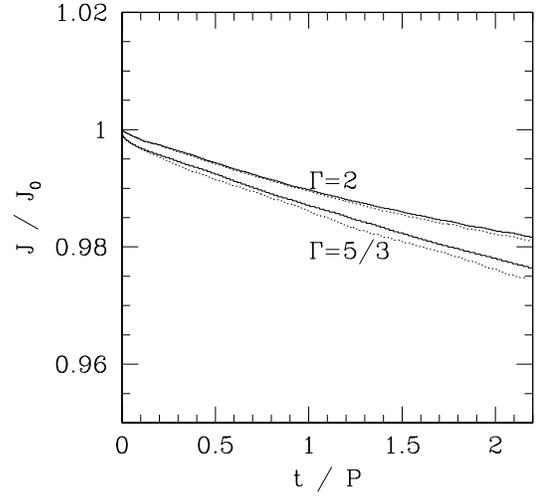}
\caption{The same as Fig. 17, but for $J/J_0$ as a function of $t/P$. 
}
\end{figure}

\clearpage

\begin{figure}[t]
\begin{center}
\epsfxsize=2.5in
\leavevmode
\epsffile{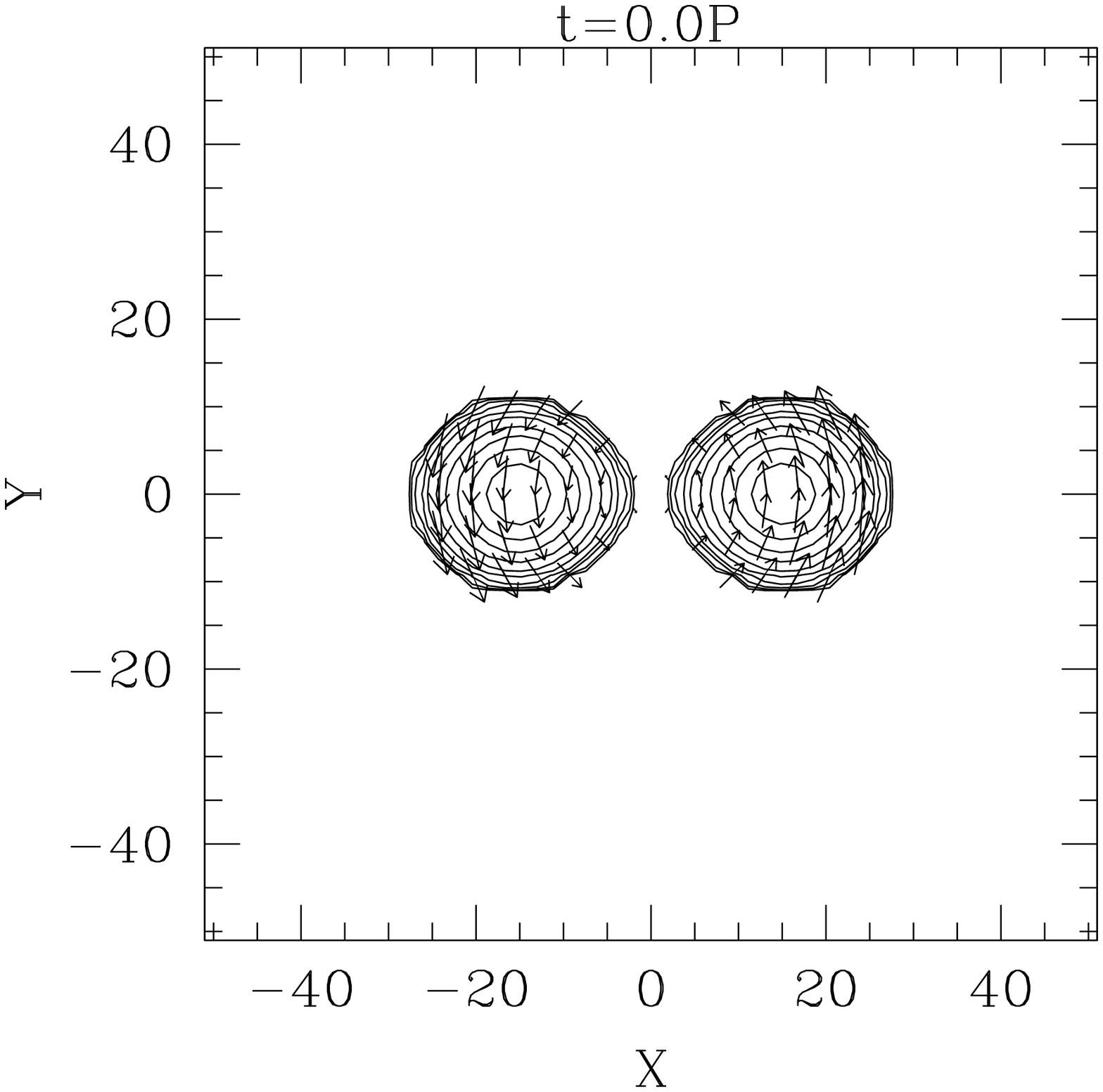}
\epsfxsize=2.5in
\leavevmode
\epsffile{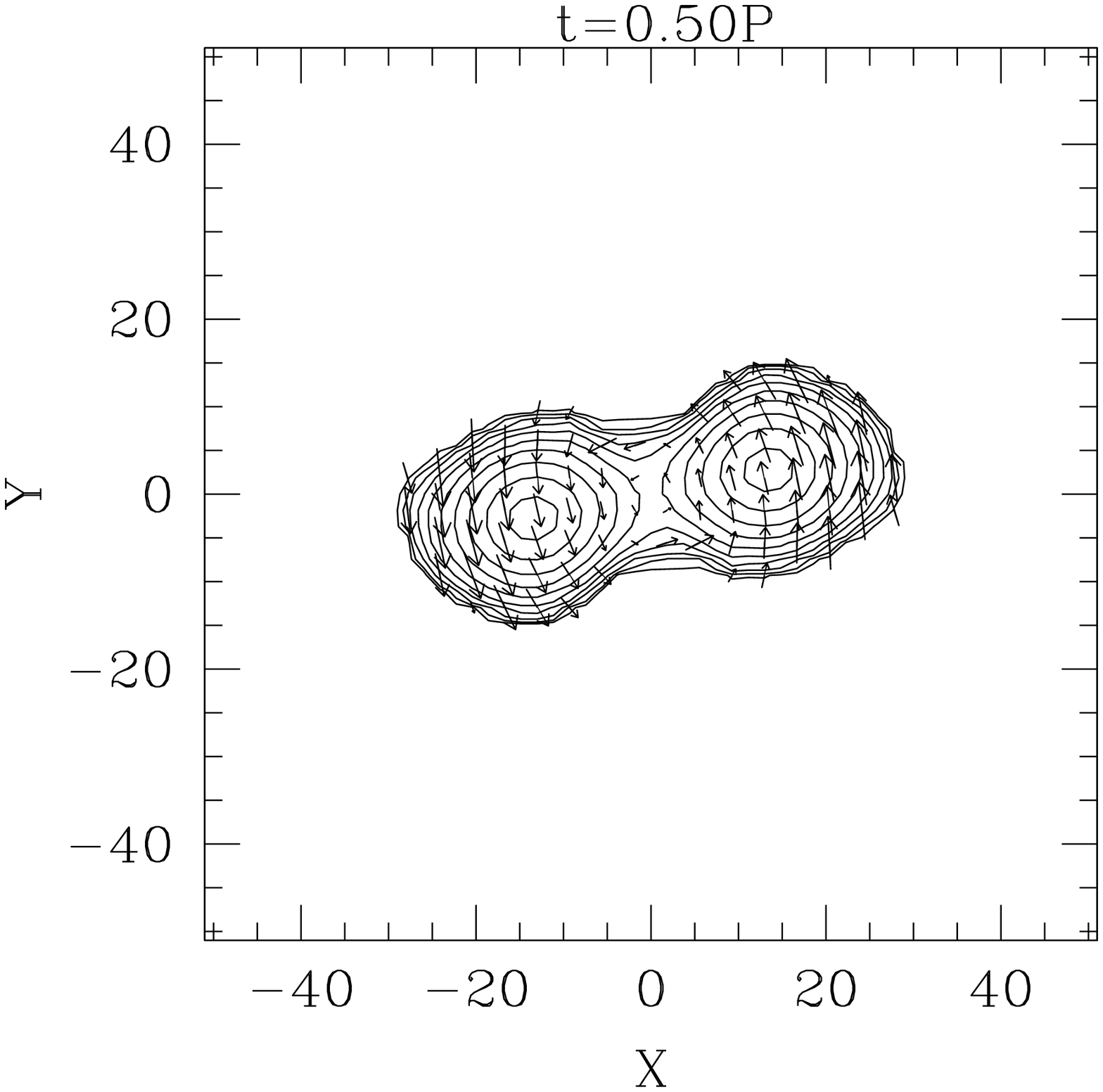}\\
\vskip 3mm
\epsfxsize=2.5in
\leavevmode
\epsffile{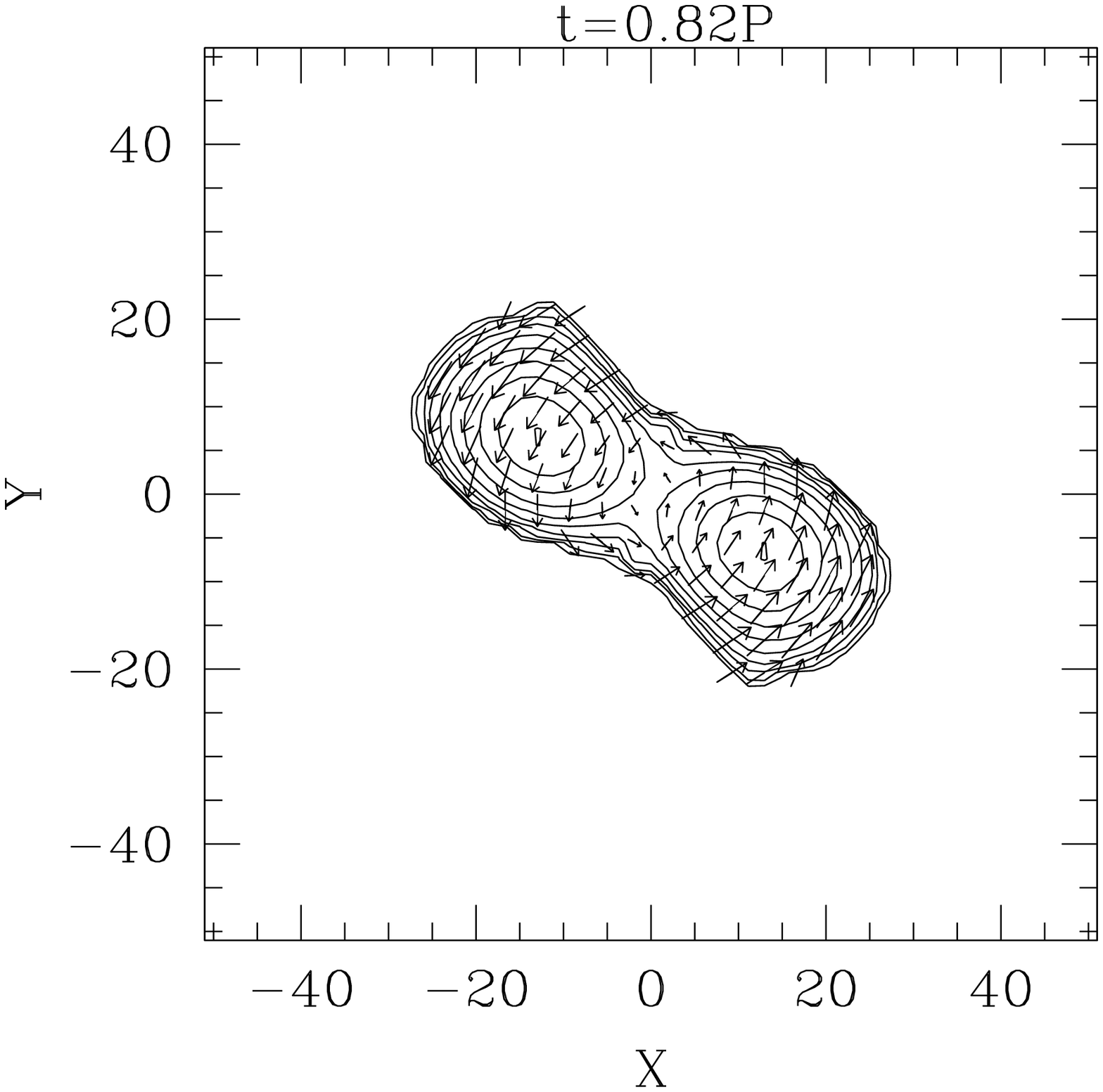}
\epsfxsize=2.5in
\leavevmode
\epsffile{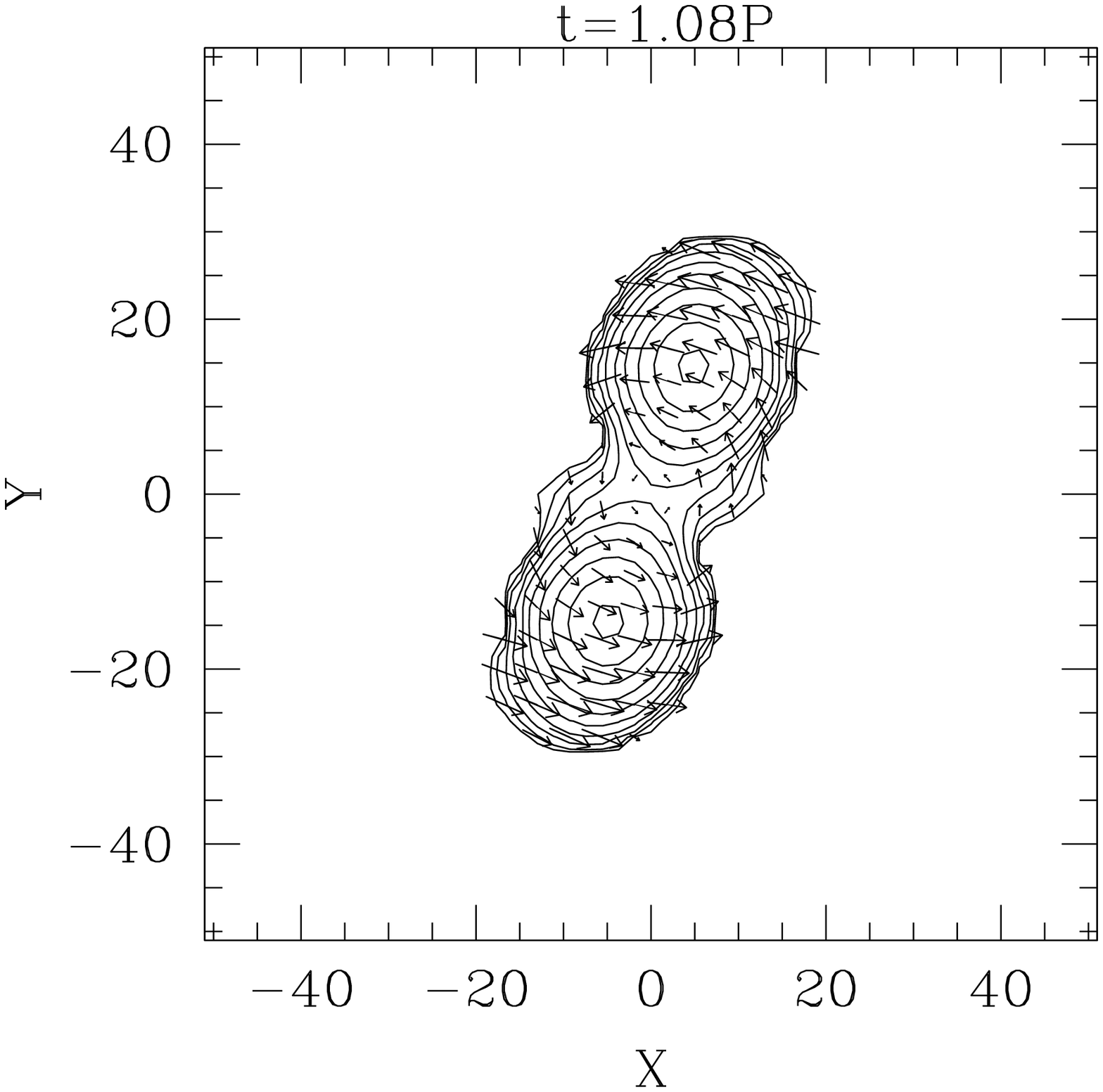}
\caption{
Snapshots of the density contour lines for $\rho_*$ and 
the velocity flow for $(v^x,v^y)$ in the equatorial plane for 
a corotating binary neutron star of $\Gamma=5/3$ 
in nearly quasi-equilibrium states. 
The contour lines are drawn for 
$\rho_*/\rho_{*~{\rm max}}=10^{-0.3j}$, where 
$\rho_{*~{\rm max}}=0.00305$ denotes the maximum value of 
$\rho_*$ at $t=0$, 
for $j=0,1,2,\cdots,10$. Vectors 
indicate the local velocity field and the maximum length denotes 
$\sim 0.23c$. 
}
\end{center}
\end{figure}

\clearpage

\begin{figure}[t]
\epsfxsize=3.in
\leavevmode
\epsffile{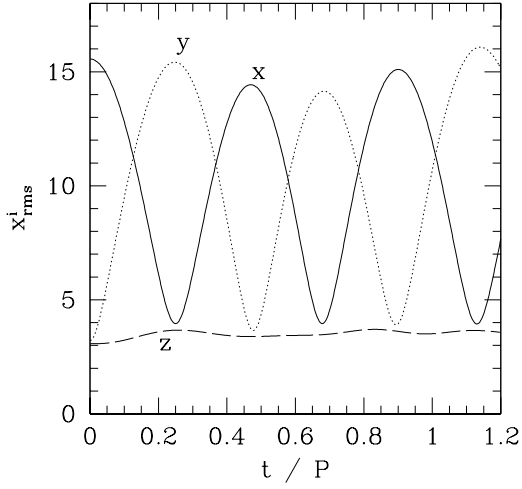}
\caption{$x^i_{\rm rms}$ as a function of $t/{\rm P}$ 
of a corotating binary neutron star in an approximate 
quasi-equilibrium state. 
The solid, dotted and dashed lines denote $x_{\rm rms}$, 
$y_{\rm rms}$, and $z_{\rm rms}$. 
}
\end{figure}

\begin{figure}[t]
\epsfxsize=3.in
\leavevmode
\epsffile{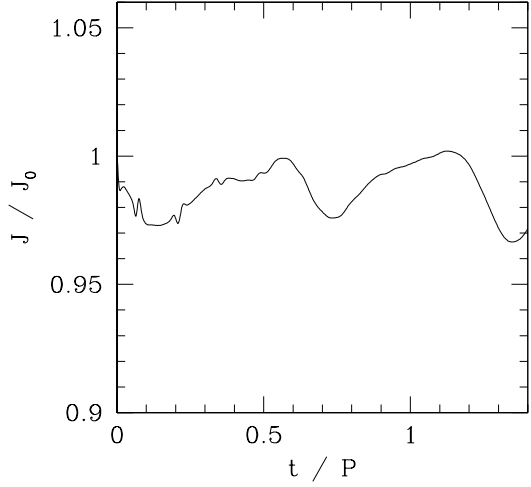}
\caption{$J/J_0$ as a function of $t/{\rm P}$
of a corotating binary neutron star in an approximate 
quasi-equilibrium state.  
}
\end{figure}

\begin{figure}[t]
\epsfxsize=3.in
\leavevmode
\epsffile{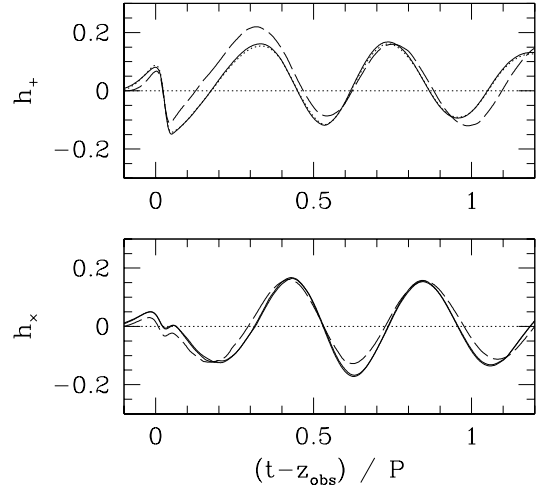}
\caption{$h_+$ and $h_{\times}$ as a function of a retarded 
time of a corotating binary neutron star in an approximate 
quasi-equilibrium state.  
The solid and dotted lines denote those extracted 
at $z_{\rm obs}=106.6$ and 85.3 for $N=116$, respectively, and 
the dashed lines those extracted at $z_{\rm obs}=69.5$ for 
$N=76$. 
}
\end{figure}

\clearpage

\begin{figure}[t]
\begin{center}
\epsfxsize=2.5in
\leavevmode
\epsffile{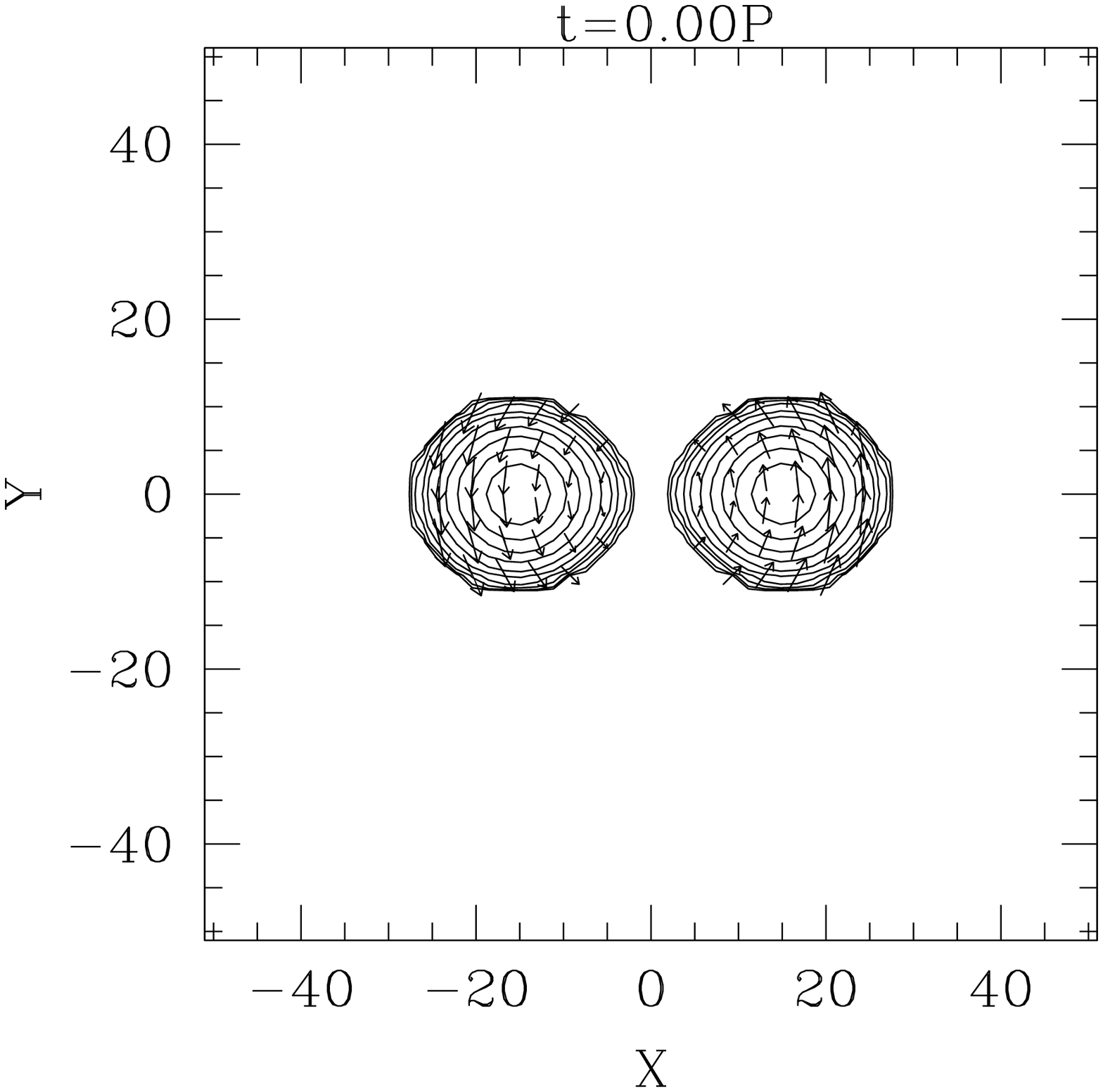}
\epsfxsize=2.5in
\leavevmode
\epsffile{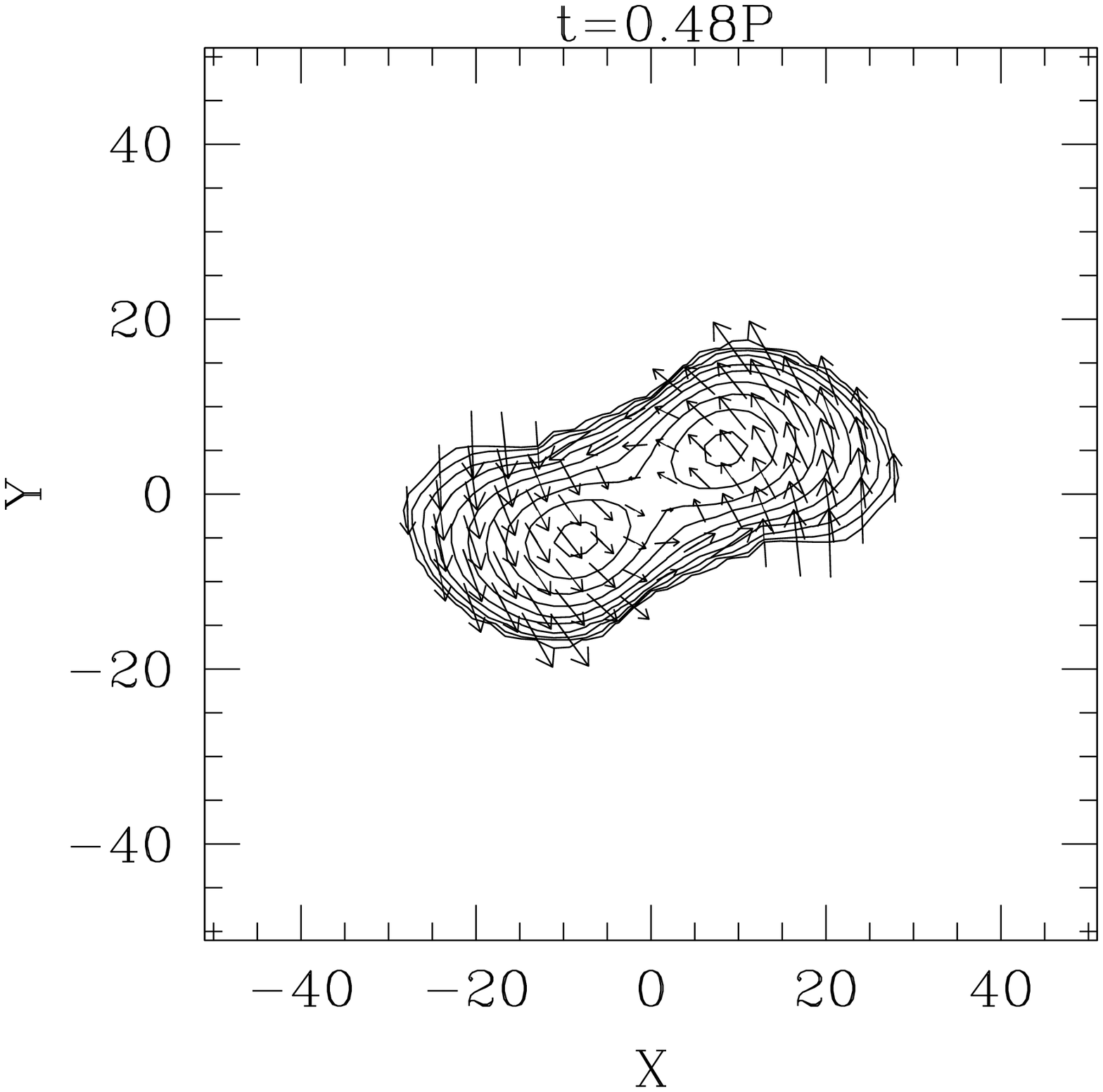}\\
\epsfxsize=2.5in
\leavevmode
\epsffile{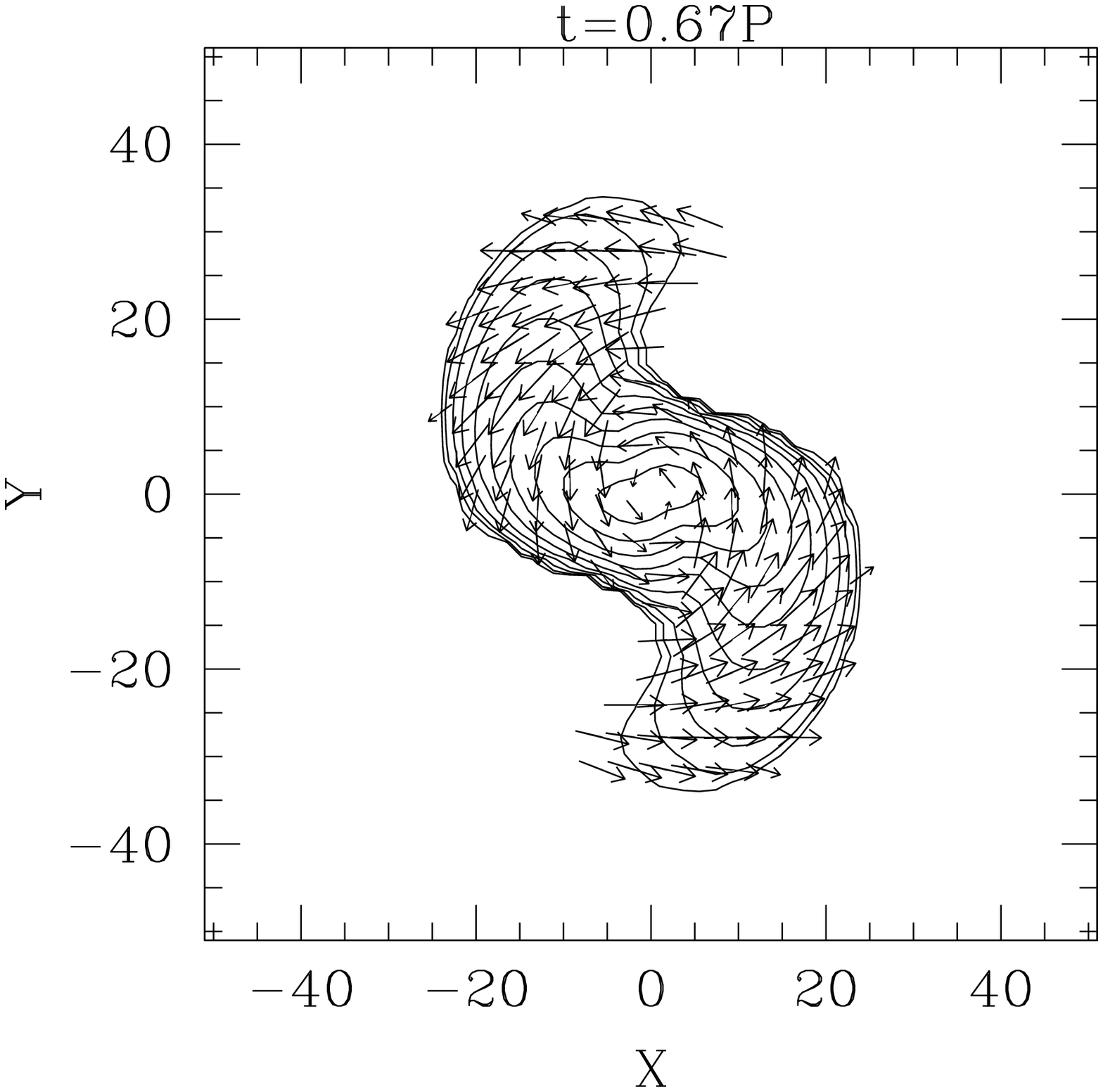}
\epsfxsize=2.5in
\leavevmode
\epsffile{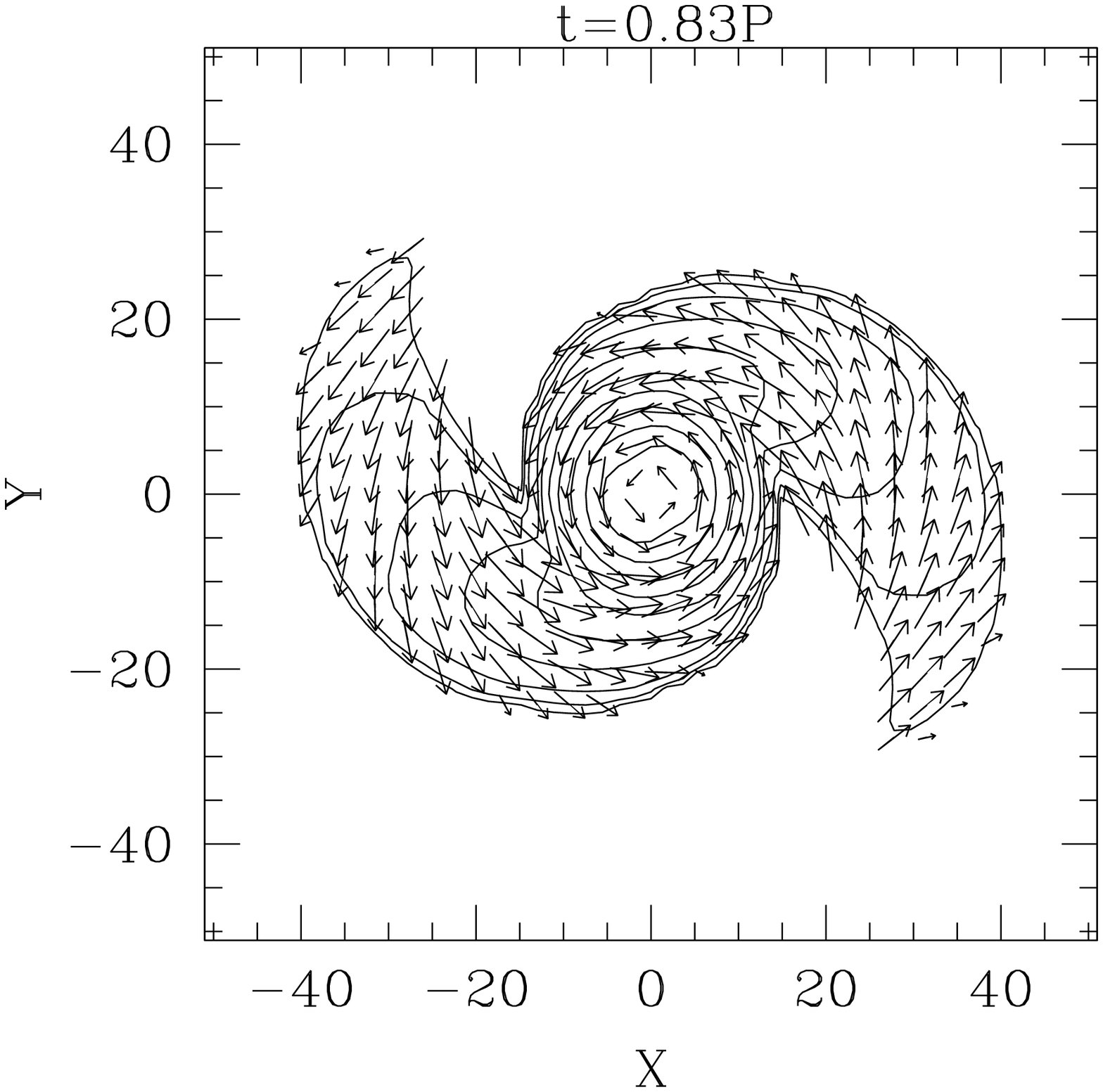}\\
\epsfxsize=2.5in
\leavevmode
\epsffile{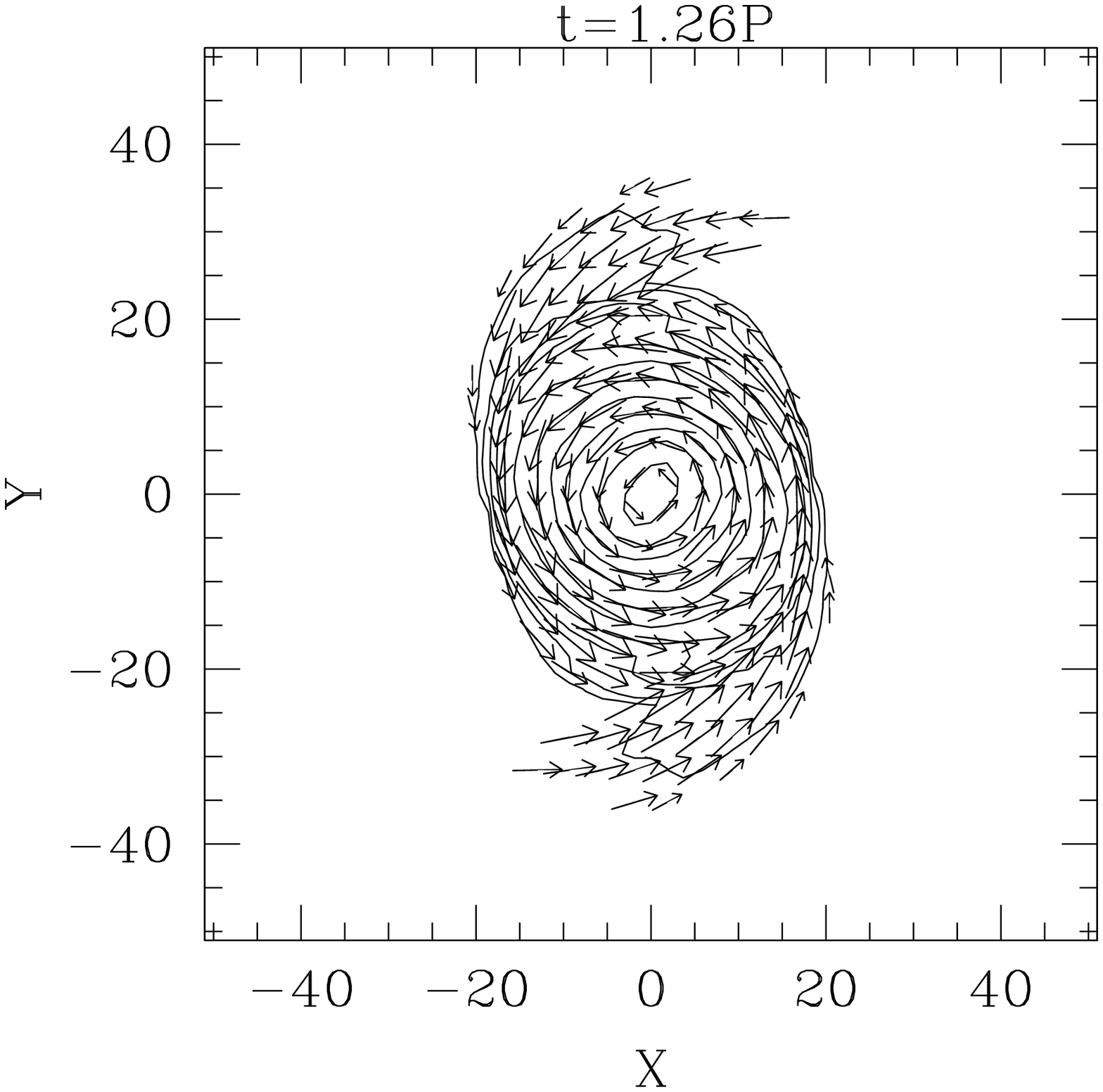}
\epsfxsize=2.5in
\leavevmode
\epsffile{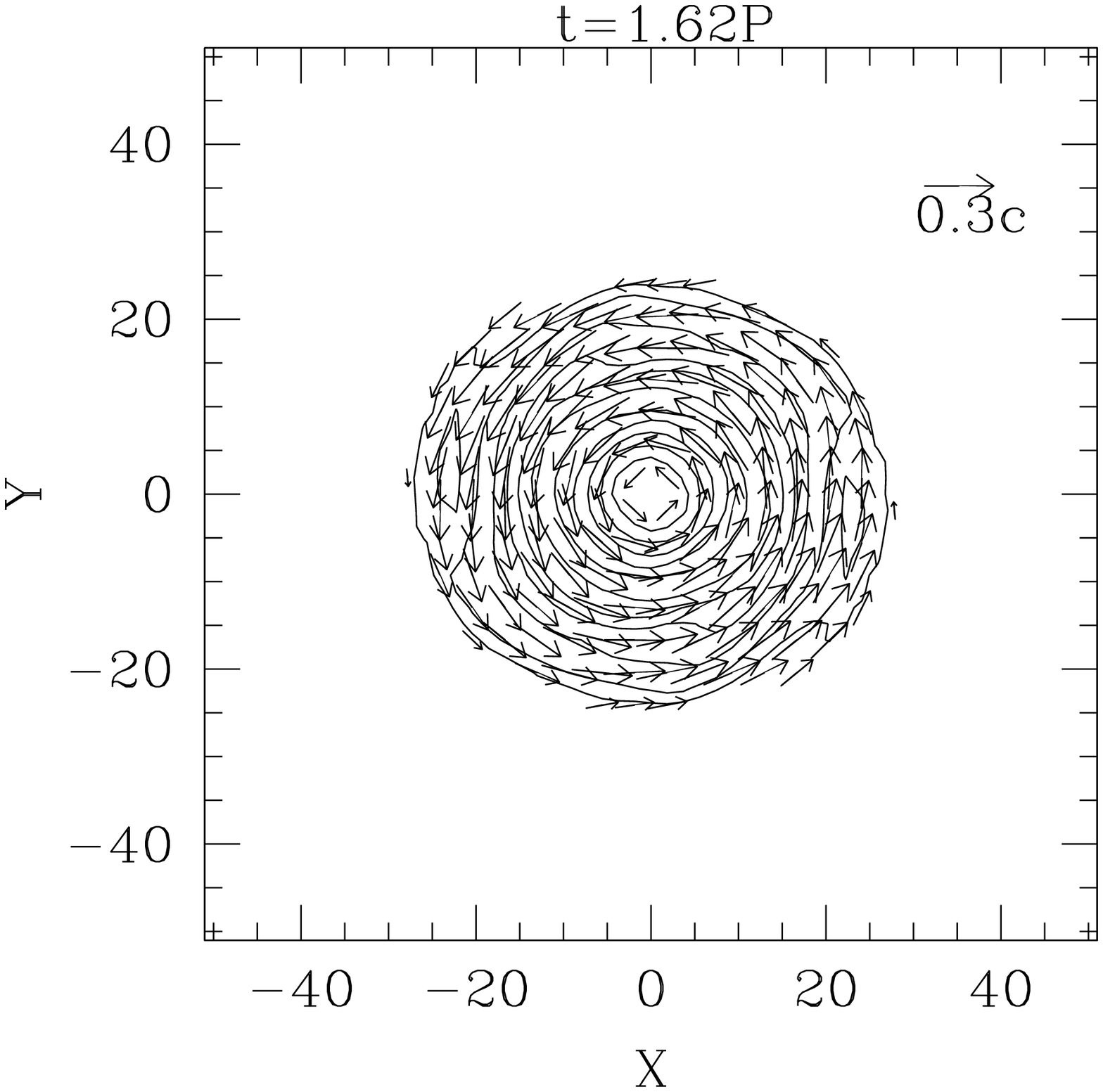}\\
\caption{
Snapshots of the density contour lines for $\rho_*$ and 
the velocity flow for $(v^x,v^y)$ in the equatorial plane for 
a corotating binary neutron star of $\Gamma=5/3$ and 
$\rho_{\rm max}(t=0)=10^{-3}$ for a merging case. 
The contour lines are drawn for 
$\rho_*/\rho_{*~{\rm max}}=10^{-0.3j}$, 
where $\rho_{*~{\rm max}}=0.00305$, for $j=0,1,2,\cdots,10$. 
Vectors indicate the local velocity field and the length 
is shown in normalization of $0.3c$. 
At $t=1.62{\rm P}$, $\rho_{*~\rm max} \simeq 0.011$ 
and $\rho_{\rm max} \simeq 0.0014$, respectively. 
}
\end{center}
\end{figure}

\clearpage

\begin{figure}[t]
\begin{center}
\epsfxsize=2.5in
\leavevmode
\epsffile{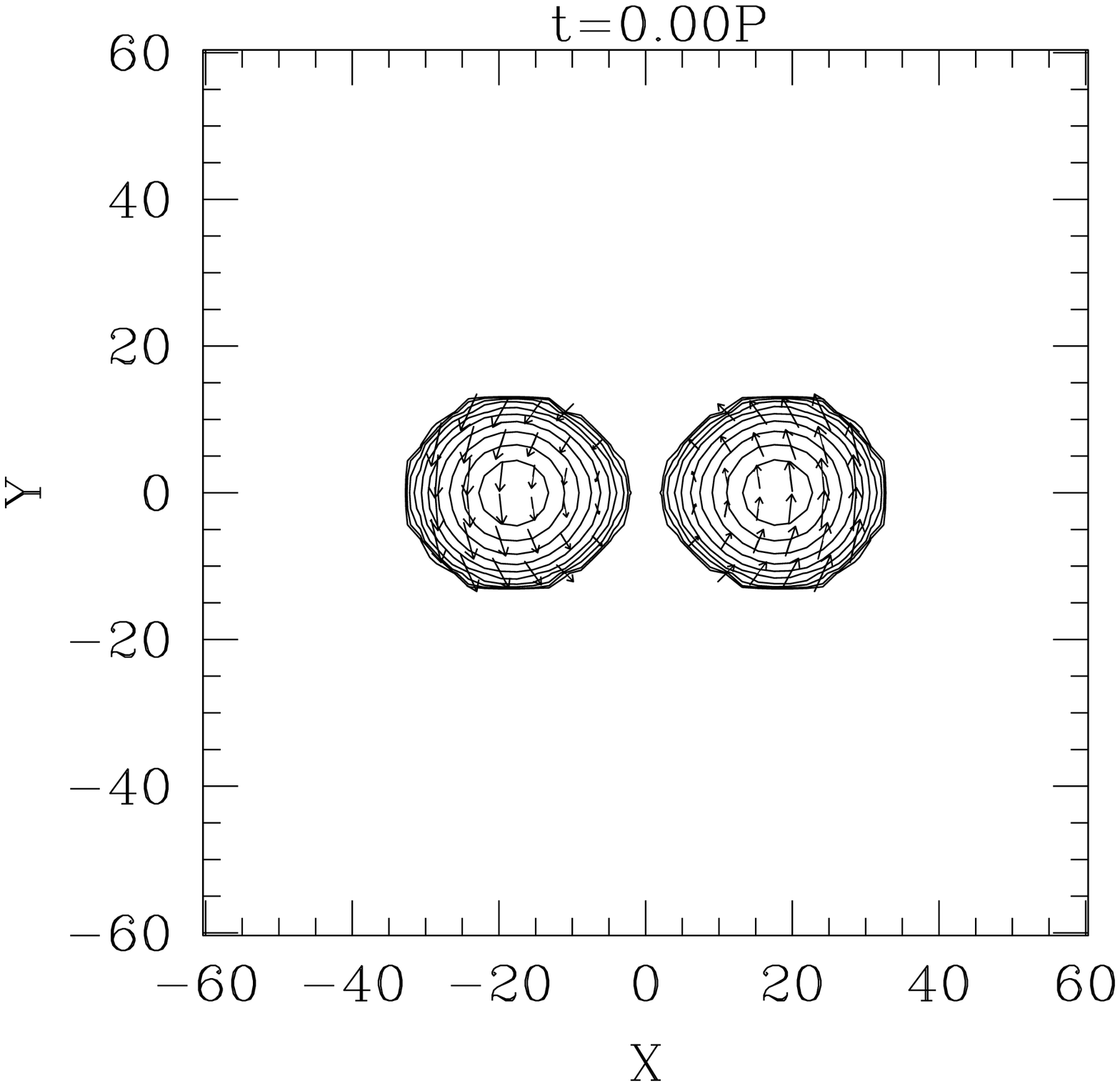}
\epsfxsize=2.5in
\leavevmode
\epsffile{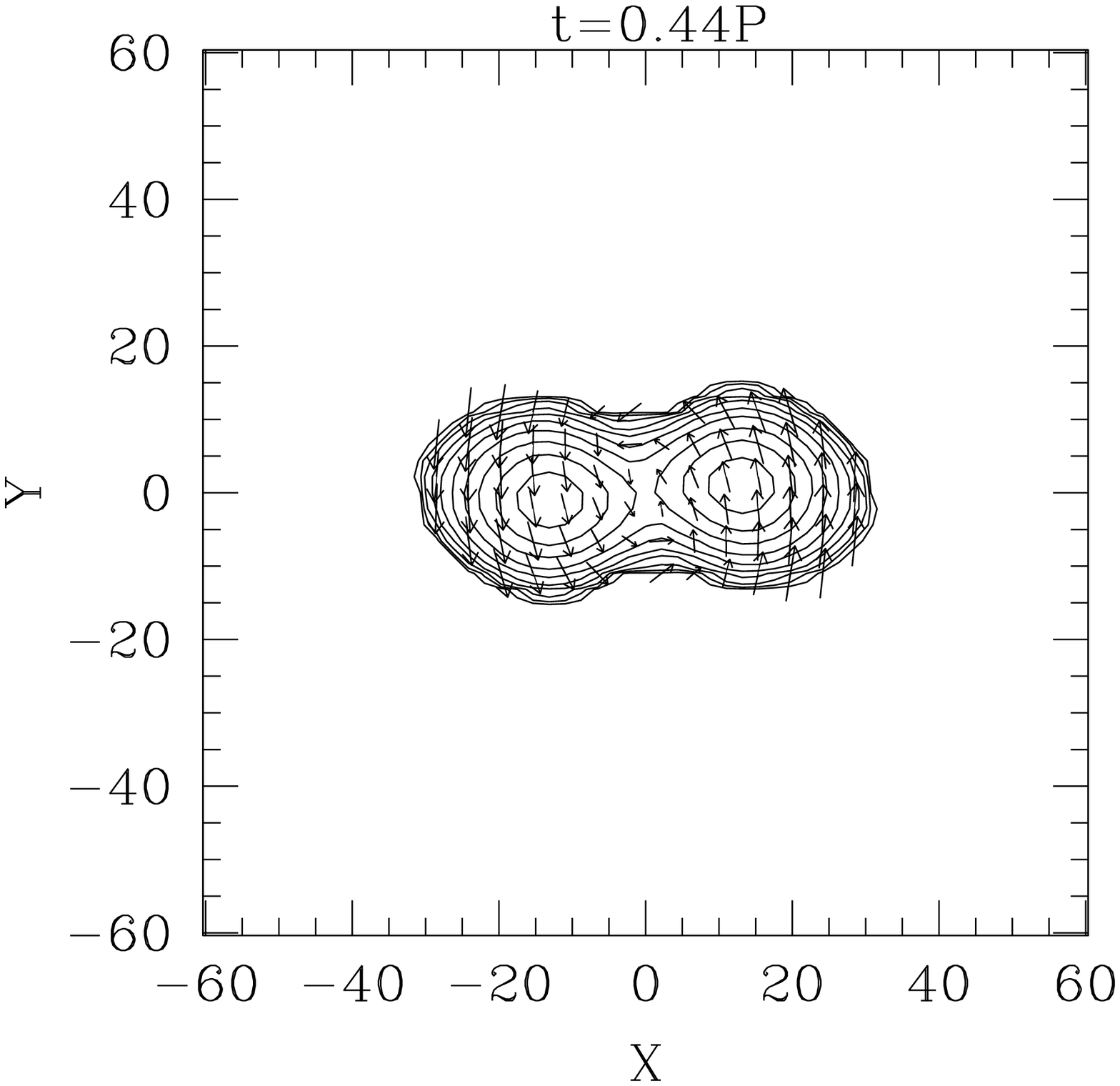}\\
\epsfxsize=2.5in
\leavevmode
\epsffile{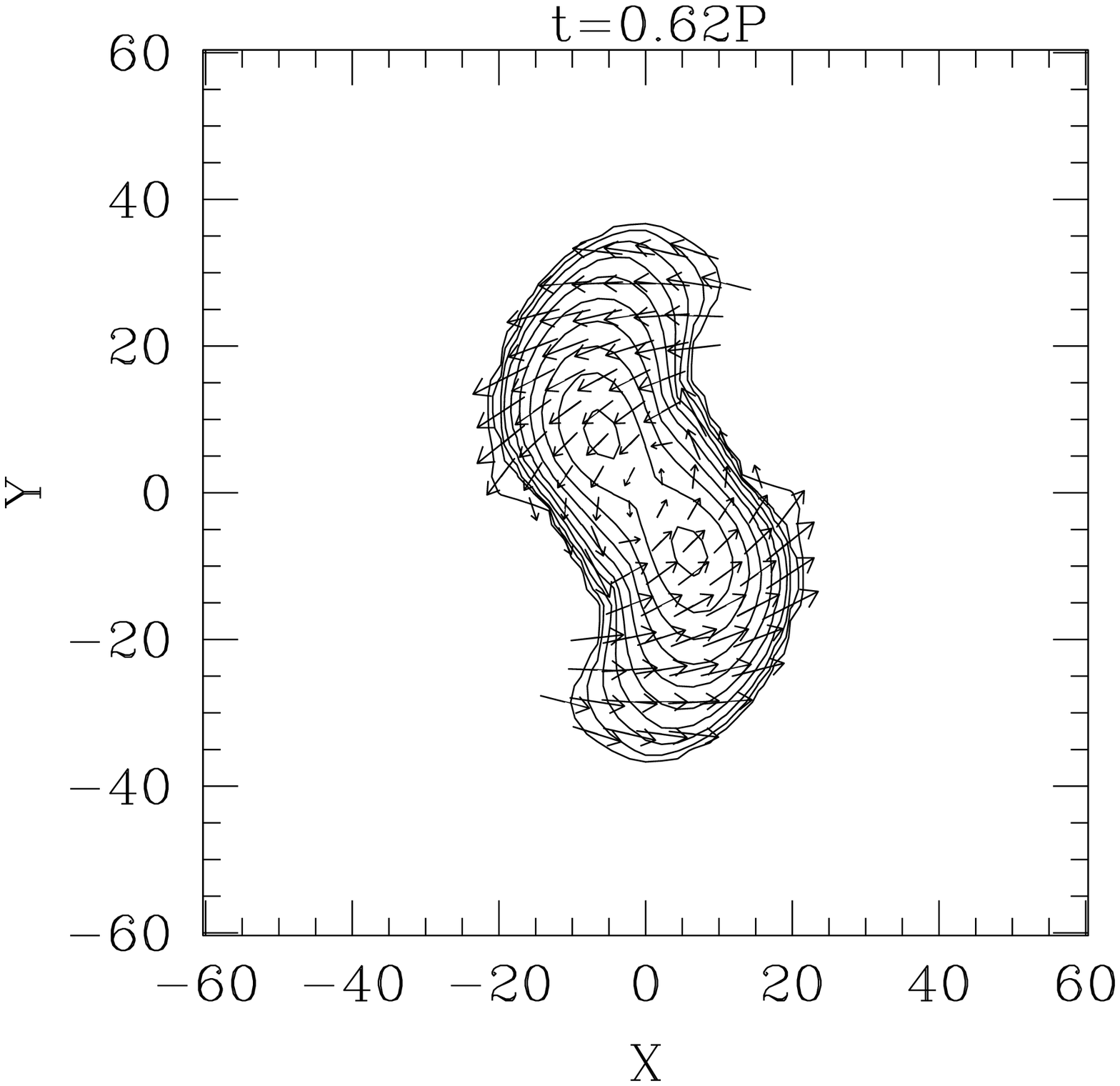}
\epsfxsize=2.5in
\leavevmode
\epsffile{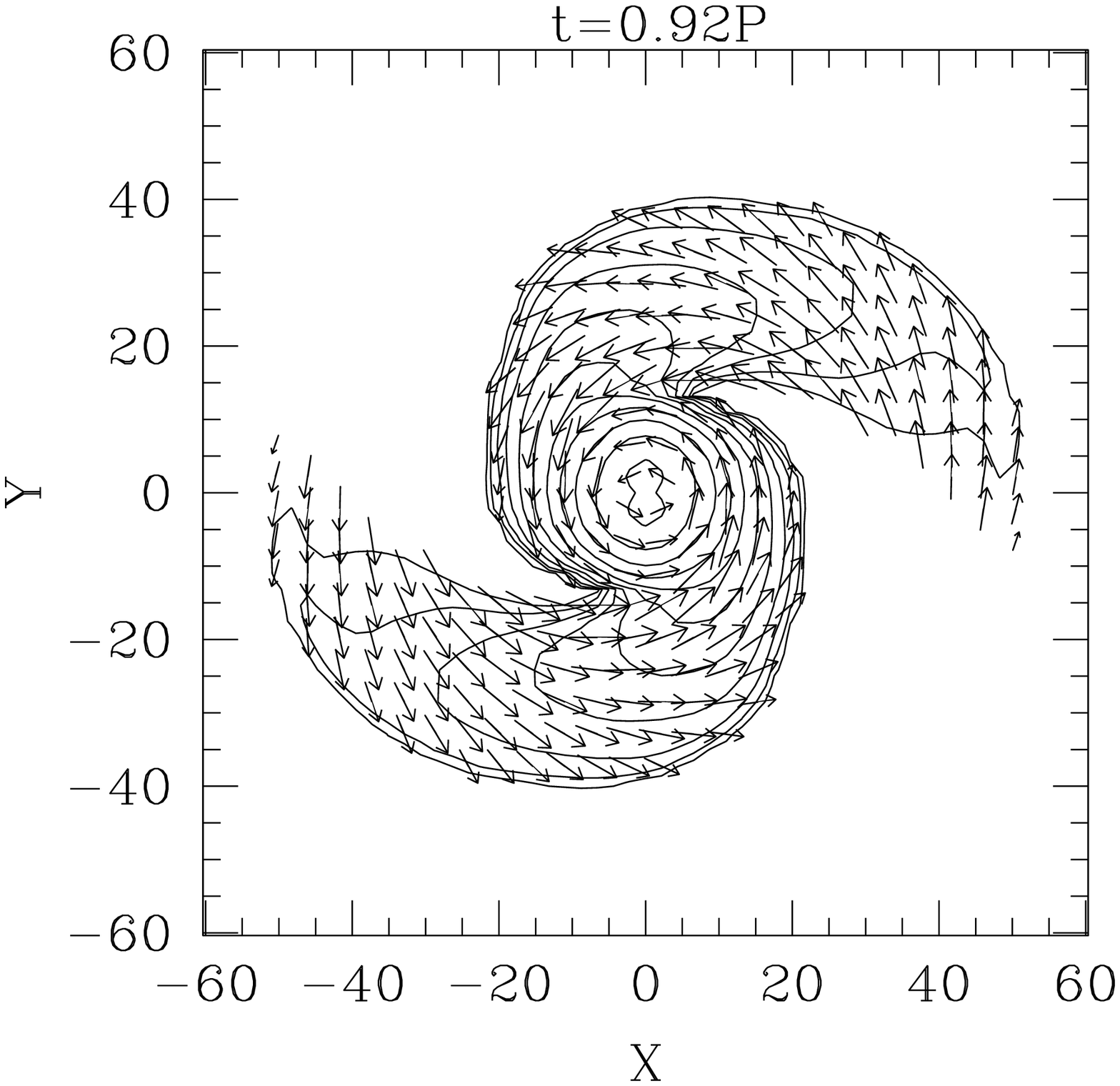}\\
\epsfxsize=2.5in
\leavevmode
\epsffile{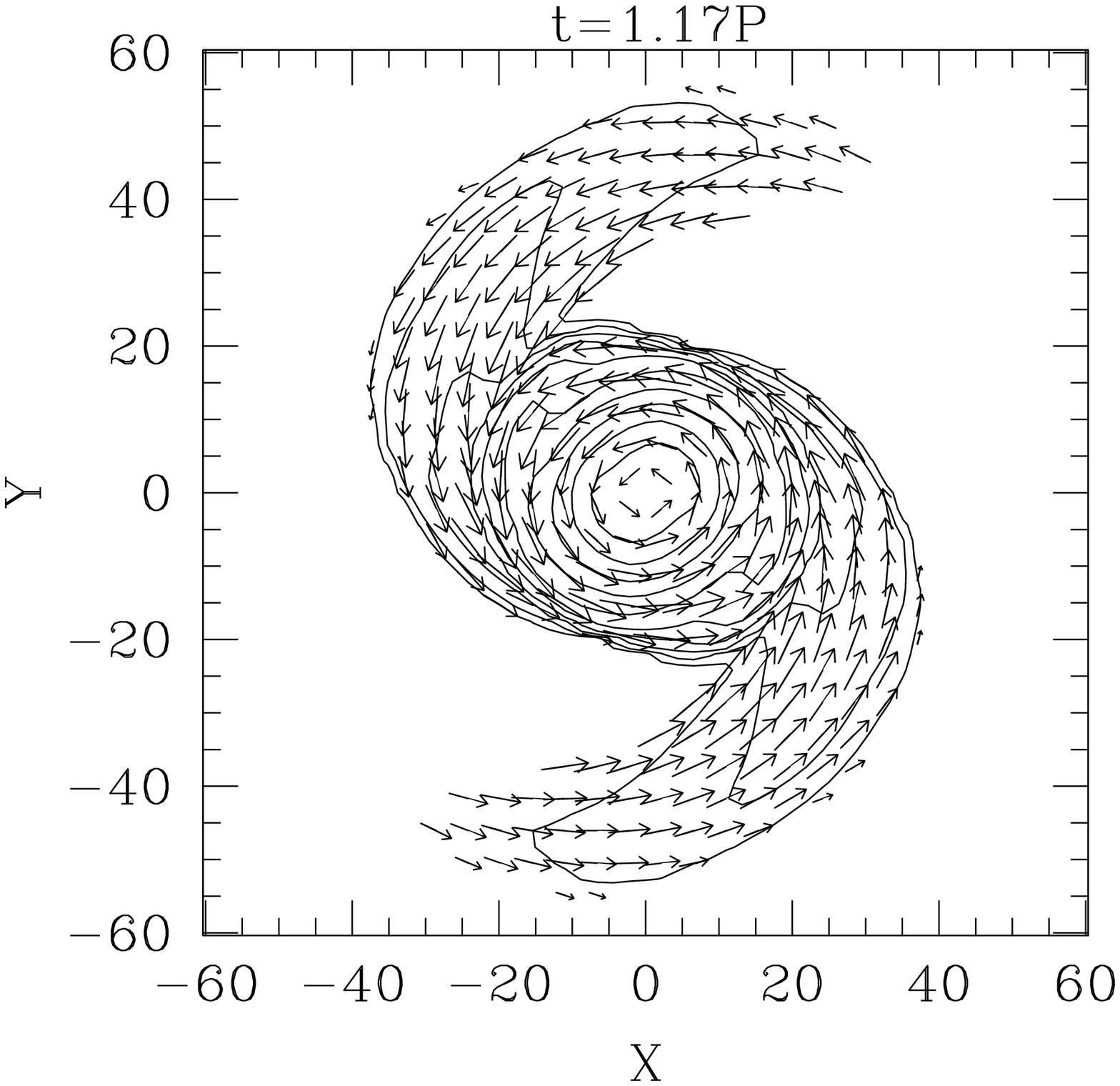}
\epsfxsize=2.5in
\leavevmode
\epsffile{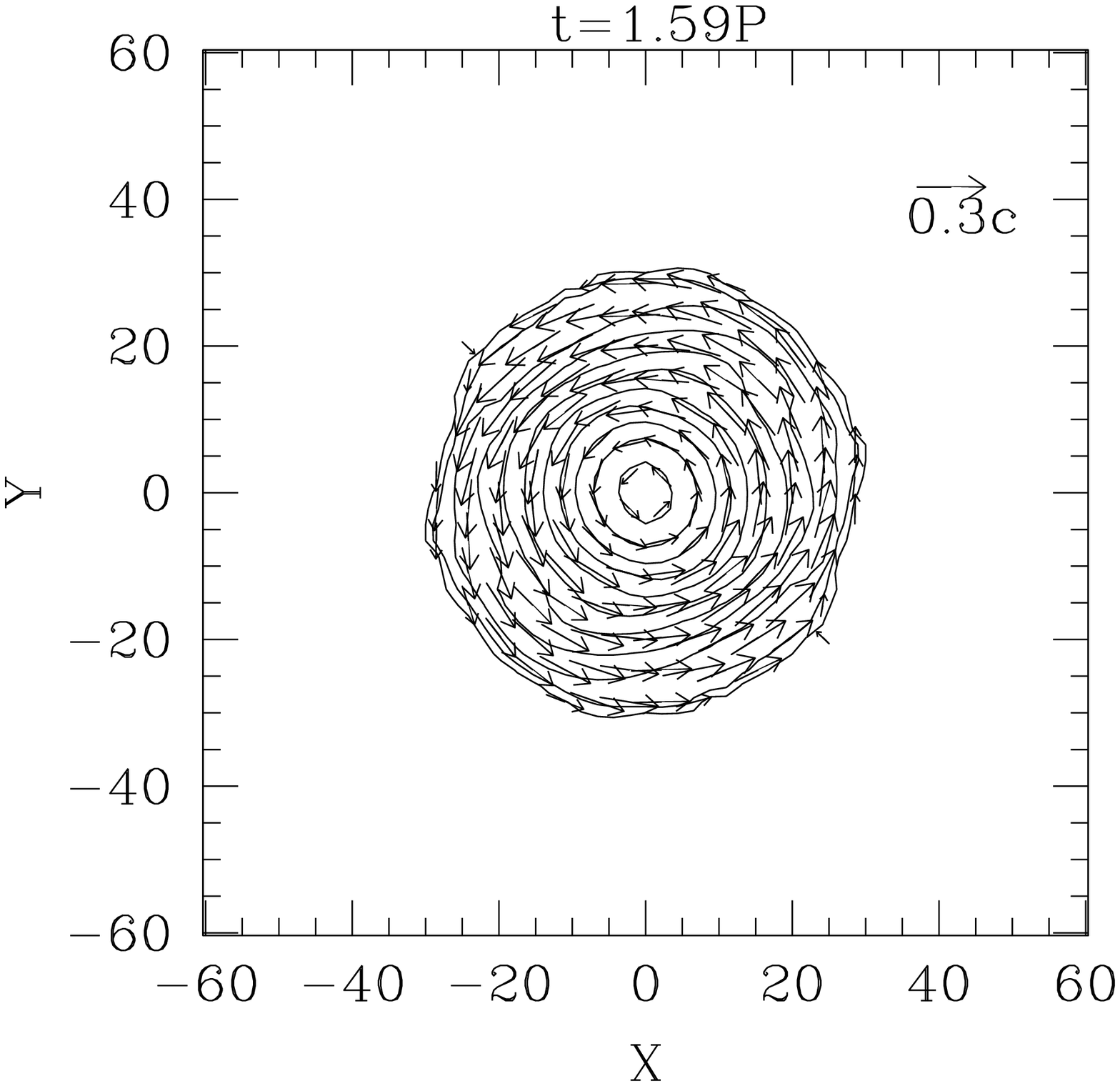}\\
\caption{The same as Fig. 24, but for 
$\rho_{\rm max}(t=0)=6\times 10^{-4}$.
The contour lines are drawn for 
$\rho_*/\rho_{*~{\rm max}}=10^{-0.3j}$, 
where $\rho_{*~{\rm max}}=0.00143$, for $j=0,1,2,\cdots,10$. 
At $t=1.59{\rm P}$, $\rho_* \simeq 0.0019$ and 
$\rho_{\rm max} \simeq 6\times 10^{-4}$, respectively. 
}
\end{center}
\end{figure}

\clearpage

\begin{figure}[t]
\begin{center}
\epsfxsize=2.6in
\leavevmode
\epsffile{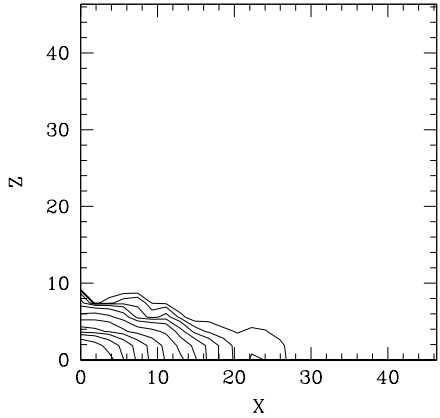}
\caption{
The density contour lines for $\rho_*$ 
in the $y=0$ plane after merger ($t=1.62{\rm P}$) of 
a corotating binary neutron star of $\Gamma=5/3$ and 
$\rho_{\rm max}(t=0)=10^{-3}$. 
The contour lines are drawn for 
$\rho_*/\rho_{*~{\rm max}}=10^{-0.3j}$, 
where $\rho_{*~{\rm max}}=0.00305$ denotes 
$\rho_*$ at $r=0$ and $t=0$, for $j=0,1,2,\cdots,10$. 
}
\end{center}
\end{figure}

\begin{figure}[t]
\begin{center}
\epsfxsize=2.6in
\leavevmode
\epsffile{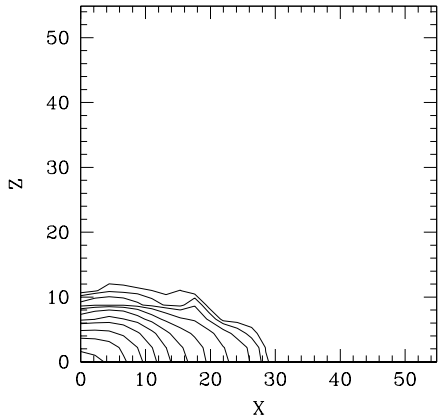}
\caption{
The same as Fig. 26, but for 
$\rho_{\rm max}(t=0)=6 \times 10^{-4}$ at $t=1.59{\rm P}$. 
The contour lines are drawn for 
$\rho_*/\rho_{*~{\rm max}}=10^{-0.3j}$, 
where $\rho_{*~{\rm max}}=0.00143$, for $j=0,1,2,\cdots,10$. 
}
\end{center}
\end{figure}

\begin{figure}[t]
\epsfxsize=3.in
\leavevmode
\epsffile{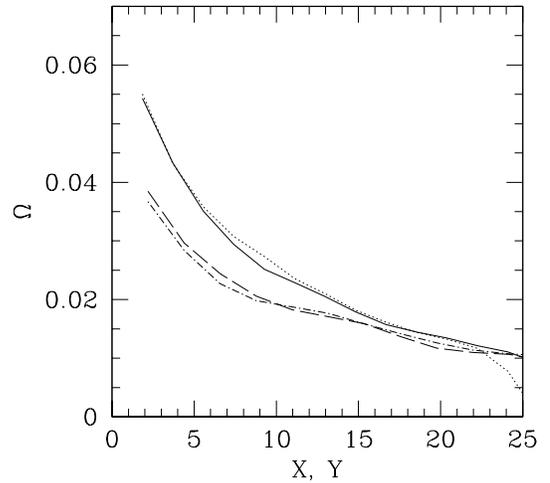}
\caption{The angular velocity along the $x$-axis 
(the solid line) and 
$y$-axis (the dotted line) at $t=1.62{\rm P}$ 
for $\rho_{\rm max}(t=0)=10^{-3}$ and 
along the $x$-axis (the dashed line) and 
$y$-axis (the dotted-dashed line) 
at $t=1.59{\rm P}$ for $\rho_{\rm max}(t=0)=6 \times 10^{-4}$. 
}
\end{figure}

\begin{figure}[t]
\epsfxsize=3.in
\leavevmode
\epsffile{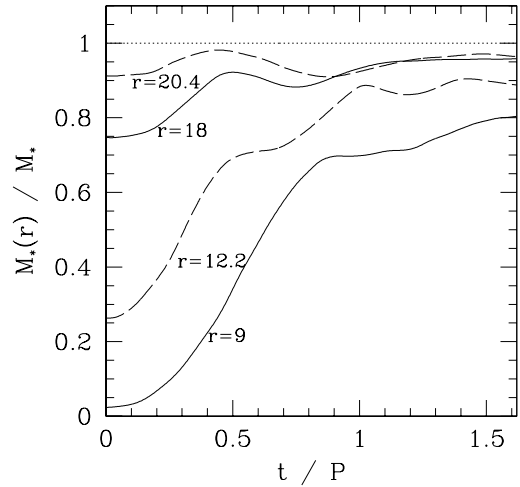}
\caption{Fraction of the rest mass inside a coordinate radius $r$ 
as a function of $t/{\rm P}$ 
for $\rho_{\rm max}(t=0)=10^{-3}$ (the solid lines) and 
$6 \times 10^{-4}$ (the dashed lines). 
}
\end{figure}

\begin{figure}[t]
\epsfxsize=3.in
\leavevmode
\epsffile{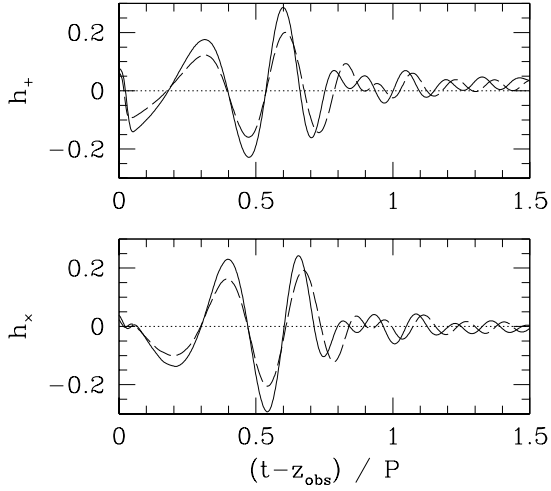}
\caption{$h_+$ and $h_{\times}$ as a function of retarded time 
in the merger of corotating binary neutron stars of 
$\rho_{\rm max}(t=0)=10^{-3}$ (the solid lines) 
and $6\times 10^{-4}$ (dashed lines). 
}
\end{figure}

\begin{figure}[t]
\epsfxsize=3.in
\leavevmode
\epsffile{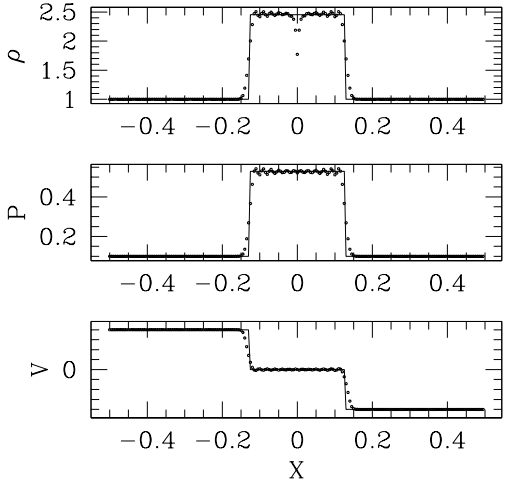}
\caption{Numerical results of $\rho$, $P$, and $v^x$ 
for the 1D wall shock problem in 
special relativity. The solid lines denote the exact solutions, 
and the open circles are numerical results. 
}
\end{figure}

\begin{figure}[t]
\epsfxsize=3.in
\leavevmode
\epsffile{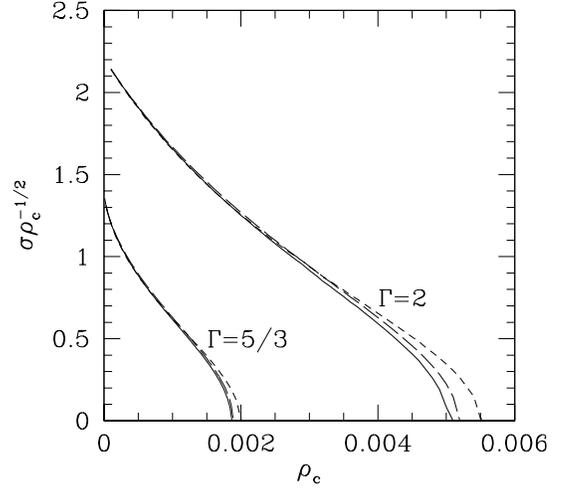}
\caption{The angular frequency ($\sigma \rho_c^{-1/2}$) of the 
fundamental radial oscillation as a function of the central density 
for spherical polytropic stars of $(K,\Gamma)=(10,5/3)$ and 
$(K,\Gamma)=(200/\pi,2)$ 
with the trial functions of the Lagrangian displacement. 
}
\end{figure}


\begin{thebibliography}{99}
\bibitem{Nakamura} 
The idea that $F_i$ should be defined and 
evolved to maintain the stability of numerical system 
was already suggested by T. Nakamura more than 10 
years ago (see \cite{supple}).
\bibitem{foot1} 
In the final phase of these simulations, 
it is found that $\rho$ decreases with time 
in the AMD gauge condition, while $\rho_*$ increases. 
These incorrect solutions are due to the fact that 
$\phi$ and $\rho_*$ inaccurately 
increases (and hence $\rho$ calculated from 
$\rho_* e^{-6\phi}/w$ may decrease).
\end{thebibliography}
\end{document}